\documentclass[a4paper, traditabstract, longauth]{aa} 
%\documentclass[referee, traditabstract, longauth]{aa} 

% for the abstract without structuration 
% (traditional abstract) 
%
\usepackage{graphicx}
\usepackage{color}
\usepackage[breaklinks, colorlinks, citecolor=blue]{hyperref}
\usepackage{natbib}
\usepackage{longtable,lscape}
\usepackage{txfonts}
\usepackage{amsfonts}
\usepackage{amssymb}
\usepackage{newlfont}
\usepackage{textcomp}
\usepackage{times}
\usepackage{units}

\def\setsymbol#1#2{\expandafter\def\csname #1\endcsname{#2}}
\def\getsymbol#1{\csname #1\endcsname}

%-----------------------------------------------------------------------
% Planck
%-----------------------------------------------------------------------
\def\Planck{{\it Planck\/}}

%-----------------------------------------------------------------------
% The Planck Helium-4 JT cooler
%-----------------------------------------------------------------------

%-----------------------------------------------------------------------
% To include all Planck Early Results papers in the reference lists
%-----------------------------------------------------------------------

%-----------------------------------------------------------------------
% Tables
%-----------------------------------------------------------------------
\newbox\tablebox    \newdimen\tablewidth
\def\leaderfil{\leaders\hbox to 5pt{\hss.\hss}\hfil}
%
% use the following definition of \endPlancktable for ApJ style notes to tables, set to the %         width of the table
%\def\endPlancktable{\tablewidth=\wd\tablebox 
%
% use the following definition of \endPlancktable instead for A&A style notes set to full 
%         column width

\def\tablenote#1 #2\par{\begingroup \parindent=0.8em
    \abovedisplayshortskip=0pt\belowdisplayshortskip=0pt
    \noindent
    $$\hss\vbox{\hsize\tablewidth \hangindent=\parindent \hangafter=1 \noindent
    \hbox to \parindent{\sup{\rm #1}\hss}\strut#2\strut\par}\hss$$
    \endgroup}

%-----------------------------------------------------------------------
% useful macros
%-----------------------------------------------------------------------
%
\def\L2{\ifmmode L_2\else $L_2$\fi}

\def\DeltaT{\ifmmode \Delta T\else $\Delta T$\fi}
\def\deltat{\ifmmode \Delta t\else $\Delta t$\fi}
\def\fknee{\ifmmode f_{\rm knee}\else $f_{\rm knee}$\fi}
\def\Fmax{\ifmmode F_{\rm max}\else $F_{\rm max}$\fi}
\def\solar{\ifmmode{\rm M}_{\mathord\odot}\else${\rm M}_{\mathord\odot}$\fi}

\def\inv{\ifmmode^{-1}\else$^{-1}$\fi}
\def\mo{\ifmmode^{-1}\else$^{-1}$\fi}
\def\sup#1{\ifmmode ^{\rm #1}\else $^{\rm #1}$\fi}
\def\expo#1{\ifmmode \times 10^{#1}\else $\times 10^{#1}$\fi}
\def\,{\thinspace}
\def\lsim{\mathrel{\raise .4ex\hbox{\rlap{$<$}\lower 1.2ex\hbox{$\sim$}}}}
\def\gsim{\mathrel{\raise .4ex\hbox{\rlap{$>$}\lower 1.2ex\hbox{$\sim$}}}}

\def\simprop{\mathrel{\raise .4ex\hbox{\rlap{$\propto$}\lower 1.2ex\hbox{$\sim$}}}}
\def\deg{\ifmmode^\circ\else$^\circ$\fi}
\def\pdeg{\ifmmode $\setbox0=\hbox{$^{\circ}$}\rlap{\hskip.11\wd0 .}$^{\circ}
          \else \setbox0=\hbox{$^{\circ}$}\rlap{\hskip.11\wd0 .}$^{\circ}$\fi}
\def\arcs{\ifmmode {^{\scriptstyle\prime\prime}}
          \else $^{\scriptstyle\prime\prime}$\fi}
\def\arcm{\ifmmode {^{\scriptstyle\prime}}
          \else $^{\scriptstyle\prime}$\fi}
\newdimen\sa  \newdimen\sb
\def\parcs{\sa=.07em \sb=.03em
     \ifmmode \hbox{\rlap{.}}^{\scriptstyle\prime\kern -\sb\prime}\hbox{\kern -\sa}
     \else \rlap{.}$^{\scriptstyle\prime\kern -\sb\prime}$\kern -\sa\fi}
\def\parcm{\sa=.08em \sb=.03em
     \ifmmode \hbox{\rlap{.}\kern\sa}^{\scriptstyle\prime}\hbox{\kern-\sb}
     \else \rlap{.}\kern\sa$^{\scriptstyle\prime}$\kern-\sb\fi}
\def\ra[#1 #2 #3.#4]{#1\sup{h}#2\sup{m}#3\sup{s}\llap.#4}
\def\dec[#1 #2 #3.#4]{#1\deg#2\arcm#3\arcs\llap.#4}
\def\deco[#1 #2 #3]{#1\deg#2\arcm#3\arcs}
\def\rra[#1 #2]{#1\sup{h}#2\sup{m}}

\def\dots{\relax\ifmmode \ldots\else $\ldots$\fi}
%
%-----------------------------------------------------------------------
% units
%-----------------------------------------------------------------------
%
\def\WHzsr{\ifmmode $W\,Hz\mo\,sr\mo$\else W\,Hz\mo\,sr\mo\fi}
\def\mHz{\ifmmode $\,mHz$\else \,mHz\fi}
\def\GHz{\ifmmode $\,GHz$\else \,GHz\fi}
\def\mKs{\ifmmode $\,mK\,s$^{1/2}\else \,mK\,s$^{1/2}$\fi}
\def\muKs{\ifmmode \,\mu$K\,s$^{1/2}\else \,$\mu$K\,s$^{1/2}$\fi}
\def\muKRJs{\ifmmode \,\mu$K$_{\rm RJ}$\,s$^{1/2}\else \,$\mu$K$_{\rm RJ}$\,s$^{1/2}$\fi}
\def\muKHz{\ifmmode \,\mu$K\,Hz$^{-1/2}\else \,$\mu$K\,Hz$^{-1/2}$\fi}
\def\MJysr{\ifmmode \,$MJy\,sr\mo$\else \,MJy\,sr\mo\fi}
\def\MJysrmK{\ifmmode \,$MJy\,sr\mo$\,mK$_{\rm CMB}\mo\else \,MJy\,sr\mo\,mK$_{\rm CMB}\mo$\fi}
\def\microns{\ifmmode \,\mu$m$\else \,$\mu$m\fi}

\def\muK{\ifmmode \,\mu$K$\else \,$\mu$\hbox{K}\fi}
\def\microK{\ifmmode \,\mu$K$\else \,$\mu$\hbox{K}\fi}
\def\muW{\ifmmode \,\mu$W$\else \,$\mu$\hbox{W}\fi}
\def\kms{\ifmmode $\,km\,s$^{-1}\else \,km\,s$^{-1}$\fi}
\def\kmsMpc{\ifmmode $\,\kms\,Mpc\mo$\else \,\kms\,Mpc\mo\fi}
%
%
%----------------------------------------------------------------------

% LFI Center Frequency

\setsymbol{LFI:center:frequency:70GHz:units}{70.3\,GHz}
\setsymbol{LFI:center:frequency:44GHz:units}{44.1\,GHz}
\setsymbol{LFI:center:frequency:30GHz:units}{28.5\,GHz}

\setsymbol{LFI:center:frequency:70GHz}{70.3}
\setsymbol{LFI:center:frequency:44GHz}{44.1}
\setsymbol{LFI:center:frequency:30GHz}{28.5}

\setsymbol{LFI:center:frequency:LFI18:Rad:M:units}{71.7\GHz}
\setsymbol{LFI:center:frequency:LFI19:Rad:M:units}{67.5\GHz}
\setsymbol{LFI:center:frequency:LFI20:Rad:M:units}{69.2\GHz}
\setsymbol{LFI:center:frequency:LFI21:Rad:M:units}{70.4\GHz}
\setsymbol{LFI:center:frequency:LFI22:Rad:M:units}{71.5\GHz}
\setsymbol{LFI:center:frequency:LFI23:Rad:M:units}{70.8\GHz}
\setsymbol{LFI:center:frequency:LFI24:Rad:M:units}{44.4\GHz}
\setsymbol{LFI:center:frequency:LFI25:Rad:M:units}{44.0\GHz}
\setsymbol{LFI:center:frequency:LFI26:Rad:M:units}{43.9\GHz}
\setsymbol{LFI:center:frequency:LFI27:Rad:M:units}{28.3\GHz}
\setsymbol{LFI:center:frequency:LFI28:Rad:M:units}{28.8\GHz}
\setsymbol{LFI:center:frequency:LFI18:Rad:S:units}{70.1\GHz}
\setsymbol{LFI:center:frequency:LFI19:Rad:S:units}{69.6\GHz}
\setsymbol{LFI:center:frequency:LFI20:Rad:S:units}{69.5\GHz}
\setsymbol{LFI:center:frequency:LFI21:Rad:S:units}{69.5\GHz}
\setsymbol{LFI:center:frequency:LFI22:Rad:S:units}{72.8\GHz}
\setsymbol{LFI:center:frequency:LFI23:Rad:S:units}{71.3\GHz}
\setsymbol{LFI:center:frequency:LFI24:Rad:S:units}{44.1\GHz}
\setsymbol{LFI:center:frequency:LFI25:Rad:S:units}{44.1\GHz}
\setsymbol{LFI:center:frequency:LFI26:Rad:S:units}{44.1\GHz}
\setsymbol{LFI:center:frequency:LFI27:Rad:S:units}{28.5\GHz}
\setsymbol{LFI:center:frequency:LFI28:Rad:S:units}{28.2\GHz}

\setsymbol{LFI:center:frequency:LFI18:Rad:M}{71.7}
\setsymbol{LFI:center:frequency:LFI19:Rad:M}{67.5}
\setsymbol{LFI:center:frequency:LFI20:Rad:M}{69.2}
\setsymbol{LFI:center:frequency:LFI21:Rad:M}{70.4}
\setsymbol{LFI:center:frequency:LFI22:Rad:M}{71.5}
\setsymbol{LFI:center:frequency:LFI23:Rad:M}{70.8}
\setsymbol{LFI:center:frequency:LFI24:Rad:M}{44.4}
\setsymbol{LFI:center:frequency:LFI25:Rad:M}{44.0}
\setsymbol{LFI:center:frequency:LFI26:Rad:M}{43.9}
\setsymbol{LFI:center:frequency:LFI27:Rad:M}{28.3}
\setsymbol{LFI:center:frequency:LFI28:Rad:M}{28.8}
\setsymbol{LFI:center:frequency:LFI18:Rad:S}{70.1}
\setsymbol{LFI:center:frequency:LFI19:Rad:S}{69.6}
\setsymbol{LFI:center:frequency:LFI20:Rad:S}{69.5}
\setsymbol{LFI:center:frequency:LFI21:Rad:S}{69.5}
\setsymbol{LFI:center:frequency:LFI22:Rad:S}{72.8}
\setsymbol{LFI:center:frequency:LFI23:Rad:S}{71.3}
\setsymbol{LFI:center:frequency:LFI24:Rad:S}{44.1}
\setsymbol{LFI:center:frequency:LFI25:Rad:S}{44.1}
\setsymbol{LFI:center:frequency:LFI26:Rad:S}{44.1}
\setsymbol{LFI:center:frequency:LFI27:Rad:S}{28.5}
\setsymbol{LFI:center:frequency:LFI28:Rad:S}{28.2}

% LFI White Noise Sensitivity, \delta T_{\rm RJ}

\setsymbol{LFI:white:noise:sensitivity:70GHz:units}{134.7\muKs}
\setsymbol{LFI:white:noise:sensitivity:44GHz:units}{164.7\muKs}
\setsymbol{LFI:white:noise:sensitivity:30GHz:units}{143.4\muKs}

\setsymbol{LFI:white:noise:sensitivity:70GHz}{134.7}
\setsymbol{LFI:white:noise:sensitivity:44GHz}{164.7}
\setsymbol{LFI:white:noise:sensitivity:30GHz}{143.4}

%***************the following are in \delta T_{\rm CMB} units!****************
%***************this segment should be revised********************************

\setsymbol{LFI:white:noise:sensitivity:LFI18:Rad:M:units}{512.0\muKs}
\setsymbol{LFI:white:noise:sensitivity:LFI19:Rad:M:units}{581.4\muKs}
\setsymbol{LFI:white:noise:sensitivity:LFI20:Rad:M:units}{590.8\muKs}
\setsymbol{LFI:white:noise:sensitivity:LFI21:Rad:M:units}{455.2\muKs}
\setsymbol{LFI:white:noise:sensitivity:LFI22:Rad:M:units}{492.0\muKs}
\setsymbol{LFI:white:noise:sensitivity:LFI23:Rad:M:units}{507.7\muKs}
\setsymbol{LFI:white:noise:sensitivity:LFI24:Rad:M:units}{462.2\muKs}
\setsymbol{LFI:white:noise:sensitivity:LFI25:Rad:M:units}{413.6\muKs}
\setsymbol{LFI:white:noise:sensitivity:LFI26:Rad:M:units}{478.6\muKs}
\setsymbol{LFI:white:noise:sensitivity:LFI27:Rad:M:units}{277.7\muKs}
\setsymbol{LFI:white:noise:sensitivity:LFI28:Rad:M:units}{312.3\muKs}
\setsymbol{LFI:white:noise:sensitivity:LFI18:Rad:S:units}{465.7\muKs}
\setsymbol{LFI:white:noise:sensitivity:LFI19:Rad:S:units}{555.6\muKs}
\setsymbol{LFI:white:noise:sensitivity:LFI20:Rad:S:units}{623.2\muKs}
\setsymbol{LFI:white:noise:sensitivity:LFI21:Rad:S:units}{564.1\muKs}
\setsymbol{LFI:white:noise:sensitivity:LFI22:Rad:S:units}{534.4\muKs}
\setsymbol{LFI:white:noise:sensitivity:LFI23:Rad:S:units}{542.4\muKs}
\setsymbol{LFI:white:noise:sensitivity:LFI24:Rad:S:units}{399.2\muKs}
\setsymbol{LFI:white:noise:sensitivity:LFI25:Rad:S:units}{392.6\muKs}
\setsymbol{LFI:white:noise:sensitivity:LFI26:Rad:S:units}{418.6\muKs}
\setsymbol{LFI:white:noise:sensitivity:LFI27:Rad:S:units}{302.9\muKs}
\setsymbol{LFI:white:noise:sensitivity:LFI28:Rad:S:units}{285.3\muKs}

\setsymbol{LFI:white:noise:sensitivity:LFI18:Rad:M}{512.0}
\setsymbol{LFI:white:noise:sensitivity:LFI19:Rad:M}{581.4}
\setsymbol{LFI:white:noise:sensitivity:LFI20:Rad:M}{590.8}
\setsymbol{LFI:white:noise:sensitivity:LFI21:Rad:M}{455.2}
\setsymbol{LFI:white:noise:sensitivity:LFI22:Rad:M}{492.0}
\setsymbol{LFI:white:noise:sensitivity:LFI23:Rad:M}{507.7}
\setsymbol{LFI:white:noise:sensitivity:LFI24:Rad:M}{462.2}
\setsymbol{LFI:white:noise:sensitivity:LFI25:Rad:M}{413.6}
\setsymbol{LFI:white:noise:sensitivity:LFI26:Rad:M}{478.6}
\setsymbol{LFI:white:noise:sensitivity:LFI27:Rad:M}{277.7}
\setsymbol{LFI:white:noise:sensitivity:LFI28:Rad:M}{312.3}
\setsymbol{LFI:white:noise:sensitivity:LFI18:Rad:S}{465.7}
\setsymbol{LFI:white:noise:sensitivity:LFI19:Rad:S}{555.6}
\setsymbol{LFI:white:noise:sensitivity:LFI20:Rad:S}{623.2}
\setsymbol{LFI:white:noise:sensitivity:LFI21:Rad:S}{564.1}
\setsymbol{LFI:white:noise:sensitivity:LFI22:Rad:S}{534.4}
\setsymbol{LFI:white:noise:sensitivity:LFI23:Rad:S}{542.4}
\setsymbol{LFI:white:noise:sensitivity:LFI24:Rad:S}{399.2}
\setsymbol{LFI:white:noise:sensitivity:LFI25:Rad:S}{392.6}
\setsymbol{LFI:white:noise:sensitivity:LFI26:Rad:S}{418.6}
\setsymbol{LFI:white:noise:sensitivity:LFI27:Rad:S}{302.9}
\setsymbol{LFI:white:noise:sensitivity:LFI28:Rad:S}{285.3}

% LFI Knee Frequency

\setsymbol{LFI:knee:frequency:70GHz:units}{29.5\mHz}
\setsymbol{LFI:knee:frequency:44GHz:units}{56.2\mHz}
\setsymbol{LFI:knee:frequency:30GHz:units}{113.7\mHz}

\setsymbol{LFI:knee:frequency:70GHz}{29.5}
\setsymbol{LFI:knee:frequency:44GHz}{56.2}
\setsymbol{LFI:knee:frequency:30GHz}{113.7}

\setsymbol{LFI:knee:frequency:LFI18:Rad:M:units}{16.3\mHz}
\setsymbol{LFI:knee:frequency:LFI19:Rad:M:units}{15.1\mHz}
\setsymbol{LFI:knee:frequency:LFI20:Rad:M:units}{18.7\mHz}
\setsymbol{LFI:knee:frequency:LFI21:Rad:M:units}{37.2\mHz}
\setsymbol{LFI:knee:frequency:LFI22:Rad:M:units}{12.7\mHz}
\setsymbol{LFI:knee:frequency:LFI23:Rad:M:units}{34.6\mHz}
\setsymbol{LFI:knee:frequency:LFI24:Rad:M:units}{46.2\mHz}
\setsymbol{LFI:knee:frequency:LFI25:Rad:M:units}{24.9\mHz}
\setsymbol{LFI:knee:frequency:LFI26:Rad:M:units}{67.6\mHz}
\setsymbol{LFI:knee:frequency:LFI27:Rad:M:units}{187.4\mHz}
\setsymbol{LFI:knee:frequency:LFI28:Rad:M:units}{122.2\mHz}
\setsymbol{LFI:knee:frequency:LFI18:Rad:S:units}{17.7\mHz}
\setsymbol{LFI:knee:frequency:LFI19:Rad:S:units}{22.0\mHz}
\setsymbol{LFI:knee:frequency:LFI20:Rad:S:units}{8.7\mHz}
\setsymbol{LFI:knee:frequency:LFI21:Rad:S:units}{25.9\mHz}
\setsymbol{LFI:knee:frequency:LFI22:Rad:S:units}{15.8\mHz}
\setsymbol{LFI:knee:frequency:LFI23:Rad:S:units}{129.8\mHz}
\setsymbol{LFI:knee:frequency:LFI24:Rad:S:units}{100.9\mHz}
\setsymbol{LFI:knee:frequency:LFI25:Rad:S:units}{38.9\mHz}
\setsymbol{LFI:knee:frequency:LFI26:Rad:S:units}{58.9\mHz}
\setsymbol{LFI:knee:frequency:LFI27:Rad:S:units}{104.4\mHz}
\setsymbol{LFI:knee:frequency:LFI28:Rad:S:units}{40.7\mHz}

\setsymbol{LFI:knee:frequency:LFI18:Rad:M}{16.3}
\setsymbol{LFI:knee:frequency:LFI19:Rad:M}{15.1}
\setsymbol{LFI:knee:frequency:LFI20:Rad:M}{18.7}
\setsymbol{LFI:knee:frequency:LFI21:Rad:M}{37.2}
\setsymbol{LFI:knee:frequency:LFI22:Rad:M}{12.7}
\setsymbol{LFI:knee:frequency:LFI23:Rad:M}{34.6}
\setsymbol{LFI:knee:frequency:LFI24:Rad:M}{46.2}
\setsymbol{LFI:knee:frequency:LFI25:Rad:M}{24.9}
\setsymbol{LFI:knee:frequency:LFI26:Rad:M}{67.6}
\setsymbol{LFI:knee:frequency:LFI27:Rad:M}{187.4}
\setsymbol{LFI:knee:frequency:LFI28:Rad:M}{122.2}
\setsymbol{LFI:knee:frequency:LFI18:Rad:S}{17.7}
\setsymbol{LFI:knee:frequency:LFI19:Rad:S}{22.0}
\setsymbol{LFI:knee:frequency:LFI20:Rad:S}{8.7}
\setsymbol{LFI:knee:frequency:LFI21:Rad:S}{25.9}
\setsymbol{LFI:knee:frequency:LFI22:Rad:S}{15.8}
\setsymbol{LFI:knee:frequency:LFI23:Rad:S}{129.8}
\setsymbol{LFI:knee:frequency:LFI24:Rad:S}{100.9}
\setsymbol{LFI:knee:frequency:LFI25:Rad:S}{38.9}
\setsymbol{LFI:knee:frequency:LFI26:Rad:S}{58.9}
\setsymbol{LFI:knee:frequency:LFI27:Rad:S}{104.4}
\setsymbol{LFI:knee:frequency:LFI28:Rad:S}{40.7}

% LFI low frequency noise slope

\setsymbol{LFI:slope:70GHz:units}{$-1.03$\mHz}
\setsymbol{LFI:slope:44GHz:units}{$-0.89$\mHz}
\setsymbol{LFI:slope:30GHz:units}{$-0.87$\mHz}

\setsymbol{LFI:slope:70GHz}{$-1.03$}
\setsymbol{LFI:slope:44GHz}{$-0.89$}
\setsymbol{LFI:slope:30GHz}{$-0.87$}

\setsymbol{LFI:slope:LFI18:Rad:M:units}{$-1.04$\mHz}
\setsymbol{LFI:slope:LFI19:Rad:M:units}{$-1.09$\mHz}
\setsymbol{LFI:slope:LFI20:Rad:M:units}{$-0.69$\mHz}
\setsymbol{LFI:slope:LFI21:Rad:M:units}{$-1.56$\mHz}
\setsymbol{LFI:slope:LFI22:Rad:M:units}{$-1.01$\mHz}
\setsymbol{LFI:slope:LFI23:Rad:M:units}{$-0.96$\mHz}
\setsymbol{LFI:slope:LFI24:Rad:M:units}{$-0.83$\mHz}
\setsymbol{LFI:slope:LFI25:Rad:M:units}{$-0.91$\mHz}
\setsymbol{LFI:slope:LFI26:Rad:M:units}{$-0.95$\mHz}
\setsymbol{LFI:slope:LFI27:Rad:M:units}{$-0.87$\mHz}
\setsymbol{LFI:slope:LFI28:Rad:M:units}{$-0.88$\mHz}
\setsymbol{LFI:slope:LFI18:Rad:S:units}{$-1.15$\mHz}
\setsymbol{LFI:slope:LFI19:Rad:S:units}{$-1.00$\mHz}
\setsymbol{LFI:slope:LFI20:Rad:S:units}{$-0.95$\mHz}
\setsymbol{LFI:slope:LFI21:Rad:S:units}{$-0.92$\mHz}
\setsymbol{LFI:slope:LFI22:Rad:S:units}{$-1.01$\mHz}
\setsymbol{LFI:slope:LFI23:Rad:S:units}{$-0.95$\mHz}
\setsymbol{LFI:slope:LFI24:Rad:S:units}{$-0.73$\mHz}
\setsymbol{LFI:slope:LFI25:Rad:S:units}{$-1.16$\mHz}
\setsymbol{LFI:slope:LFI26:Rad:S:units}{$-0.79$\mHz}
\setsymbol{LFI:slope:LFI27:Rad:S:units}{$-0.82$\mHz}
\setsymbol{LFI:slope:LFI28:Rad:S:units}{$-0.91$\mHz}

\setsymbol{LFI:slope:LFI18:Rad:M}{$-1.04$}
\setsymbol{LFI:slope:LFI19:Rad:M}{$-1.09$}
\setsymbol{LFI:slope:LFI20:Rad:M}{$-0.69$}
\setsymbol{LFI:slope:LFI21:Rad:M}{$-1.56$}
\setsymbol{LFI:slope:LFI22:Rad:M}{$-1.01$}
\setsymbol{LFI:slope:LFI23:Rad:M}{$-0.96$}
\setsymbol{LFI:slope:LFI24:Rad:M}{$-0.83$}
\setsymbol{LFI:slope:LFI25:Rad:M}{$-0.91$}
\setsymbol{LFI:slope:LFI26:Rad:M}{$-0.95$}
\setsymbol{LFI:slope:LFI27:Rad:M}{$-0.87$}
\setsymbol{LFI:slope:LFI28:Rad:M}{$-0.88$}
\setsymbol{LFI:slope:LFI18:Rad:S}{$-1.15$}
\setsymbol{LFI:slope:LFI19:Rad:S}{$-1.00$}
\setsymbol{LFI:slope:LFI20:Rad:S}{$-0.95$}
\setsymbol{LFI:slope:LFI21:Rad:S}{$-0.92$}
\setsymbol{LFI:slope:LFI22:Rad:S}{$-1.01$}
\setsymbol{LFI:slope:LFI23:Rad:S}{$-0.95$}
\setsymbol{LFI:slope:LFI24:Rad:S}{$-0.73$}
\setsymbol{LFI:slope:LFI25:Rad:S}{$-1.16$}
\setsymbol{LFI:slope:LFI26:Rad:S}{$-0.79$}
\setsymbol{LFI:slope:LFI27:Rad:S}{$-0.82$}
\setsymbol{LFI:slope:LFI28:Rad:S}{$-0.91$}

% LFI Beam FWHM

\setsymbol{LFI:FWHM:70GHz:units}{13\parcm01}
\setsymbol{LFI:FWHM:44GHz:units}{27\parcm92}
\setsymbol{LFI:FWHM:30GHz:units}{32\parcm65}

\setsymbol{LFI:FWHM:70GHz}{13.01}
\setsymbol{LFI:FWHM:44GHz}{27.92}
\setsymbol{LFI:FWHM:30GHz}{32.65}

\setsymbol{LFI:FWHM:LFI18:units}{13\parcm39}
\setsymbol{LFI:FWHM:LFI19:units}{13\parcm01}
\setsymbol{LFI:FWHM:LFI20:units}{12\parcm75}
\setsymbol{LFI:FWHM:LFI21:units}{12\parcm74}
\setsymbol{LFI:FWHM:LFI22:units}{12\parcm87}
\setsymbol{LFI:FWHM:LFI23:units}{13\parcm27}
\setsymbol{LFI:FWHM:LFI24:units}{22\parcm98}
\setsymbol{LFI:FWHM:LFI25:units}{30\parcm46}
\setsymbol{LFI:FWHM:LFI26:units}{30\parcm31}
\setsymbol{LFI:FWHM:LFI27:units}{32\parcm65}
\setsymbol{LFI:FWHM:LFI28:units}{32\parcm66}

\setsymbol{LFI:FWHM:LFI18}{13.39}
\setsymbol{LFI:FWHM:LFI19}{13.01}
\setsymbol{LFI:FWHM:LFI20}{12.75}
\setsymbol{LFI:FWHM:LFI21}{12.74}
\setsymbol{LFI:FWHM:LFI22}{12.87}
\setsymbol{LFI:FWHM:LFI23}{13.27}
\setsymbol{LFI:FWHM:LFI24}{22.98}
\setsymbol{LFI:FWHM:LFI25}{30.46}
\setsymbol{LFI:FWHM:LFI26}{30.31}
\setsymbol{LFI:FWHM:LFI27}{32.65}
\setsymbol{LFI:FWHM:LFI28}{32.66}

% LFI Beam FWHM Uncertainty
% When uncertainties are routinely available for all quantities, we'll likely change the format to build them into 
% the \setsymbol command.  For now, this will be a bit easier.

%\setsymbol{LFI:FWHM:uncertainty:70GHz}{TBD\arcm}
%\setsymbol{LFI:FWHM:uncertainty:44GHz}{TBD\arcm}
%\setsymbol{LFI:FWHM:uncertainty:30GHz}{TBD\arcm}

\setsymbol{LFI:FWHM:uncertainty:LFI18:units}{0.170\arcm}
\setsymbol{LFI:FWHM:uncertainty:LFI19:units}{0.174\arcm}
\setsymbol{LFI:FWHM:uncertainty:LFI20:units}{0.170\arcm}
\setsymbol{LFI:FWHM:uncertainty:LFI21:units}{0.156\arcm}
\setsymbol{LFI:FWHM:uncertainty:LFI22:units}{0.164\arcm}
\setsymbol{LFI:FWHM:uncertainty:LFI23:units}{0.171\arcm}
\setsymbol{LFI:FWHM:uncertainty:LFI24:units}{0.652\arcm}
\setsymbol{LFI:FWHM:uncertainty:LFI25:units}{1.075\arcm}
\setsymbol{LFI:FWHM:uncertainty:LFI26:units}{1.131\arcm}
\setsymbol{LFI:FWHM:uncertainty:LFI27:units}{1.266\arcm}
\setsymbol{LFI:FWHM:uncertainty:LFI28:units}{1.287\arcm}

\setsymbol{LFI:FWHM:uncertainty:LFI18}{0.170}
\setsymbol{LFI:FWHM:uncertainty:LFI19}{0.174}
\setsymbol{LFI:FWHM:uncertainty:LFI20}{0.170}
\setsymbol{LFI:FWHM:uncertainty:LFI21}{0.156}
\setsymbol{LFI:FWHM:uncertainty:LFI22}{0.164}
\setsymbol{LFI:FWHM:uncertainty:LFI23}{0.171}
\setsymbol{LFI:FWHM:uncertainty:LFI24}{0.652}
\setsymbol{LFI:FWHM:uncertainty:LFI25}{1.075}
\setsymbol{LFI:FWHM:uncertainty:LFI26}{1.131}
\setsymbol{LFI:FWHM:uncertainty:LFI27}{1.266}
\setsymbol{LFI:FWHM:uncertainty:LFI28}{1.287}

% HFI Center Frequency

\setsymbol{HFI:center:frequency:100GHz:units}{100\,GHz}
\setsymbol{HFI:center:frequency:143GHz:units}{143\,GHz}
\setsymbol{HFI:center:frequency:217GHz:units}{217\,GHz}
\setsymbol{HFI:center:frequency:353GHz:units}{353\,GHz}
\setsymbol{HFI:center:frequency:545GHz:units}{545\,GHz}
\setsymbol{HFI:center:frequency:857GHz:units}{857\,GHz}

\setsymbol{HFI:center:frequency:100GHz}{100}
\setsymbol{HFI:center:frequency:143GHz}{143}
\setsymbol{HFI:center:frequency:217GHz}{217}
\setsymbol{HFI:center:frequency:353GHz}{353}
\setsymbol{HFI:center:frequency:545GHz}{545}
\setsymbol{HFI:center:frequency:857GHz}{857}

% HFI Number of Detectors

\setsymbol{HFI:Ndetectors:100GHz}{8}
\setsymbol{HFI:Ndetectors:143GHz}{11}
\setsymbol{HFI:Ndetectors:217GHz}{12}
\setsymbol{HFI:Ndetectors:353GHz}{12}
\setsymbol{HFI:Ndetectors:545GHz}{3}
\setsymbol{HFI:Ndetectors:857GHz}{4}

% HFI FWHM on maps

\setsymbol{HFI:FWHM:Maps:100GHz:units}{9\parcm88}
\setsymbol{HFI:FWHM:Maps:143GHz:units}{7\parcm18}
\setsymbol{HFI:FWHM:Maps:217GHz:units}{4\parcm87}
\setsymbol{HFI:FWHM:Maps:353GHz:units}{4\parcm65}
\setsymbol{HFI:FWHM:Maps:545GHz:units}{4\parcm72}
\setsymbol{HFI:FWHM:Maps:857GHz:units}{4\parcm39}
\setsymbol{HFI:FWHM:Maps:100GHz}{9.88}
\setsymbol{HFI:FWHM:Maps:143GHz}{7.18}
\setsymbol{HFI:FWHM:Maps:217GHz}{4.87}
\setsymbol{HFI:FWHM:Maps:353GHz}{4.65}
\setsymbol{HFI:FWHM:Maps:545GHz}{4.72}
\setsymbol{HFI:FWHM:Maps:857GHz}{4.39}

% HFI Beam Ellipticity on maps

\setsymbol{HFI:beam:ellipticity:Maps:100GHz}{1.15}
\setsymbol{HFI:beam:ellipticity:Maps:143GHz}{1.01}
\setsymbol{HFI:beam:ellipticity:Maps:217GHz}{1.06}
\setsymbol{HFI:beam:ellipticity:Maps:353GHz}{1.05}
\setsymbol{HFI:beam:ellipticity:Maps:545GHz}{1.14}
\setsymbol{HFI:beam:ellipticity:Maps:857GHz}{1.19}

% HFI optical Beam FWHM from Mars; time response deconvolved: frequency  average of values in table 4 in HFI instrument paper

\setsymbol{HFI:FWHM:Mars:100GHz:units}{9\parcm37}
\setsymbol{HFI:FWHM:Mars:143GHz:units}{7\parcm04}
\setsymbol{HFI:FWHM:Mars:217GHz:units}{4\parcm68}
\setsymbol{HFI:FWHM:Mars:353GHz:units}{4\parcm43}
\setsymbol{HFI:FWHM:Mars:545GHz:units}{3\parcm80}
\setsymbol{HFI:FWHM:Mars:857GHz:units}{3\parcm67}

\setsymbol{HFI:FWHM:Mars:100GHz}{9.37}
\setsymbol{HFI:FWHM:Mars:143GHz}{7.04}
\setsymbol{HFI:FWHM:Mars:217GHz}{4.68}
\setsymbol{HFI:FWHM:Mars:353GHz}{4.43}
\setsymbol{HFI:FWHM:Mars:545GHz}{3.80}
\setsymbol{HFI:FWHM:Mars:857GHz}{3.67}

% HFI optical Beam Ellipticity from Mars; time response deconvolved: frequency average of values in table 4 in HFI instrument paper

\setsymbol{HFI:beam:ellipticity:Mars:100GHz}{1.18}
\setsymbol{HFI:beam:ellipticity:Mars:143GHz}{1.03}
\setsymbol{HFI:beam:ellipticity:Mars:217GHz}{1.14}
\setsymbol{HFI:beam:ellipticity:Mars:353GHz}{1.09}
\setsymbol{HFI:beam:ellipticity:Mars:545GHz}{1.25}
\setsymbol{HFI:beam:ellipticity:Mars:857GHz}{1.03}

% HFI CMB relative calibration accuracy

\setsymbol{HFI:CMB:relative:calibration:100GHz}{$\lsim 1\%$}
\setsymbol{HFI:CMB:relative:calibration:143GHz}{$\lsim 1\%$}
\setsymbol{HFI:CMB:relative:calibration:217GHz}{$\lsim 1\%$}
\setsymbol{HFI:CMB:relative:calibration:353GHz}{$\lsim 1\%$}
\setsymbol{HFI:CMB:relative:calibration:545GHz}{}
\setsymbol{HFI:CMB:relative:calibration:857GHz}{}

% HFI CMB absolute calibration accuracy

\setsymbol{HFI:CMB:absolute:calibration:100GHz}{$\lsim 2\%$}
\setsymbol{HFI:CMB:absolute:calibration:143GHz}{$\lsim 2\%$}
\setsymbol{HFI:CMB:absolute:calibration:217GHz}{$\lsim 2\%$}
\setsymbol{HFI:CMB:absolute:calibration:353GHz}{$\lsim 2\%$}
\setsymbol{HFI:CMB:absolute:calibration:545GHz}{}
\setsymbol{HFI:CMB:absolute:calibration:857GHz}{}

% HFI FIRAS gain calibration accuracy: statistical

\setsymbol{HFI:FIRAS:gain:calibration:accuracy:statistical:100GHz}{}
\setsymbol{HFI:FIRAS:gain:calibration:accuracy:statistical:143GHz}{}
\setsymbol{HFI:FIRAS:gain:calibration:accuracy:statistical:217GHz}{}
\setsymbol{HFI:FIRAS:gain:calibration:accuracy:statistical:353GHz}{2.5\%}
\setsymbol{HFI:FIRAS:gain:calibration:accuracy:statistical:545GHz}{1\%}
\setsymbol{HFI:FIRAS:gain:calibration:accuracy:statistical:857GHz}{0.5\%}

% HFI FIRAS gain calibration accuracy: systematic

\setsymbol{HFI:FIRAS:gain:calibration:accuracy:systematic:100GHz}{}
\setsymbol{HFI:FIRAS:gain:calibration:accuracy:systematic:143GHz}{}
\setsymbol{HFI:FIRAS:gain:calibration:accuracy:systematic:217GHz}{}
\setsymbol{HFI:FIRAS:gain:calibration:accuracy:systematic:353GHz}{}
\setsymbol{HFI:FIRAS:gain:calibration:accuracy:systematic:545GHz}{7\%}
\setsymbol{HFI:FIRAS:gain:calibration:accuracy:systematic:857GHz}{7\%}

% HFI FIRAS zero point accuracy:

\setsymbol{HFI:FIRAS:zero:point:accuracy:100GHz:units}{0.8\MJysr}
\setsymbol{HFI:FIRAS:zero:point:accuracy:143GHz:units}{}
\setsymbol{HFI:FIRAS:zero:point:accuracy:217GHz:units}{}
\setsymbol{HFI:FIRAS:zero:point:accuracy:353GHz:units}{1.4\MJysr}
\setsymbol{HFI:FIRAS:zero:point:accuracy:545GHz:units}{2.2\MJysr}
\setsymbol{HFI:FIRAS:zero:point:accuracy:857GHz:units}{1.7\MJysr}

\setsymbol{HFI:FIRAS:zero:point:accuracy:100GHz}{0.8}
\setsymbol{HFI:FIRAS:zero:point:accuracy:143GHz}{}
\setsymbol{HFI:FIRAS:zero:point:accuracy:217GHz}{}
\setsymbol{HFI:FIRAS:zero:point:accuracy:353GHz}{1.4}
\setsymbol{HFI:FIRAS:zero:point:accuracy:545GHz}{2.2}
\setsymbol{HFI:FIRAS:zero:point:accuracy:857GHz}{1.7}

% HFI diffuse source sensitivity unit conversion

\setsymbol{HFI:unit:conversion:100GHz:units}{0.2415\MJysrmK}
\setsymbol{HFI:unit:conversion:143GHz:units}{0.3694\MJysrmK}
\setsymbol{HFI:unit:conversion:217GHz:units}{0.4811\MJysrmK}
\setsymbol{HFI:unit:conversion:353GHz:units}{0.2883\MJysrmK}
\setsymbol{HFI:unit:conversion:545GHz:units}{0.05826\MJysrmK}
\setsymbol{HFI:unit:conversion:857GHz:units}{0.002238\MJysrmK}

\setsymbol{HFI:unit:conversion:100GHz}{0.2415}
\setsymbol{HFI:unit:conversion:143GHz}{0.3694}
\setsymbol{HFI:unit:conversion:217GHz}{0.4811}
\setsymbol{HFI:unit:conversion:353GHz}{0.2883}
\setsymbol{HFI:unit:conversion:545GHz}{0.05826}
\setsymbol{HFI:unit:conversion:857GHz}{0.002238}

% HFI Colour Correction for \alpha = -2, for V1.01 of the spectral bands

\setsymbol{HFI:colour:correction:alpha=-2:V1.01:100GHz}{0.9893}
\setsymbol{HFI:colour:correction:alpha=-2:V1.01:143GHz}{0.9759}
\setsymbol{HFI:colour:correction:alpha=-2:V1.01:217GHz}{1.0007}
\setsymbol{HFI:colour:correction:alpha=-2:V1.01:353GHz}{1.0028}
\setsymbol{HFI:colour:correction:alpha=-2:V1.01:545GHz}{1.0019}
\setsymbol{HFI:colour:correction:alpha=-2:V1.01:857GHz}{0.9889}

% HFI Colour Correction for \alpha = 0, for V1.01 of the spectral bands

\setsymbol{HFI:colour:correction:alpha=0:V1.01:100GHz}{1.0008}
\setsymbol{HFI:colour:correction:alpha=0:V1.01:143GHz}{1.0148}
\setsymbol{HFI:colour:correction:alpha=0:V1.01:217GHz}{0.9909}
\setsymbol{HFI:colour:correction:alpha=0:V1.01:353GHz}{0.9888}
\setsymbol{HFI:colour:correction:alpha=0:V1.01:545GHz}{0.9878}
\setsymbol{HFI:colour:correction:alpha=0:V1.01:857GHz}{1.0014}

\def\reff@jnl#1{{\rm#1\/}}
\def\apj{\reff@jnl{ApJ}}       % Astrophysical Journal
\def\apjs{\reff@jnl{ApJS}}     % Astrophysical Journal, Supplement
\def\aaps{\reff@jnl{A\&AS}}    % Astronomy and Astrophysics, Supplement
\def\mnras{\reff@jnl{MNRAS}}   % Monthly Notices of the RAS
\def\prd{\reff@jnl{Phys.\ Rev.\ D}}    % Physical Review D

 % N_side
   % N_pix
   % N_tau: one sided length of kernel

 % transpose
\newcommand{\beq}{\begin{equation}}
\newcommand{\eeq}{\end{equation}}

\newcommand{\be}{\begin{equation}}
\newcommand{\ee}{\end{equation}}
\newcommand{\bea}{\begin{eq}}
\newcommand{\eea}{\end{equation}}

{\begin{enumerate}\setlength{\itemsep}{0mm}}%
{\end{enumerate}}
{\begin{enumerate}\setlength{\itemsep}{0mm}}%
{\end{enumerate}}

\newcommand{\bc}{\begin{center}}
\newcommand{\ec}{\end{center}}
\newcommand{\bi}{\begin{itemize}}
\newcommand{\ei}{\end{itemize}}
\newcommand{\ben}{\begin{enumerate}}
\newcommand{\een}{\end{enumerate}}

\usepackage{mathrsfs}
\usepackage{mathbbol}

\newfont{\gwpfont}{cmssq8 scaled 1000}
\newcommand{\rexcess}{{\gwpfont REXCESS}}

\def\xmm{{\it XMM-Newton}}

\def\msol {\mathrm{M}_{\odot}}

%%%%%%%%%%%%%%%%%%%%%%%%%%%%%%%%%%%%%%%%%%%%%%%%%%%%%%%%%%%%

\begin{document}

   \title{\textit{Planck} early results VIII: The all-sky early
Sunyaev-Zeldovich cluster sample}

%This author list corresponds to \title{Author list for Proj. Ref. 5.1: The all-sky early Sunyaev-Zeldovich cluster sample}
%Prepared by R. Leonardi (rleonardi@sciops.esa.int), ESAC/ESA, on 19APR2011
%This version is from 19 Apr 2011 at 12:00 CET
%\subtitle{There are 237 co-authors in this list}
\author{\small
Planck Collaboration:
P.~A.~R.~Ade\inst{77}
\and
N.~Aghanim\inst{50}
\and
M.~Arnaud\inst{62}
\and
M.~Ashdown\inst{60, 4}
\and
J.~Aumont\inst{50}
\and
C.~Baccigalupi\inst{75}
\and
A.~Balbi\inst{31}
\and
A.~J.~Banday\inst{85, 7, 68}
\and
R.~B.~Barreiro\inst{57}
\and
M.~Bartelmann\inst{83, 68}
% ...
}
\institute{\small
Aalto University Mets\"{a}hovi Radio Observatory, Mets\"{a}hovintie 114, FIN-02540 Kylm\"{a}l\"{a}, Finland\\
\and
Agenzia Spaziale Italiana Science Data Center, c/o ESRIN, via Galileo Galilei, Frascati, Italy\\
\and
Astroparticule et Cosmologie, CNRS (UMR7164), Universit\'{e} Denis Diderot Paris 7, B\^{a}timent Condorcet, 10 rue A. Domon et L\'{e}onie Duquet, Paris, France\\
\and
Astrophysics Group, Cavendish Laboratory, University of Cambridge, J J Thomson Avenue, Cambridge CB3 0HE, U.K.\\
\and
Atacama Large Millimeter/submillimeter Array, ALMA Santiago Central Offices, Alonso de Cordova 3107, Vitacura, Casilla 763 0355, Santiago, Chile\\
\and
CITA, University of Toronto, 60 St. George St., Toronto, ON M5S 3H8, Canada\\
\and
CNRS, IRAP, 9 Av. colonel Roche, BP 44346, F-31028 Toulouse cedex 4, France\\
\and
California Institute of Technology, Pasadena, California, U.S.A.\\
\and
Centre of Mathematics for Applications, University of Oslo, Blindern, Oslo, Norway\\
\and
Centro de Astrof\'{\i}sica, Universidade do Porto, Rua das Estrelas, 4150-762 Porto, Portugal\\
\and
DAMTP, University of Cambridge, Centre for Mathematical Sciences, Wilberforce Road, Cambridge CB3 0WA, U.K.\\
\and
DSM/Irfu/SPP, CEA-Saclay, F-91191 Gif-sur-Yvette Cedex, France\\
\and
DTU Space, National Space Institute, Juliane Mariesvej 30, Copenhagen, Denmark\\
\and
Departamento de F\'{\i}sica, Universidad de Oviedo, Avda. Calvo Sotelo s/n, Oviedo, Spain\\
\and
Department of Astronomy and Astrophysics, University of Toronto, 50 Saint George Street, Toronto, Ontario, Canada\\
\and
Department of Physics \& Astronomy, University of British Columbia, 6224 Agricultural Road, Vancouver, British Columbia, Canada\\
\and
Department of Physics and Astronomy, University of Southern California, Los Angeles, California, U.S.A.\\
\and
Department of Physics and Astronomy, University of Sussex, Brighton BN1 9QH, U.K.\\
\and
Department of Physics, Gustaf H\"{a}llstr\"{o}min katu 2a, University of Helsinki, Helsinki, Finland\\
\and
Department of Physics, Princeton University, Princeton, New Jersey, U.S.A.\\
\and
Department of Physics, Purdue University, 525 Northwestern Avenue, West Lafayette, Indiana, U.S.A.\\
\and
Department of Physics, University of California, Berkeley, California, U.S.A.\\
\and
Department of Physics, University of California, One Shields Avenue, Davis, California, U.S.A.\\
\and
Department of Physics, University of California, Santa Barbara, California, U.S.A.\\
\and
Department of Physics, University of Illinois at Urbana-Champaign, 1110 West Green Street, Urbana, Illinois, U.S.A.\\
\and
Dipartimento di Fisica G. Galilei, Universit\`{a} degli Studi di Padova, via Marzolo 8, 35131 Padova, Italy\\
\and
Dipartimento di Fisica, Universit\`{a} La Sapienza, P. le A. Moro 2, Roma, Italy\\
\and
Dipartimento di Fisica, Universit\`{a} degli Studi di Milano, Via Celoria, 16, Milano, Italy\\
\and
Dipartimento di Fisica, Universit\`{a} degli Studi di Trieste, via A. Valerio 2, Trieste, Italy\\
\and
Dipartimento di Fisica, Universit\`{a} di Ferrara, Via Saragat 1, 44122 Ferrara, Italy\\
\and
Dipartimento di Fisica, Universit\`{a} di Roma Tor Vergata, Via della Ricerca Scientifica, 1, Roma, Italy\\
\and
Discovery Center, Niels Bohr Institute, Blegdamsvej 17, Copenhagen, Denmark\\
\and
Dpto. Astrof\'{i}sica, Universidad de La Laguna (ULL), E-38206 La Laguna, Tenerife, Spain\\
\and
European Southern Observatory, ESO Vitacura, Alonso de Cordova 3107, Vitacura, Casilla 19001, Santiago, Chile\\
\and
European Space Agency, ESAC, Camino bajo del Castillo, s/n, Urbanizaci\'{o}n Villafranca del Castillo, Villanueva de la Ca\~{n}ada, Madrid, Spain\\
\and
European Space Agency, ESAC, Planck Science Office, Camino bajo del Castillo, s/n, Urbanizaci\'{o}n Villafranca del Castillo, Villanueva de la Ca\~{n}ada, Madrid, Spain\\
\and
European Space Agency, ESTEC, Keplerlaan 1, 2201 AZ Noordwijk, The Netherlands\\
\and
GEPI, Observatoire de Paris, Section de Meudon, 5 Place J. Janssen, 92195 Meudon Cedex, France\\
\and
Helsinki Institute of Physics, Gustaf H\"{a}llstr\"{o}min katu 2, University of Helsinki, Helsinki, Finland\\
\and
INAF - Osservatorio Astronomico di Padova, Vicolo dell'Osservatorio 5, Padova, Italy\\
\and
INAF - Osservatorio Astronomico di Roma, via di Frascati 33, Monte Porzio Catone, Italy\\
\and
INAF - Osservatorio Astronomico di Trieste, Via G.B. Tiepolo 11, Trieste, Italy\\
\and
INAF/IASF Bologna, Via Gobetti 101, Bologna, Italy\\
\and
INAF/IASF Milano, Via E. Bassini 15, Milano, Italy\\
\and
INRIA, Laboratoire de Recherche en Informatique, Universit\'{e} Paris-Sud 11, B\^{a}timent 490, 91405 Orsay Cedex, France\\
\and
IPAG: Institut de Plan\'{e}tologie et d'Astrophysique de Grenoble, Universit\'{e} Joseph Fourier, Grenoble 1 / CNRS-INSU, UMR 5274, Grenoble, F-38041, France\\
\and
Imperial College London, Astrophysics group, Blackett Laboratory, Prince Consort Road, London, SW7 2AZ, U.K.\\
\and
Infrared Processing and Analysis Center, California Institute of Technology, Pasadena, CA 91125, U.S.A.\\
\and
Institut N\'{e}el, CNRS, Universit\'{e} Joseph Fourier Grenoble I, 25 rue des Martyrs, Grenoble, France\\
\and
Institut d'Astrophysique Spatiale, CNRS (UMR8617) Universit\'{e} Paris-Sud 11, B\^{a}timent 121, Orsay, France\\
\and
Institut d'Astrophysique de Paris, CNRS UMR7095, Universit\'{e} Pierre \& Marie Curie, 98 bis boulevard Arago, Paris, France\\
\and
Institut de Ci\`{e}ncies de l'Espai, CSIC/IEEC, Facultat de Ci\`{e}ncies, Campus UAB, Torre C5 par-2, Bellaterra 08193, Spain\\
\and
Institute of Astronomy and Astrophysics, Academia Sinica, Taipei, Taiwan\\
\and
Institute of Astronomy, University of Cambridge, Madingley Road, Cambridge CB3 0HA, U.K.\\
\and
Institute of Theoretical Astrophysics, University of Oslo, Blindern, Oslo, Norway\\
\and
Instituto de Astrof\'{\i}sica de Canarias, C/V\'{\i}a L\'{a}ctea s/n, La Laguna, Tenerife, Spain\\
\and
Instituto de F\'{\i}sica de Cantabria (CSIC-Universidad de Cantabria), Avda. de los Castros s/n, Santander, Spain\\
\and
Jet Propulsion Laboratory, California Institute of Technology, 4800 Oak Grove Drive, Pasadena, California, U.S.A.\\
\and
Jodrell Bank Centre for Astrophysics, Alan Turing Building, School of Physics and Astronomy, The University of Manchester, Oxford Road, Manchester, M13 9PL, U.K.\\
\and
Kavli Institute for Cosmology Cambridge, Madingley Road, Cambridge, CB3 0HA, U.K.\\
\and
LERMA, CNRS, Observatoire de Paris, 61 Avenue de l'Observatoire, Paris, France\\
\and
Laboratoire AIM, IRFU/Service d'Astrophysique - CEA/DSM - CNRS - Universit\'{e} Paris Diderot, B\^{a}t. 709, CEA-Saclay, F-91191 Gif-sur-Yvette Cedex, France\\
\and
Laboratoire Traitement et Communication de l'Information, CNRS (UMR 5141) and T\'{e}l\'{e}com ParisTech, 46 rue Barrault F-75634 Paris Cedex 13, France\\
\and
Laboratoire de Physique Subatomique et de Cosmologie, CNRS/IN2P3, Universit\'{e} Joseph Fourier Grenoble I, Institut National Polytechnique de Grenoble, 53 rue des Martyrs, 38026 Grenoble cedex, France\\
\and
Laboratoire de l'Acc\'{e}l\'{e}rateur Lin\'{e}aire, Universit\'{e} Paris-Sud 11, CNRS/IN2P3, Orsay, France\\
\and
Lawrence Berkeley National Laboratory, Berkeley, California, U.S.A.\\
\and
MPA Partner Group, Key Laboratory for Research in Galaxies and Cosmology, Shanghai Astronomical Observatory, Chinese Academy of Sciences, Nandan Road 80, Shanghai 200030, China\\
\and
Max-Planck-Institut f\"{u}r Astrophysik, Karl-Schwarzschild-Str. 1, 85741 Garching, Germany\\
\and
Max-Planck-Institut f\"{u}r Extraterrestrische Physik, Giessenbachstra{\ss}e, 85748 Garching, Germany\\
\and
MilliLab, VTT Technical Research Centre of Finland, Tietotie 3, Espoo, Finland\\
\and
National University of Ireland, Department of Experimental Physics, Maynooth, Co. Kildare, Ireland\\
\and
Niels Bohr Institute, Blegdamsvej 17, Copenhagen, Denmark\\
\and
Observational Cosmology, Mail Stop 367-17, California Institute of Technology, Pasadena, CA, 91125, U.S.A.\\
\and
Optical Science Laboratory, University College London, Gower Street, London, U.K.\\
\and
SISSA, Astrophysics Sector, via Bonomea 265, 34136, Trieste, Italy\\
\and
SUPA, Institute for Astronomy, University of Edinburgh, Royal Observatory, Blackford Hill, Edinburgh EH9 3HJ, U.K.\\
\and
School of Physics and Astronomy, Cardiff University, Queens Buildings, The Parade, Cardiff, CF24 3AA, U.K.\\
\and
Space Research Institute (IKI), Russian Academy of Sciences, Profsoyuznaya Str, 84/32, Moscow, 117997, Russia\\
\and
Space Sciences Laboratory, University of California, Berkeley, California, U.S.A.\\
\and
Stanford University, Dept of Physics, Varian Physics Bldg, 382 Via Pueblo Mall, Stanford, California, U.S.A.\\
\and
Tartu Observatory, Toravere, Tartumaa, 61602, Estonia\\
\and
Tuorla Observatory, Department of Physics and Astronomy, University of Turku, V\"ais\"al\"antie 20, FIN-21500, Piikki\"o, Finland\\
\and
Universit\"{a}t Heidelberg, Institut f\"{u}r Theoretische Astrophysik, Albert-\"{U}berle-Str. 2, 69120, Heidelberg, Germany\\
\and
Universit\'{e} Denis Diderot (Paris 7), 75205 Paris Cedex 13, France\\
\and
Universit\'{e} de Toulouse, UPS-OMP, IRAP, F-31028 Toulouse cedex 4, France\\
% ...
}

%   \date{Received January 5, 2010}

  \abstract { We present the first all-sky sample of galaxy clusters
    detected blindly by the \Planck\ satellite through the
    Sunyaev-Zeldovich (SZ) effect from its six highest frequencies.
    This early SZ (ESZ) sample is comprised of 189 candidates, which
    have a high signal-to-noise ratio ranging from 6 to 29. Its high
    reliability (purity above 95\%) is further ensured by an extensive
    validation process based on \Planck\ internal quality assessments
    and by external cross-identification and follow-up
    observations. \Planck\ provides the first measured SZ signal for
    about 80\% of the 169 previously-known ESZ
    clusters. \Planck\ furthermore releases 30 new cluster candidates,
    amongst which 20 meet the ESZ signal-to-noise selection
    criterion. At the submission date, twelve of the 20 ESZ candidates 
    were confirmed as new clusters, with eleven confirmed using
    XMM-Newton snapshot observations, most of them
    with disturbed morphologies and low luminosities. The ESZ clusters
    are mostly at moderate redshifts (86\% with $z$ below 0.3) and
    span more than a decade in mass, up to the rarest and most massive
    clusters with masses above  $1\times 10^{15}\, M_{\odot}$.}


\authorrunning{Planck Collaboration}
\titlerunning{The \Planck\ all-sky Early
Sunyaev-Zeldovich cluster sample}

   \maketitle
%\allearlypapers
%________________________________________________________________

\section{Introduction}

Galaxy clusters provide valuable information on cosmology, from the
nature of dark energy to the physics that drives galaxy and structure
formation. The main baryonic component in these dark matter dominated
objects is a hot, ionised intra-cluster medium (ICM). The ICM can be
studied both in the X-ray and through the Sunyaev-Zeldovich effect
(SZ) \citep{sun72, sun80}, a fairly new and highly promising technique
that has made tremendous progress in recent years since its first
observations \citep{bir78}; see also \citet{rep95,bir99,car02}.

\begin{figure}
\centering

\includegraphics[width=8cm]{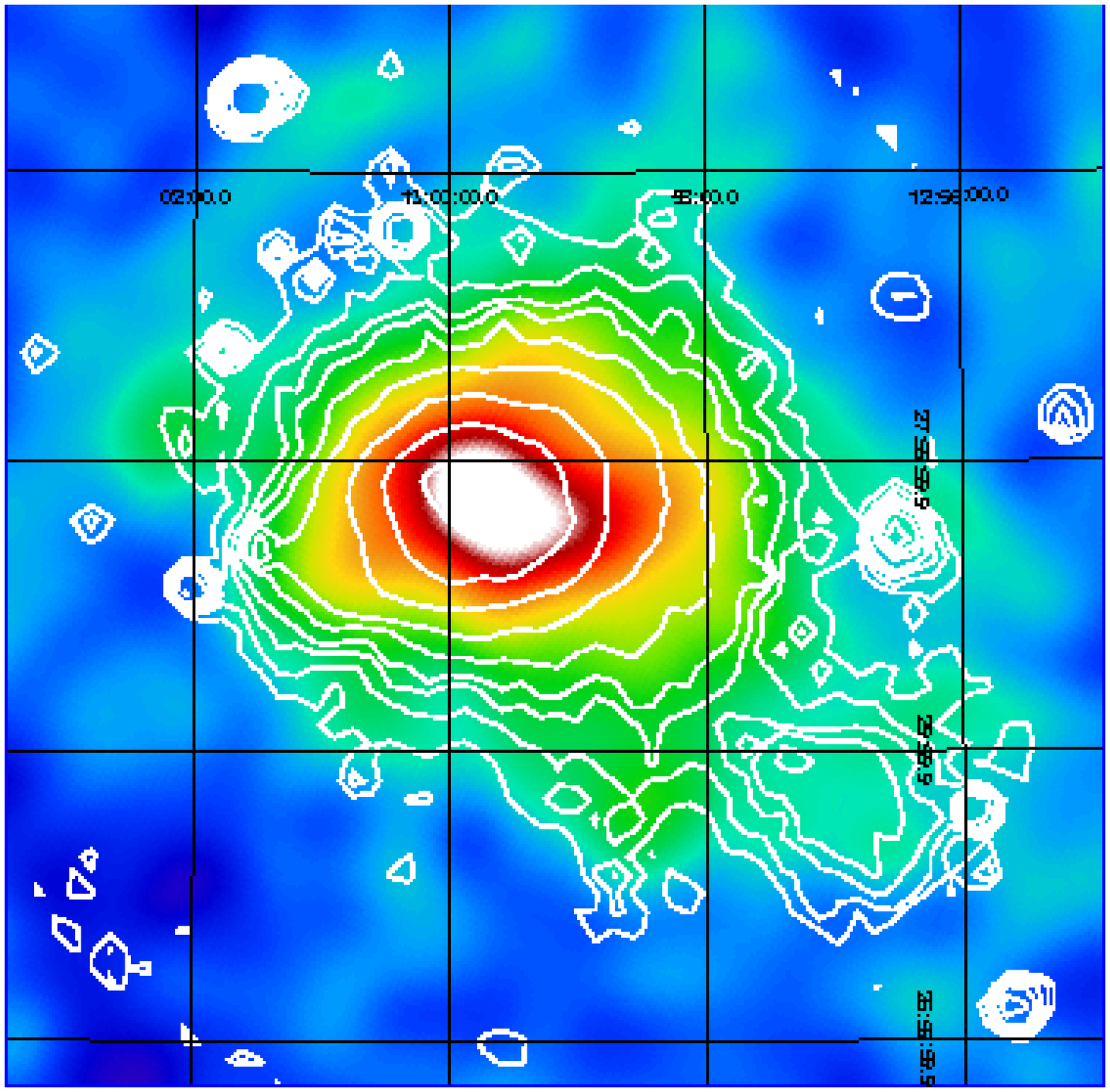}
\caption{\Planck\ $y$-map of Coma on a $\sim 3\deg\times3\deg$ patch with the
\emph{ROSAT-PSPC} iso-luminosity contours overlaid.}
\label{fig:coma.ps}
\end{figure}

The SZ effect is undoubtedly the best known and most studied secondary
contribution, due to cosmic structure, that is imprinted on the cosmic
microwave background after decoupling [for a review of secondary
  anisotropies see \cite{agh08}]. It is caused by the inverse Compton
interaction between the cosmic microwave background (CMB) photons and
the free electrons of the hot ICM. It can be broadly subdivided into
the thermal SZ (TSZ) effect, where the photons are scattered by the
random motion of thermal electrons, and the kinetic SZ (KSZ) effect
caused by the bulk motion of the electrons. In the former case,
the scattered CMB photons have a unique spectral dependence, whereas
the final spectrum remains Planckian in the case of the KSZ effect.

The SZ effect offers a number of advantages for cluster
studies. First, the Compton $y$ parameter, which measures the integral
of the gas pressure along the line of sight and sets the amplitude of
the SZ signal, does not suffer from cosmological surface-brightness
dimming. This implies that the SZ effect is an efficient method for
finding high-redshift clusters. Second, the total SZ signal $Y$,
integrated over the cluster's angular extent, directly measures the
total thermal energy of the gas and as such is expected to correlate
closely (i.e., with a tight scatter in the scaling relation) with
total cluster mass.  This fact is borne out both by numerical
simulations \citep{bor06,das01,mot05,pfr07} and indirectly from X-ray
observations \citep{nag07,arn07,vik09} using $Y_{\mathrm{X}}$, the
product of the gas mass and mean temperature giving an X-ray analogue
of the integrated SZ Compton parameter first introduced by
\citet{kra06}.  This contrasts with the X-ray luminosity which, at a
given mass, is very sensitive to the cluster's thermodynamical state,
for instance due to a recent merger event or in the presence of a
strong cooling core. Hence SZ surveys are expected to provide clean
cluster samples over a wide range of redshifts, in the sense of being
close to an unbiased mass-limited selection. These are key properties
for statistical studies with clusters, either to constrain
cosmological models (e.g., from the evolution of the mass function) or
to probe the physics of structure formation (e.g., from cluster
scaling and structural properties).

For these reasons, alongside the efforts developed to measure CMB
anisotropies many pioneering instruments were used or developed to
observe the SZ effect and use it as new observational probe of cluster
physics, large-scale structure, and the cosmological model. The first
observations of the SZ effect, targeted at specific X-ray selected
clusters, were performed using interferometric or single-dish
experiments mostly observing in the Rayleigh--Jeans part of the
spectrum: the Ryle Telescope at 15 GHz \citep{jon93}, the OVRO 5 meter
telescope at 32 GHz \citep{bir94}, the SuZIE array at 140 GHz
\citep{hol97}, BOLOCAM at 143 and 265 GHz \citep{gle98}, the Diabolo
array on IRAM 30 meter telescope at 140 GHz \citep{poi99}, MITO at
143, 214, 272, and 353 GHz \citep{dep99}, the Nobeyama 45 metre
telescope at 21 GHZ, 43 GHz and 150 GHz \citep{kom99}, the BIMA array
at 30 GHz \citep{daw01}, ACBAR at 150 and 220 GHz \citep{gom03}, CBI
working between 25 and 36 GHz \citep{udo04}, VSA at 30 GHz
\citep{lan05}, the Atacama Pathfinder Experiment (APEX) SZ Camera at
150 GHz \citep{dob06}, the SZ Array at 30 GHz \citep{muc07}, AMI at 15
GHz \citep{ami}, and AMIBA at 90GHz \citep{wu08}; see \cite{bir05} for
a review of observational techniques. Measurements of the SZ effect
were further made or attempted in the Wien part of the spectrum with
PRONAOS \citep{lam98}, SCUBA \citep{zem07}, and more recently with the
Herschel Space Observatory \citep{zem10}.

These experiments have not only allowed us to accumulate SZ
measurements for about a hundred clusters, but have also laid the
groundwork for SZ-based studies of clusters and of cosmology. In
combination with other observations, especially in X-rays, they were
used to measure cosmological parameters such as the Hubble constant,
and to probe the distance-duality relation between the
angular-diameter and luminosity distances,  bulk flows, and the
cluster gas mass fraction [e.g.,
  \cite{sil78,kob96,gre01,ree02,uza04,ame06,bon06,kas08}]. The SZ
effect has also been used to characterise the clusters themselves, as
it can potentially measure their radial peculiar velocities
\citep{ben03}.  The relativistic corrections to the SZ effect (e.g.,
\cite{ito98}) can be used to measure the gas temperature directly for
massive clusters \citep{poi98}. The spectral signature of the SZ
effect can in principle even probe the electron gas distribution and
constrain any non-thermal electron population in the intracluster
medium \citep{col03,shi04}. The SZ effect can also be used as a tracer
of the WHIM diffuse gas \citep{gen05,bat06}. Moreover multi-frequency
SZ measurements might provide a novel way of constraining the CMB
temperature and its evolution with redshift \citep{bat02,hor05,luz09}.

Deep surveys covering hundreds of square degrees and capable of
detecting many tens to hundreds of clusters, performed by the South
Pole Telescope (SPT) \citep{spt} and the Atacama Cosmology Telescope
(ACT) \citep{mar10}, are accumulating, and already delivering, data. One
of their goals is to use SZ cluster counts and
the SZ angular correlation function as cosmological tools
\citep{hai01,wel02,lev02,maj04,dou06}. Such surveys are
particularly powerful for detecting distant clusters, as was recently
proven by results from \citet{van10}. 

In this context ESA's
\Planck\footnote{\Planck\ (http://www.rssd.esa.int/Planck) is a
  project of the European Space Agency (ESA) with instruments provided
  by two scientific consortia funded by ESA member states (in
  particular the lead countries France and Italy), with contributions
  from NASA (USA) and telescope reflectors provided by a collaboration
  between ESA and a scientific consortium led and funded by
  Denmark.}\ mission, launched on 14 May 2009, carries a scientific
payload consisting of an array of 74 detectors sensitive to a range of
frequencies between roughly $25$ and $1000\GHz$, which scan the sky
simultaneously and continuously with an angular resolution varying
between about 30 arcmin (FWHM) at the lowest frequencies and about
four arcmin at the highest.  The array is arranged into two
instruments. The detectors of the Low Frequency Instrument (LFI) are
pseudo-correlation radiometers covering three bands centred at 30, 44,
and 70 GHz. The detectors of the High Frequency Instrument
\citep[HFI;]{Lamarre2010,planck2011-1.5} are bolometers covering six
bands centred at 100, 143, 217, 353, 545, and 857 GHz with bolometers
cooled to 0.1\,\hbox{K}. The design of \Planck\ allows it to image the
whole sky approximately twice per year, with an unprecedented
combination of sensitivity, angular resolution, and frequency
coverage.  The \Planck\ satellite, its payload, and its performance as
predicted at the time of launch are described in 13 articles included
in a special issue (Volume 520) of Astronomy \& Astrophysics.  The
main objective of \Planck\ is to measure the spatial anisotropies of
the temperature of the Cosmic Microwave Background (CMB) with an
accuracy set by fundamental astrophysical limits. Its level of
performance will enable \Planck\ to extract essentially all the
information in the CMB temperature anisotropies. \Planck\ will also
measure to high accuracy the polarisation of the CMB anisotropies,
which not only encodes a wealth of cosmological information but also
provides a unique probe of the thermal history of the Universe during
the time when the first stars and galaxies formed. In addition, the
\Planck\ sky surveys will produce a wealth of information on the dust
and gas in our own galaxy and on the properties of extragalactic
sources.

\Planck\ was specifically designed from the beginning to measure the
SZ effect \citep{agh97} and provide us with an all-sky SZ cluster
catalogue.  The first galaxy cluster searched for in the HFI data,
Abell 2163 (Fig.~\ref{fig:a2219} and \ref{FigVibStab2}), was indeed
found from 100 GHz to 353 GHz shortly after the First Light Survey
(FLS) was performed and observations in routine mode by
\Planck\ started. Three other known clusters falling in the FLS region
were seen across the positive and negative parts of the SZ
spectrum. The scanning strategy soon allowed us to map extended
clusters such as Coma on wide patches of the sky
(Fig.~\ref{fig:coma.ps}).  SZ detection techniques were then applied
to the data and the first blind detections were performed.

The \Planck\ all-sky SZ cluster catalogue, with clusters out to
redshifts $z\sim1$, that will be delivered to the community at the end
of the mission will be the first all-sky cluster survey since the
ROSAT All-Sky Survey (RASS), which was at much lower depth (the median
redshift of the NORAS/REFLEX cluster catalogue is $z \simeq
0.1$). Thanks to its all-sky nature, \Planck\ will detect the rarest
clusters, i.e., the most massive clusters in the exponential tail of
the mass function which are the best clusters for cosmological
studies. The \Planck\ Early SZ (ESZ) sample is delivered alongside the
Early Release Compact Source Catalogue (ERCSC)
\citep{planck2011-1.10}, the nine-band source catalogue, and the Early
Cold Core (ECC) catalogue \citep{planck2011-7.7b} at {\it
  http://www.rssd.esa.int/Planck} \citep{planck2011-1.10sup}.  The ESZ is
a high-reliability sample of 189 SZ clusters or candidates detected over
the whole sky from the first ten months of the \Planck\ survey of the
sky.

The present article details the process by which \Planck\ ESZ sample was
constructed and validated. The \Planck\ data and the specific SZ
extraction methods used to detect the SZ candidates are
presented in Sections \ref{sec:data} and \ref{sec:method}.  \Planck's
measurements provide an estimate of the integrated Compton parameter,
$Y$, of detected SZ cluster ``candidates''. A subsequent validation
process is needed to identify which among the candidates are
previously known clusters, and an additional follow-up programme is
required to scientifically exploit \Planck\ cluster data. This includes
cluster confirmation (catalogue validation) and the measurement of
relevant physical parameters. These different steps of the ESZ
construction and validation are presented in Section
\ref{sec:validgene} and the subsequent results are given in Section
\ref{sec:validext}. Finally, Sections \ref{sec:budg}, \ref{sec:pur},
and \ref{sec:stat} present the general properties of the ESZ cluster
sample. \Planck\ early results on clusters of galaxies are presented
here and in a set of accompanying articles
\citep{planck2011-5.1b,planck2011-5.2a,planck2011-5.2b,planck2011-5.2c}.

Throughout the article, and in all the above cited \Planck\ SZ early
result papers, the adopted cosmological model is a $\Lambda$CDM
cosmology with Hubble constant, $H_0 = 70 \kmsMpc$, matter density
parameter $\Omega_{\mathrm{m}} = 0.3$ and dark energy density
parameter $\Omega_\Lambda = 0.7$. The quantity $E(z)$ is the ratio of
the Hubble constant at redshift $z$ to its present value, $H_0$ ,
i.e., $E^2(z) = \Omega_{\mathrm{m}}(1 + z)^3 + \Omega_\Lambda$.
\section{\Planck\ data description}
\label{sec:data}

The ESZ sample was constructed out of the \Planck\ channel maps of the
HFI instrument, as described in detail in \cite{planck2011-1.7}. These
maps correspond to the observations of the temperature in the first
ten months of the survey by \Planck, which give complete sky coverage.
Raw data were first processed to produce cleaned time-lines
(time-ordered information, TOI) and associated flags correcting for
different systematic effects. This includes a low-pass filter, glitch
treatment, conversion to units of absorbed power, and a decorrelation
of thermal stage fluctuations. For cluster detection, and more
generally for source detection, one data flag of special importance is
associated with Solar System Objects (SSO).  These objects were
identified in TOI data using the publicly-available Horizon ephemeris,
and the SSO flag was created to ensure that they are not projected
onto the sky, in order to avoid possible false detections, ringing,
etc.

Focal-plane reconstruction and beam-shape estimates were obtained using
observations of Mars. Beams are described by an elliptical Gaussian
parameterisation leading to FWHM $\theta_{\mathrm{S}}$ given in
\cite{planck2011-1.7}. The attitude of the satellite as a function
of time is provided by the two star trackers installed on the \Planck\
spacecraft. The pointing for each bolometer was computed by combining
the attitude with the location of the bolometer in the focal plane
reconstructed from Mars observations.

From the cleaned TOI and the pointing, channel maps have been made
by co-adding bolometers at a given frequency. The path from TOI to
maps in the HFI data processing is schematically divided into three
steps: ring-making, destriping, and map-making. The first step
averages circles within a pointing period to make rings with higher
signal-to-noise (S/N) ratio, taking advantage of the redundancy of
observations provided by the \Planck\ scanning strategy. The low
amplitude $1/f$ component is accounted for in the second step using a
destriping technique. Finally, cleaned maps are produced using a
simple co-addition of the Healpix-based
rings\footnote{http://healpix.jpl.nasa.gov/ \citep{gor05}}. SSO flag
channel maps, used in the internal validation of the ESZ sample, were
also made following the same procedure.

The noise in the channel maps is essentially white with a standard
deviation of $1.6, 0.9, 1.4, 5.0, 70, 1180 \muK\,
\mathrm{degree}$\footnote{In the following and unless otherwise
  stated, $\muK$ refers to equivalent CMB temperature fluctuations in
  $\muK$.} from low to high frequencies \citep{planck2011-1.7}.
Photometric calibration is performed for the lower frequency channels
at the ring level using the CMB dipole (from WMAP \citep{hin09}), and
at the map level using FIRAS data \citep{fix94} for the higher
frequency channels at 545 and 857 GHz. The absolute gain calibration
of HFI \Planck\ maps is known to better than 2\%
\citep{planck2011-1.7}.

\section{Detection and Cluster extraction}
\label{sec:method}

In order to generate a cluster candidate list, a suitable extraction
algorithm must be run on the maps.
SZ clusters can be considered as compact sources with respect to
the \Planck\ beam, but they are definitely not point sources. Their
extension thus merits a special adapted processing. For this reason,
several extraction methods were developed within the
\Planck\ collaboration, and those were tested and compared using the
\Planck\ Sky Model Simulation (PSM\footnote{`\Planck\ Sky Model',
  http://www.apc.univ-paris7.fr/APC{\textunderscore}CS/Recherche/Adamis/PSM/
  psky-en.php}). The details of the comparison of the cluster
extraction algorithms, called the ``SZ challenge'', can be found in Melin et al. (in prep.).%\cite{szchal}.

Methods fall into two classes: ``direct'' methods use individual
channel maps to extract the clusters, while ``indirect'' methods use
sky $y$-maps obtained via component separation algorithms.  The
methods used in this article are direct methods, with the reference
method chosen on the basis of the SZ challenge.  The direct detection
algorithms used to construct and validate the ESZ sample incorporate
prior assumptions on the cluster signal, specifically its spectral and
spatial [i.e., the shape of the Intra-Cluster Medium (ICM) pressure
profile] characteristics (see Sect.~\ref{sec:aetalp}). This enhances
the cluster contrast over a set of observations containing
contaminating signals.

Most of the methods developed prior to the launch were applied to the
\Planck\ data, but only direct methods were favoured for
implementation in the pipeline infrastructure. The following three
were used to construct and validate the ESZ sample:
\begin{itemize}
\item A matched multi-frequency filter (MMF) algorithm, referred to
  henceforth as MMF3, was the reference method used for
  the blind detection of SZ candidates, and the construction of the
  ESZ list.
\item Two other methods (Sections \ref{sec:pws} and \ref{sec:mmf1})
  were used to confirm the blind detections of the ESZ candidates.
\end{itemize}

In addition, a slightly different version of MMF3 was run as part of
ESZ validation, in order to re-extract the Compton $Y$ parameter of
the SZ clusters incorporating fixed cluster sizes and positions taken
from X-ray observations (see Sect.~\ref{sec:sizeF}).

\subsection{Baseline cluster model}
\label{sec:aetalp}

The ICM pressure profile has historically been described by an
isothermal $\beta$-model \citep[e.g.,]{cav78,gre01,ree02}. However,
recent X-ray observations have shown that a $\beta$-model is a poor
description of the gas distribution in clusters, leading several
authors to propose more realistic analytical functions based on a
Generalised Navarro-Frenk-and-White (GNFW) profile
\citep{nag07,arn10}.

The baseline pressure profile used in the present work is the standard
``universal'' pressure profile derived by \cite{arn10}. It is
constructed by combining the observed X-ray pressure profile within
$R_{500}$, from 31 galaxy clusters of the \rexcess\ sample
\citep{boe07}, with data from state-of-the-art numerical simulations
\citep{bor04,nag07,pif08} out to $5\,R_{500}$.  In the following,
$R_{500}$ is the cluster size defined as the radius where the mean
enclosed density is 500 times the critical density. It relates to the
characteristic cluster scale $R_{\mathrm{s}}$ through the NFW
concentration parameter $c_{500}$ ($R_{\mathrm{s}}=R_{500}/c_{500}$).

The pressure profile model used in the present article is equivalent
to the standard self-similar case described in Appendix~B of
\cite{arn10}.\footnote{More details on the pressure profile can be
  found in \cite{planck2011-5.2a}} It is equivalent to a shape
function characterised by two free parameters, a central value and a
characteristic scale $\theta_{\mathrm{s}}$.

The SZ effect from the hot ICM is due to the first-order correction
for energy transfer in Thomson scattering.  There is a spectral
distortion, energy being transferred from photons in the
Rayleigh--Jeans tail of the cosmic blackbody radiation to the Wien
tail. In the non-relativistic limit the frequency dependence of the
distortion is universal (the same for all clusters), characterised by
a distinct frequency, $\nu \sim 220$ GHz, where the TSZ effect
vanishes. Below this frequency there is a decrement of the CMB
intensity, giving an apparent {\it decrease} in the sky brightness,
and above lies an enhancement.

The magnitude of the SZ effect, known as the Compton parameter $y$,
depends only on the cluster's characteristics, electronic temperature
$T_e$ and density $n_e$, as
\[
y=\frac{k\sigma_T}{m_ec^2}\int^lT_e(l)n_e(l)\,\mathrm{d}l
\]
\noindent
where $k$ is the Boltzmann constant, $\sigma_{\mathrm T}$ the Thomson
cross section, $m_ec^2$ the electron rest mass and $l$ is the distance
along the line of sight. The total SZ signal is characterised by the
integrated Compton parameter denoted $Y=\int y\,{\mathrm d}\Omega$,
where $\Omega$ is solid angle. It can be written as $D_{\mathrm{A}}^2
Y = (\sigma_{\mathrm T}/m_{\mathrm e} c^2)\int P dV$, where
$D_{\mathrm A}$ is the angular-diameter distance to the system and
$P=n_e k T_e$ the electron pressure. In the following, the integral
performed over the sphere of radius $R_{500}$ ($5R_{500}$) is denoted
$Y_{500}$ ($Y_{5R500}$).Thus, as defined here, $Y_{500}$ and
$Y_{5R500}$ have units of solid angle, e.g., arcmin$^2$.

\subsection{Reference extraction method (Matched Multi-Filter, MMF3)}
\label{sec:mmf3}

The ESZ sample is the result of a blind multi-frequency search in the
all-sky \Planck-HFI maps, i.e., no prior positional information on
detected known clusters was used as input to the detection
algorithm. The ESZ sample is produced by running the MMF3 algorithm,
which is an all-sky extension of the matched multi-frequency filter
algorithm described in \cite{mel06}, over the six HFI frequency
maps. The spectral distortion of the CMB due to the ICM can in
principle be detected down to the lowest frequencies at which
\Planck\ operates; however, the beam at the lowest frequencies is large
compared to typical cluster sizes.  Since clusters at moderate
redshifts typically span angular scales of $\sim5 \textrm{arcmin}$,
the large beam of \Planck\ at the LFI bands results in beam dilution
of the SZ signal. The inclusion of the lowest \Planck\ frequencies
using the current algorithm therefore results in a lower
S/N for the detected sources than if only the
HFI bands were used. This reduces the efficiency of SZ cluster
detection, which can potentially be improved in the future with
refinements to the algorithm.  As a consequence, for the generation of
the ESZ list, only the \Planck\ all-sky maps at frequencies of 100 GHz
and above are considered.

The MMF algorithm, studied extensively by \cite{her02} and
\cite{mel06}, enhances the contrast, and thus S/N, of objects of known
shape and known spectral emission profile over a set of observations
containing contaminating signals.  In its application for SZ, the
method makes use of the universal frequency dependence of the thermal
SZ effect.  The filter optimises the detectability using a linear
combination of maps (which requires an estimate of the statistics of
the contamination) and uses spatial filtering to suppress both
foregrounds and noise (making use of the prior knowledge of the
cluster profile). The filter optimises cluster detection but it is not
immune to contamination by false, non-SZ, detections which calls for
an extensive validation procedure described in Section
\ref{sec:validgene}.

MMF3 first divides the all-sky maps into a set of 504 overlapping
square patches of area $10\times10$ square degrees. Holes in the maps
due to unsampled or badly sampled pixels are
identified to construct an effective detection mask and are then
filled in with the median value of the adjacent pixels.
The matched multi-frequency filter then combines optimally the six
frequencies of each patch assuming the SZ frequency spectrum and using
the reference pressure profile presented in Section \ref{sec:aetalp}.

Auto- and cross-power spectra used by MMF3 are directly estimated
from the data and are adapted to the local instrumental noise and
astrophysical contamination.  For each patch, the position and the
scale radius (chosen to be $5R_{500}$) of the cluster profile,
i.e., the cluster size $5\theta_{500}$, are varied to maximise the
S/N of each detection. The algorithm hence assigns
to each detected source a position, an estimated cluster size,
$5\theta_{500}$, and an integrated Compton parameter, $Y_{5R500}$. In
the present article and unless otherwise stated the {\it measured}
integrated Compton parameter, noted $Y_{5R500}$, is thus computed by
integrating the GNFW profile within a sphere of $5R_{500}$\footnote{In
  the spherical assumption with this profile, $Y_{500}$ the integrated
  Compton parameter within $R_{500}$ relates to $Y_{5R500}$ by
  $Y_{5R500}=1.81 \times Y_{500}$.}  encompassing most of the SZ 
signal. The detected sources extracted from the
individual patches, with their assigned sizes and integrated Compton
parameters, are finally merged into an all-sky cluster list. In
practice the MMF3 algorithm is run in an iterative way; after a first
detection of the SZ candidates, consecutive runs centred on the
positions of the candidates refine the estimated S/N and
candidate properties. At this stage, the uncertainty on $Y_{5R500}$ is
provided and takes into account the uncertainty in the cluster size
estimate. The MMF3 algorithm can also be performed with fixed cluster
size and position to estimate the SZ signal. This version of the
algorithm was used to measure the integrated Compton parameters of
known X-ray clusters in the ESZ sample, as explained in Section
\ref{sec:sizeF}.

In order to address contamination by point sources, MMF3 uses a
built-in source detection algorithm to reject point sources with
S/N above ten which are then masked. This step avoids most of
the false SZ detections associated with point sources. 
However, some residual contamination by non-SZ sources
captured by the MMF3 algorithm may still be present and requires additional
validation of the detection candidates (see
Section \ref{sec:validgene}).
\subsection{Other extraction methods}

The two other ``direct'' SZ detection methods used to confirm the
blind detections of the ESZ candidates by MMF3 are discussed
below. These methods previously compared rather well to each other
within the SZ challenge match in terms of the detection properties
(especially for high S/N sources). Their estimated
sizes and SZ signals agree on average as well, though they differ on a
case by case basis.

\subsubsection{The Matched Multi-Filter Method, MMF1}
\label{sec:mmf1}

The MMF1 algorithm is a completely independent implementation of the
multi-frequency matched filter integrated within the \Planck-HFI
pipeline and infrastructure. A more detailed description of MMF1 is
given in \cite{szchal}. The full-sky \Planck\ frequency maps are
divided into 640 flat patches, each 14.66 degrees on a side (corresponding to
512 by 512 pixels), with overlapping regions of six degrees.  The
performance of the MMF algorithm is extremely sensitive to the quality
of the estimated auto- and cross-power spectra of the background
component in each frequency map. The size of the patches thus needs to
be large enough to ensure a representative assessment of the
background. The large overlap between patches was chosen so that all
detections in a two-degree border around the edge of the patch may be
discarded.

The detection of the SZ-candidates is performed on all the
patches, and the resultant sub-catalogues are merged together to
produce a single SZ-candidate catalogue. Similarly to MMF3, the
candidate size is estimated by filtering the patches over the range of
potential scales, from point-source sized
objects and larger, and finding the scale which maximises the
S/N of the detection of the candidate. In the
version used on the \Planck\ data, when merging sub-catalogues produced
from the analysis of individual patches, it is also the
S/N of the detection which is used when deciding
which detection of the candidate is kept.

\subsubsection{PowellSnakes (PwS) for SZ}
\label{sec:pws}

PowellSnakes (PwS) is quite different from the MMF methods. It is a
fast Bayesian multi-frequency detection algorithm designed to identify
and characterise compact objects buried in a diffuse background. The
detection process is grounded in a statistical model comparison test
where two competing hypotheses are compared: the detection hypothesis
and the null hypothesis. The statistical foundations of PwS are
described in \cite{car09}.

Similarly to the MMF algorithms, a template parameterised SZ pressure
profile is assumed known and representative of the majority of the
cluster population observable with the resolution and noise
characteristics of the instrumental setup.  According to our data
model, the pixel intensities result from the contribution of three
independent components: the SZ signal, the astronomical background
component, and the instrumental pixel noise. The last is assumed to be
a realisation of a homogeneous stationary Gaussian random white noise
process. The background astronomical components and the pixel noise
are assumed uncorrelated and can each be modelled locally by a
homogeneous Gaussian process.

The algorithm starts by minimising the model's likelihood ratio with
respect to the model's parameters by using a Powell minimiser
iteratively one source at a time. We assume that the sources are
well separated and the fields not too crowded. The parameter
estimation and the acceptance/rejection threshold is defined using
Bayesian approach with priors adjusted on
the \Planck\ Sky Model SZ Catalogue.

PwS performs on flat $512 \times 512$ pixel patches of 14.66 degrees
on a side.  When applying a Galactic cut of $|b|> 14$ degrees, PwS
splits the sphere into 2324 patches. However, only detections lying
inside the inner $256\times 256$ pixels are considered. So, on average
PwS detects each cluster more than three times (usually four times),
increasing the reliability of the detection.
The selection of the candidate detection that goes into the final
catalogue uses the Bayesian mode of PwS, based on the highest ratio of
model posteriors.
\section{Validation of the ESZ sample}
\label{sec:validgene}

\begin{figure}
\centering
\includegraphics[width=8cm]{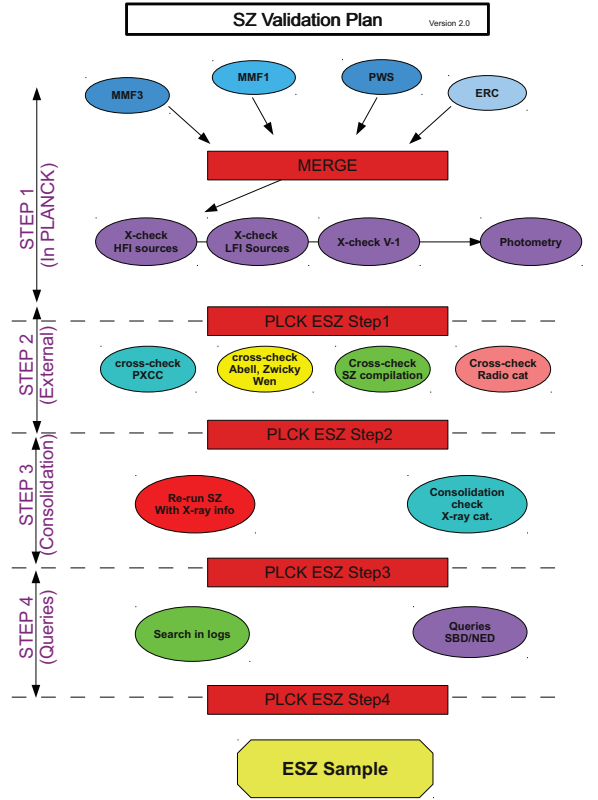}
\caption{Flowchart of the SZ validation process applied to
        the \Planck\ cluster sample.}
\label{fig:szvp}
\end{figure}

The SZ validation process, Fig.~\ref{fig:szvp}, is an integrated
HFI-LFI effort within \Planck\ Working Group 5
(WG5\footnote{http://www.ita.uni-heidelberg.de/collaborations/planck/})
``Clusters and Secondary anisotropies''.  It has been established in
order to validate the full SZ candidate lists obtained from the
extraction methods developed by the \Planck\ collaboration.  It relies
mainly on a three-stage process detailed in the following subsections:
\begin{itemize}
\item {\bf Internal validation} steps based on \Planck\ data:
  \begin{itemize}
    \item search for and rejection of associations with SSOs and artefacts;
    \item rejection of sources with rising spectral energy
      distribution in the high HFI frequency bands;
    \item cross-check with other \Planck\ source catalogues to reject
      SZ candidates identified with cold cores (CC) and other Galactic
      sources; and
    \item redundant detections of the same candidates by methods other
      than the reference one.
  \end{itemize}
\item {\bf Candidate identification} steps based on ancillary data:
  \begin{itemize}
    \item identification of SZ candidates with known clusters from
      existing X-ray, optical/near infrared (NIR), and SZ catalogues
      and lists; and
    \item search in NED and SIMBAD databases.
\end{itemize}
\item {\bf Follow-up programmes} for verification and confirmation of
  SZ candidates.
\end{itemize}

\subsection{Construction of ESZ sample and internal validation}
\label{sec:const}
 	 
The construction of the ESZ list of SZ candidates starts with the
blind detection of candidates using the implementation of the MMF3
algorithm at the US Planck Data Center applied to \Planck-HFI channel
maps at Galactic latitudes $\left|b\right| > 14 $ deg. A total of
about 1000 blind SZ candidates are detected with S/N $\ge 4$.  As
discussed above, the MMF3 algorithm uses prior information on the SZ
spectrum and on the cluster shape. However, especially due to the
beam-sizes of the order of a few arcminutes, the resulting list of SZ
candidates is not immune from false detection due mainly to dust
emission at high frequencies from the interstellar medium (ISM) or
infrared sources, and very moderately to the CMB fluctuations at low
frequencies (see illustrations of channel maps in upper panels of
Fig.~\ref{fig:a2219}). In the following, we do not explicitly check
for association with extragalactic point sources emitting at
\Planck-HFI frequencies, which is essentially dealt with internally by
the MMF3 algorithm (Sect.~\ref{sec:mmf3}). Some residual contamination
of the SZ Compton $Y$ parameter by point sources may, however, still
be present (see Section \ref{sec:contam} for a specific discussion).

The internal validation process starts by removing spurious detections
from the output list of blind SZ candidates, which is achieved in two
steps. We first reject the candidates showing rising spectral energy
distributions in the highest \Planck-HFI frequency bands. They
represent around 14\% of the initial blind SZ candidates. Second, the
remaining blind SZ candidate sample is further cleaned by rejecting
all objects associated with either Galactic sources, or CC detected
using the CoCoCodet algorithm \citep{mon10,planck2011-7.7b} within a 7
arcmin radius of SZ candidates. This step further reduces the sample
of remaining SZ candidates by about 17\%.

After this two-step process, the initial blind SZ candidate sample has
been reduced to around 770 blind SZ candidates with S/N $\ge
4$. However at this S/N level many candidates will not correspond to
actual clusters. Theoretical predictions based on the PSM simulations
indicate that the purity (ratio of true to all detections) is
expected to be of the order of 70\% at S/N = 4
(Fig.~\ref{fig:purity}). The simulations do not account fully for the
complexity of the true sky nor for the inhomogeneity of the noise
across the sky. The actual purity is thus likely to be worse than the
prediction. In order to ensure a high level of purity in the ESZ
sample and based on lessons learnt from the XMM-Newton observations of
low S/N candidates (see \citet{planck2011-5.1b}), an early decision
was made to cut at a higher S/N level of $\mathrm{S/N}
\geq 6$ for this first \Planck\ data release.
This more stringent condition retains 201 SZ
cluster candidates. Taking advantage of the outcome of the
follow-up programme for cluster confirmation by XMM-Newton, we further
retain only the SZ candidates detected blindly by the MMF3 algorithm
{\it and} at least one other method, be it MMF1 or PowellSnakes.  This
results in 190 SZ cluster candidates; these constitute the baseline
ESZ sample. A detailed inspection of the SZ maps and spectra of the 11
discarded SZ candidates was performed (see Section \ref{sec:cand}) and
confirmed that these sources were false detections.

A final internal check consisted of searching for associations of the
obtained 190 SZ candidates with possible artefacts such as
low-frequency noise stripes, ringing from neighbouring bright sources,
hot pixels, non-observed pixels or poorly sampled pixels in the
vicinity of SSO-flagged regions. None of the 190 ESZ candidates was
associated with such artefacts.

\subsection{Candidate identification with ancillary data}

The second stage of the SZ validation process consisted of
cross-matching the obtained list of 190 blind SZ candidates with
external cluster catalogues in X-rays, optical and SZ domains. This allowed us
to identify the SZ candidates associated with previously known clusters
and consequently isolate the \Planck\ candidate new clusters.
\subsubsection{With X-ray cluster catalogues}
\label{sec:idX}

For the association of \Planck\ SZ candidates from the blind
extraction with known X-ray clusters, we have used the Meta-Catalogue
of X-ray detected Clusters of galaxies \citep[MCXC,][]{pif10}. This
homogenised compilation of X-ray detected clusters of galaxies
comprises approximately 1800 clusters from publicly available ROSAT
All Sky Survey-based (NORAS, REFLEX, BCS, SGP, NEP, MACS, and CIZA)
and serendipitous (160SD, 400SD, SHARC, WARPS, EMSS, etc) cluster
catalogues.

For each X-ray cluster in the MCXC several properties are available,
amongst which are the X-ray centroid coordinates, redshift,
identifiers, and $L_{500}$.\footnote{The X-ray luminosities as
  measured within an aperture of radius $R_{500}$} The luminosities
are adopted as proxies to estimate the total mass $M_{500}$ using the
$L$--$M$ relation from \rexcess\ \citep{pra09}, and radius $R_{500}$,
and to predict the integrated Compton $Y_{5R500}^{L_X}$, or
alternatively $Y_{500}^{L_X}$, as detailed in \cite{planck2011-5.2a}
as well as other \Planck-related quantities.

Because the MCXC compilation includes only clusters with available
luminosity (redshift) information, we supplement it with about 150
clusters where this information is missing. This implies that for the
latter only centroid positions are available.  The resulting
meta-catalogue, for simplicity referred to as MCXC in the reminder of
the article, is extensively used during the external validation
process. For a given \Planck\ candidate cluster we identify the
closest MCXC cluster. The reliability of the association is checked based
on the distance, as compared to the cluster size, and on the measured
$Y_{5R500}$ (or S/N) values, as compared to the expected values
$Y_{5R500}^{L_X}$ (or S/N) for the MCXC clusters.
\subsubsection{With optical cluster catalogues}
\label{sec:idOp}

The baseline for the identification of blind SZ candidates from the
ESZ with clusters known in the optical is the cross-match with the
Abell cluster catalogue \citep[][5250 clusters of which 1026 have a
  redshift]{abe58} and the Zwicky cluster catalogue \citep[][9134
  objects]{zwi61}. The association criterion here was a positional
match with a search radius for both catalogues set to five arcminutes.

Furthermore, the ESZ sample was cross-checked against the MaxBCG
\citep{koe07} and \cite{wen09} catalogues with a search radius of 5
arcmin.
\subsubsection{With known SZ clusters}
\label{sec:idSZ}

The identification of SZ candidates is also performed at millimetre
wavelengths by cross-matching the SZ candidate list with a compilation
of SZ observed galaxy clusters from the literature undertaken by Douspis et al. (in prep.). This compilation is based on SZ
observations conducted with the numerous experiments developed
during the last 30 years (Ryle, OVRO, BIMA, MITO, Nobeyama, SZA,
APEX-SZ, AMI, Diabolo, Suzie, Ryle, AMIBA, ACBAR, etc). It also
includes the new clusters recently discovered through their SZ
signature by ACT and SPT. In total the compilation comprises 111 SZ
clusters including 28 newly discovered by ACT and SPT
\citep{men10,van10}.  The association of the \Planck\ SZ candidates was
based on positional matching with a search radius of five arcminutes.

\subsubsection{Queries in SIMBAD and NED databases}
\label{sec:idDB}

The information provided from querying databases is mainly redundant
with cross-checks with cluster catalogues in X-ray or
optical. However, running both cross-matches is important to avoid
missing a few associations. It is also important to retrieve the
information on redshifts for those identified clusters not included in
the MCXC. We therefore performed a systematic query in SIMBAD.  The
adopted search radius was set to five arcminutes. For NED, no systematic
query was implemented.  Cluster candidates within the same search
radius were rather checked against a list of objects retrieved from NED
flagged as ``Clusters of galaxies''. Finally the
candidates were also checked against the X-ray cluster database
\citep[BAX:]{sad04}.
\subsection{Follow-up programme for validation and confirmation}
\label{sec:fu}

In parallel to the effort of SZ candidate cross-identification, a
coherent follow-up programme targeted towards the
verification/validation of the cluster candidates in the SZ catalogue
was put into place in the form of an internal roadmap. The main goals
of this follow-up programme are to confirm \Planck\ candidates as new
clusters, and as a consequence to better understand both the SZ
selection criteria in the \Planck\ survey and the reliability of
selected sources.

Considering the complementarity of X-ray, optical and IR/SZ,
observational follow-ups have been coordinated to optimise the
validation and the understanding of the \Planck\ selection. In
practice, this took the form of a confirmation programme
relying on observations with XMM-Newton\footnote{\xmm\ is an ESA
  science mission with instruments and contributions directly funded
  by ESA Member States and the USA (NASA)} making use of Director
Discretionary Time (DDT) as detailed in \cite{planck2011-5.1b}.  This
is complemented by observations in the optical using the European
Northern Observatory facilities (ENO), the European Southern
Observatory 2.2m-telescope, and two pilot programmes, one with the WISE
experiment \citep{wri10} for the search of overdensities in the IR
data, and one with the Arcminute MicroKelvin Imager\footnote{AMI is a
  pair of interferometer arrays located near Cambridge, UK, operating
  in six bands between 13.5 and 18 GHz, with sensitivity to angular
  scales 30 arcsec -- 10 arcmin.} (AMI, \cite{ami}) for the
confirmation of \Planck\ candidates with SZ observations.

An ensemble of SZ candidates spanning a range of S/N 
between four and eleven was selected from earlier versions of
the HFI channel maps and sent to the above-mentioned facilities. The
targets were selected from a list of SZ candidates after the external
validation stage (i.e., identification of known clusters). They went
through visual inspection of their maps and spectra produced by all
the available methods described in Section \ref{sec:cand}. Furthermore,
in order to avoid duplicating existing observations of candidates with
the same or similar facilities, the cluster candidates were further
cross-matched with logs of X-ray, optical, and NIR observatories.

The search in X-ray observatories (ROSAT, Suzaku, XMM-Newton, and
Chandra) was performed using the HEASARC
tool.\footnote{http://heasarc.gsfc.nasa.gov/cgi-bin/W3Browse/w3browse.pl}
For XMM-Newton and Chandra both master catalogues and accepted GO
(Guest Observer) targets were used in the search. For Suzaku, only the
master catalogue was used. In the case of optical and NIR
observatories, the search was performed in the public logs of several
optical/infrared observatories. In some cases, this search was
completed using VO (Virtual Observatory) tools\footnote{VO command
  line tools http://iraf-nvo.noao.edu/vo-cli/}. The checked resources
were: ING Archive, UKIRT Archive, ESO Archive, HST Archive (at ESO),
CFHT Archive, AAT Archive, NOAO Science Archive, Multimission Archive
at STScI (MAST), Gemini Science Archive, and SMOKA (Subaru Mitaka
Okayama Kiso Archive). In addition, a search in the footprint of the
covered area for known surveys was performed. The searched areas
considered were those of SDSS, UKIDSS, and HST (ACS-WFC) as they are
described in the VO footprint
service\footnote{http://www.voservices.net/footprint}
\citep{vo-footprint-paper}, as well as those of SPT and ACT
experiments.

The details and results of the confirmation follow-up with XMM-Newton
are given in \citet{planck2011-5.1b}. A total of 25 targets were
observed with short snapshot exposures (i.e., 10~ks nominal EPN) out
of which 21 were confirmed as clusters or systems of multiple extended
X-ray sources (i.e., double or triple). Complying with
\Planck\ policies and following the agreement between the \Planck\ and
XMM-{\it Newton} ESA project scientist, all the data are made public
with the publication of the \Planck\ early results and the
\Planck\ ERCSC. Of the 21 confirmed \Planck\ SZ sources, 11 are found
in the ESZ sample and are discussed in Section \ref{sec:conf}. The
remaining clusters with S/N $< 6$ are discussed in
\cite{planck2011-5.1b}.
One candidate cluster in the ESZ sample was confirmed by AMI and
WISE. None of the targets sent for observation in the optical with the
ENO telescopes met the ESZ selection criteria.

\section{Results of the validation}
\label{sec:validext}

\begin{figure*}
\centering
\includegraphics[angle=0,width=8cm]{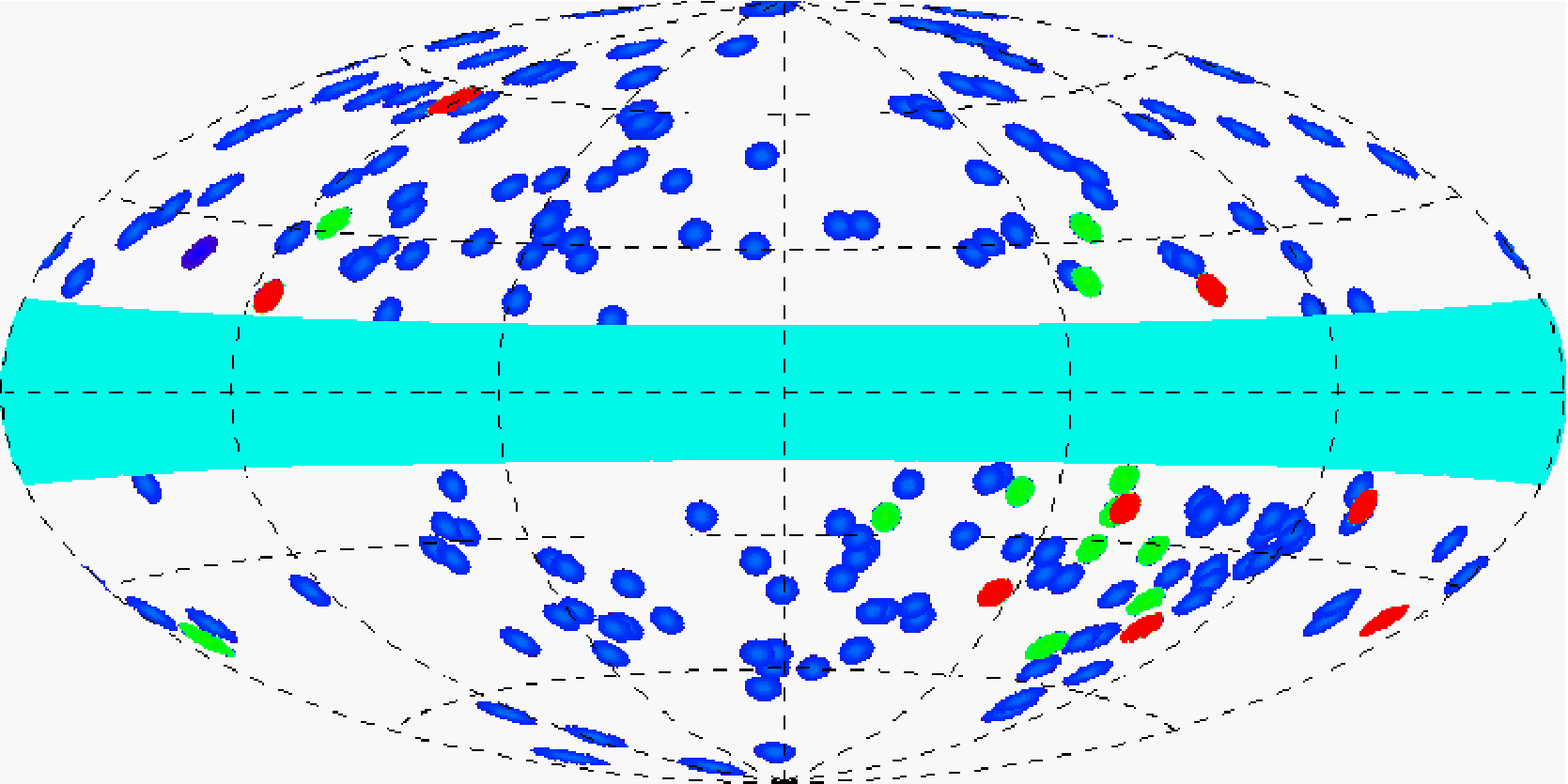}
\includegraphics[angle=0,width=8cm]{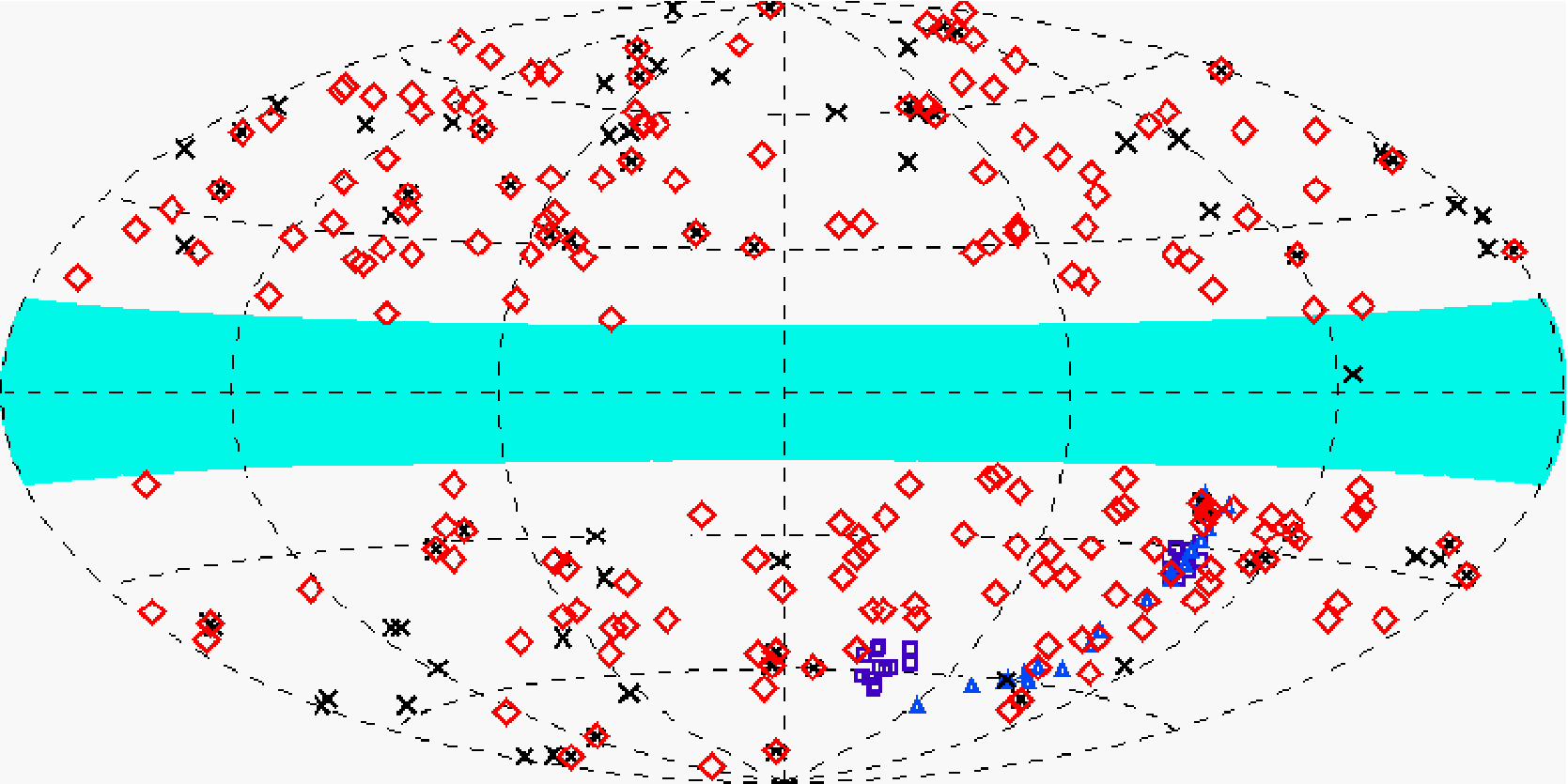}
\caption{Distribution of ESZ clusters and candidate clusters on the
  sky (Galactic Aitoff projection). Left panel: In blue are ESZ
  clusters identified with known clusters, in green the ESZ confirmed
  candidates, and in red the ESZ candidate new clusters yet to be
  confirmed.  Right panel: In red diamonds the ESZ sample, in black
  crosses the compilation of SZ observations prior to 2010, in dark
  blue triangles ACT clusters from \cite{men10}, and in purple squares
  SPT clusters from \cite{van10}. The blue area represents the masked
  area of $\left|b\right| < 14 $ deg.}
\label{Figskymap}
\end{figure*}

In the following we will detail the outcome of the external validation
of the 190 SZ candidate clusters retained after the internal
validation. We find that they are distributed between known clusters
(169 in total, Fig.~\ref{Figskymap} blue) and 21 candidate new
clusters. Among those 21, twelve have been confirmed
(Fig.~\ref{Figskymap} yellow) and these are discussed in Section
\ref{sec:conf}.  Nine remain as candidate new clusters requiring
confirmation (Fig.~\ref{Figskymap} red); they are described in Section
\ref{sec:cand}. The further checks performed on the candidate new
clusters resulted in the rejection of one of the nine candidates.

\begin{figure*}
   \centering
\includegraphics[width=8cm ]{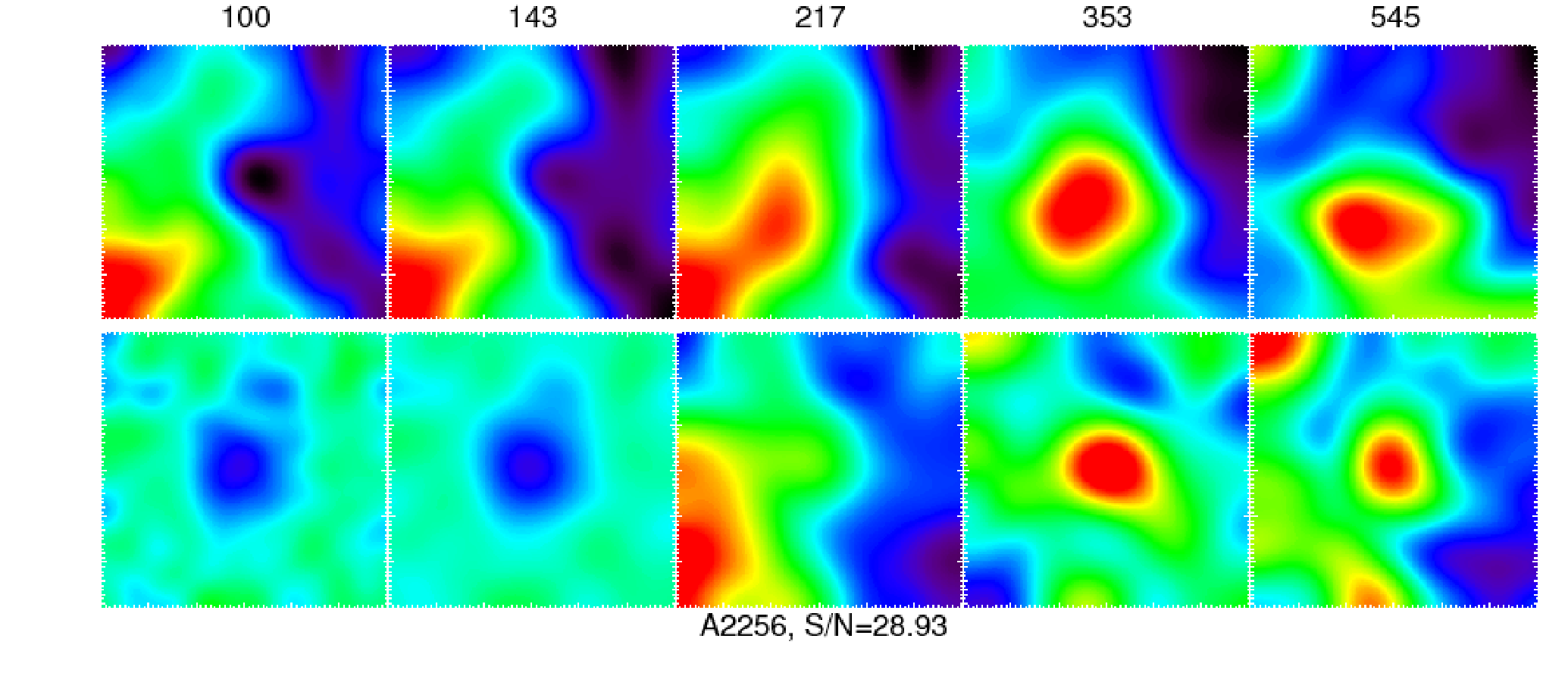}
\includegraphics[width=8cm ]{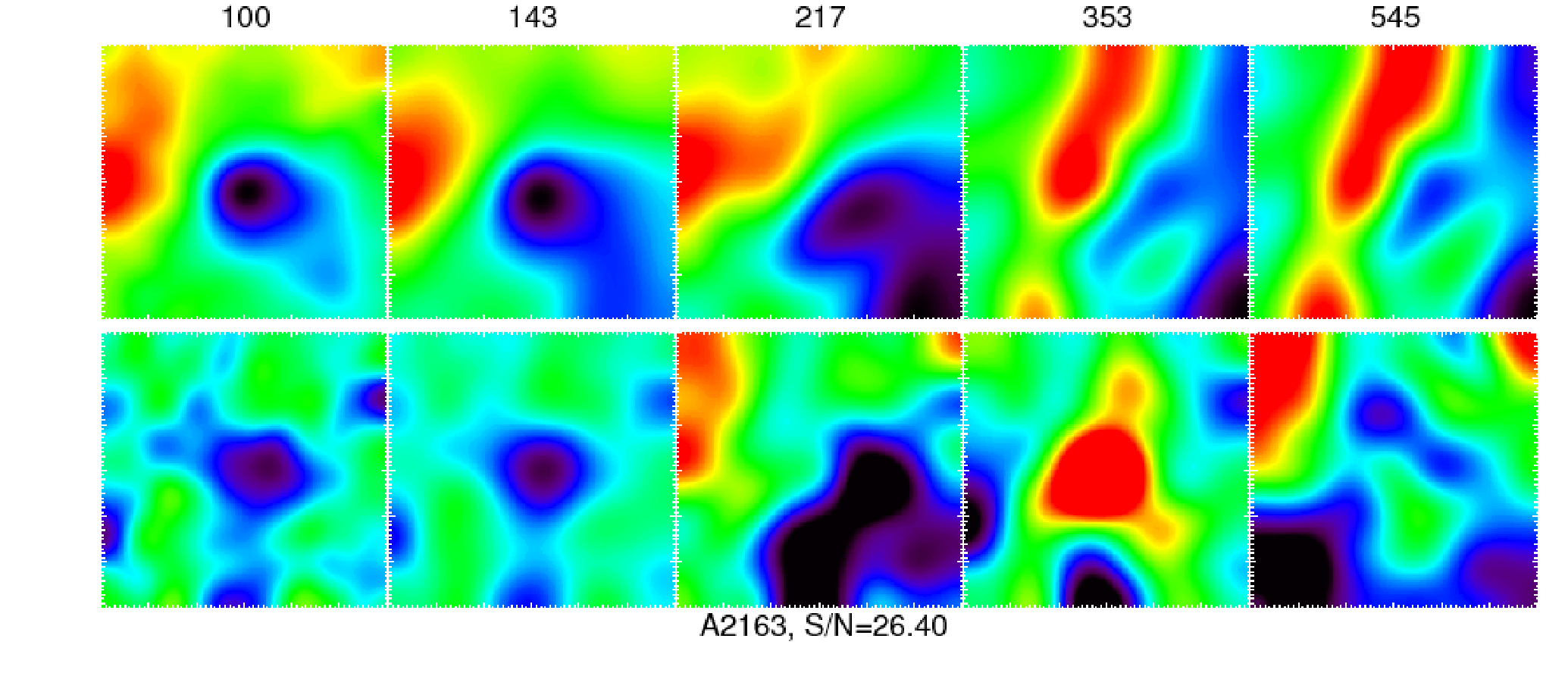}\\
\includegraphics[width=8cm ]{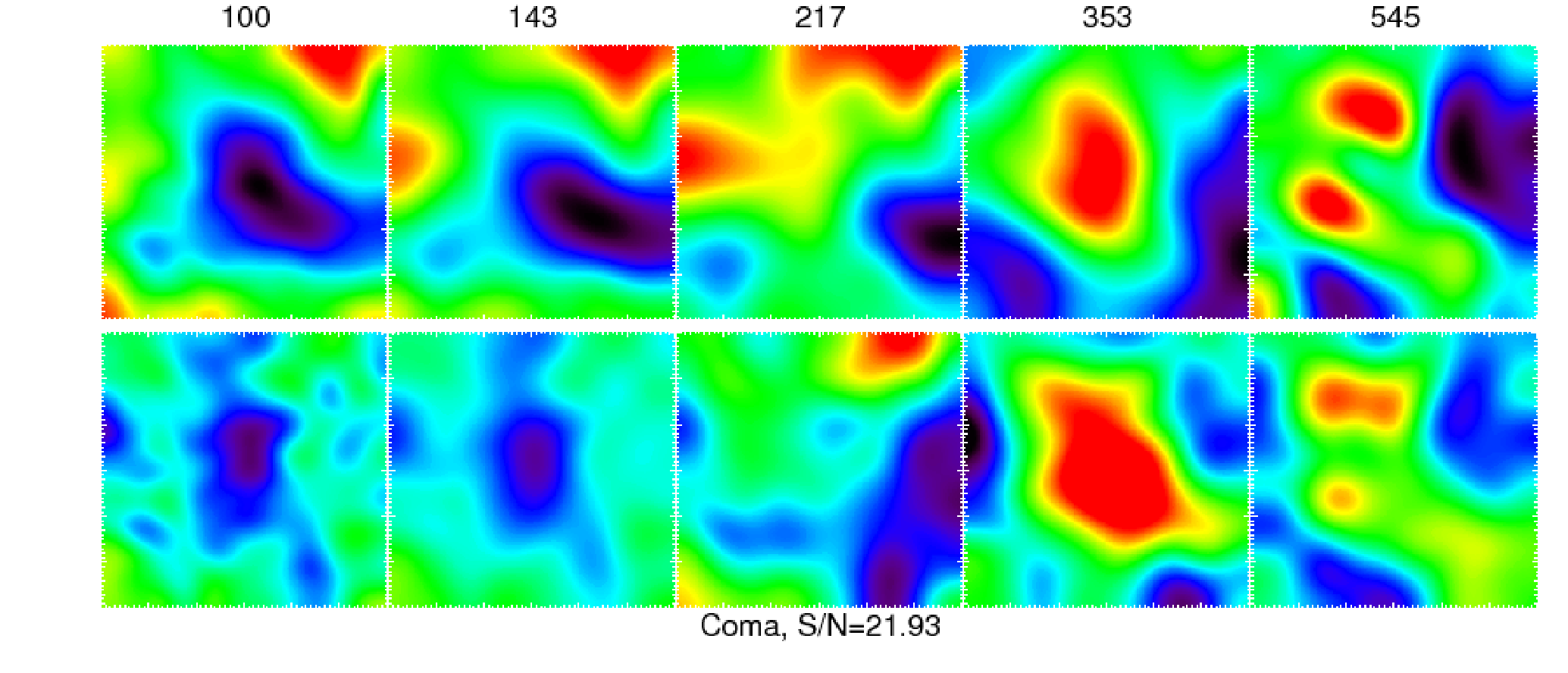}
\includegraphics[width=8cm ]{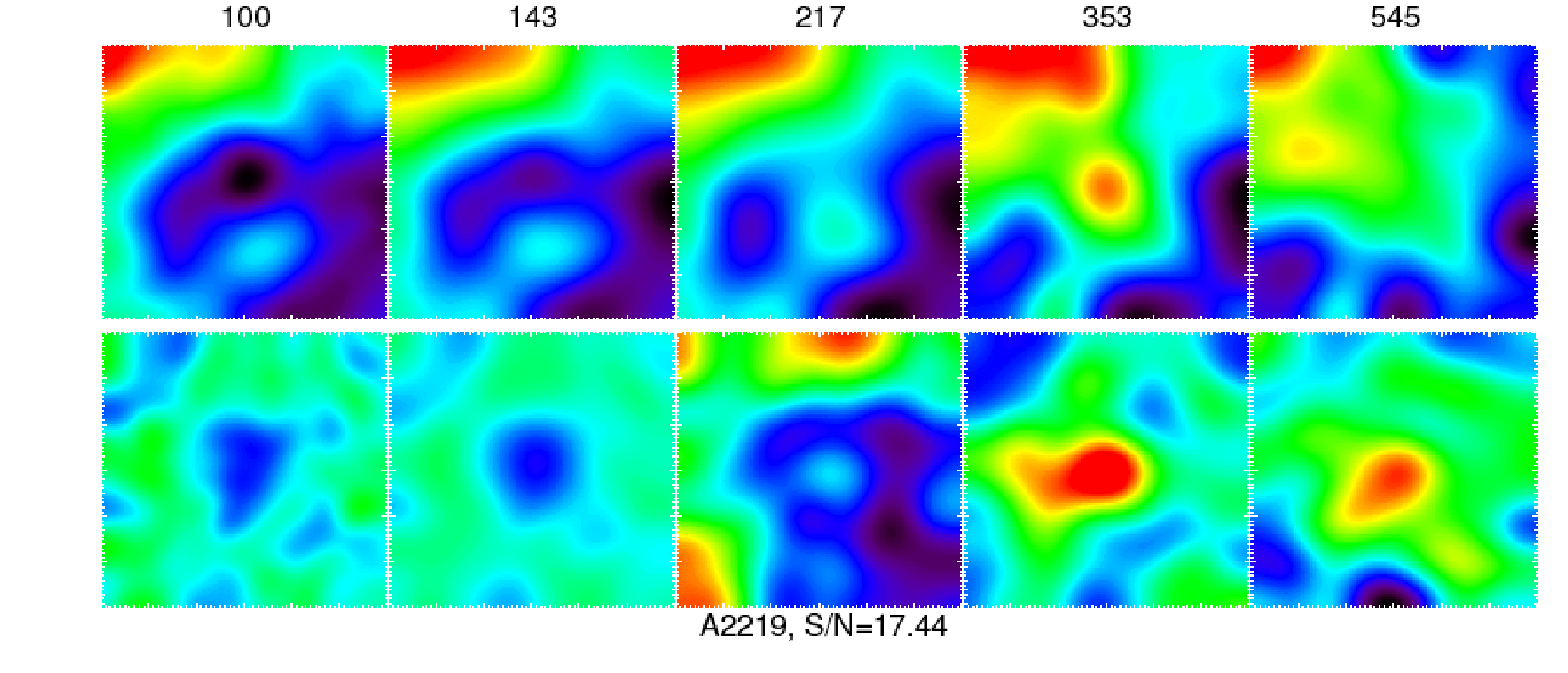}\\
\includegraphics[width=8cm ]{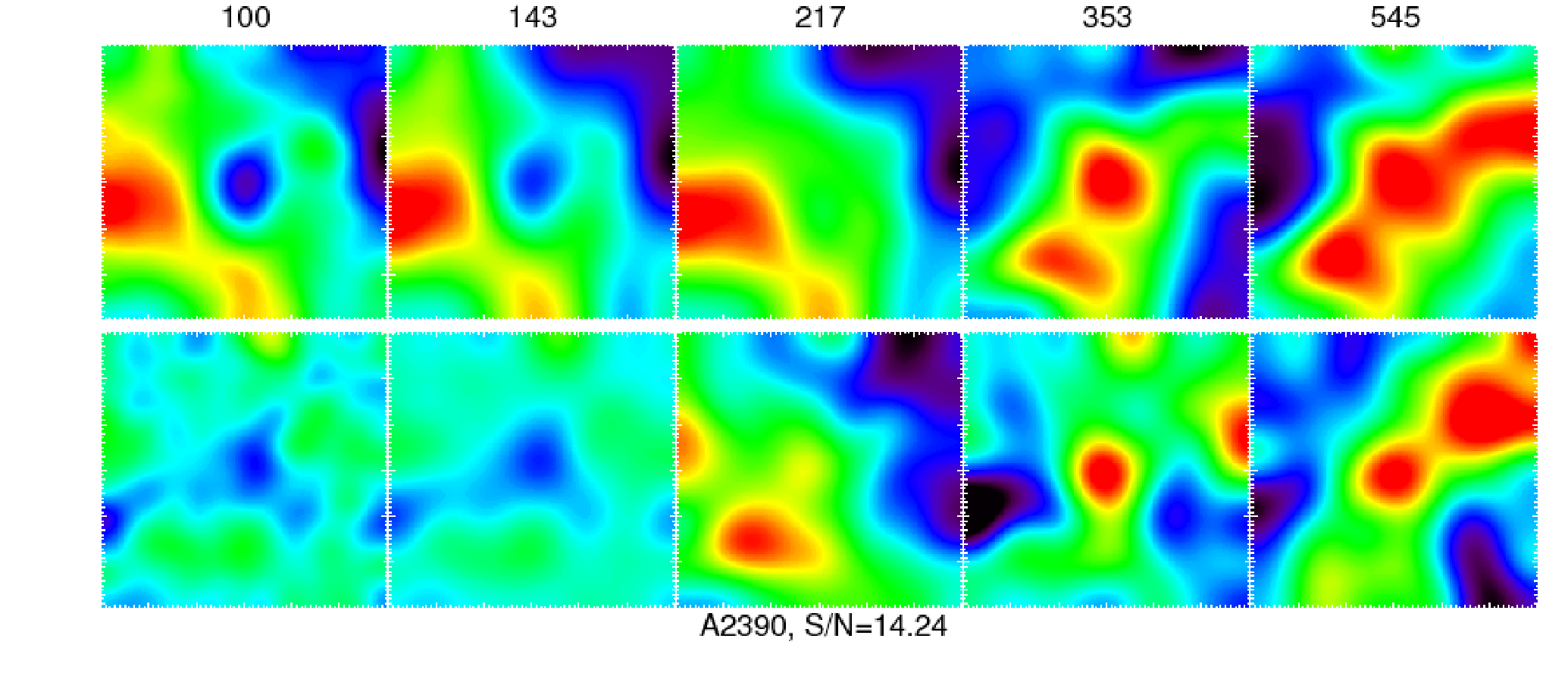}    
\includegraphics[width=8cm ]{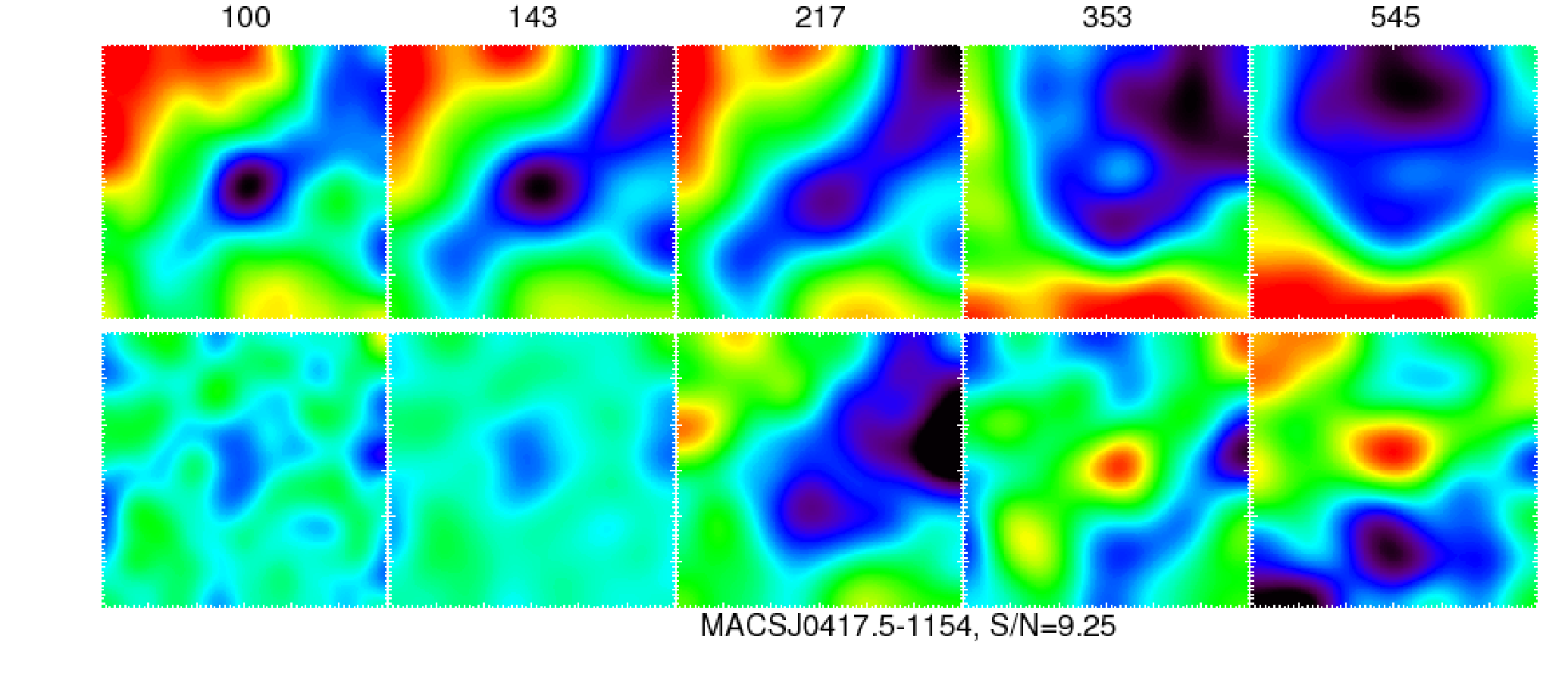} \\  
\includegraphics[width=8cm ]{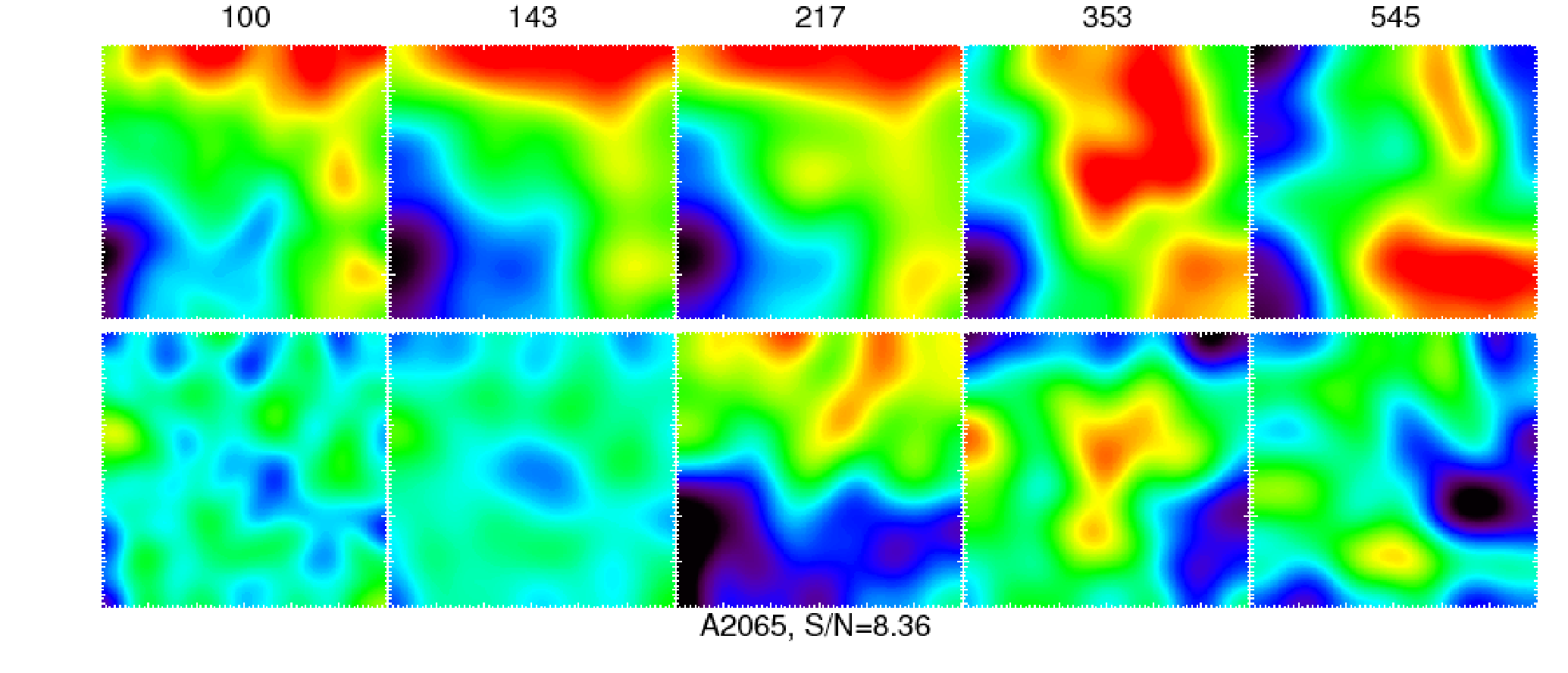}
\includegraphics[width=8cm ]{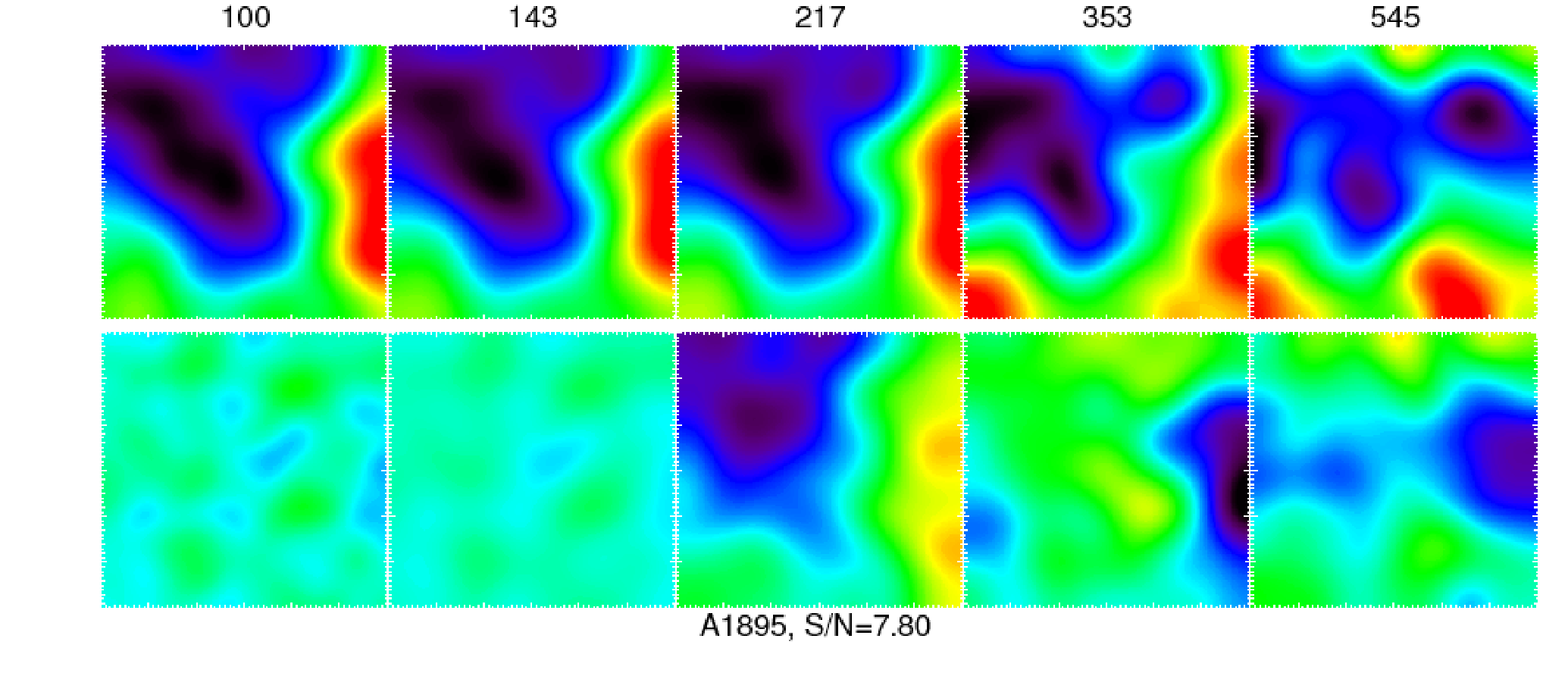}\\
\includegraphics[width=8cm ]{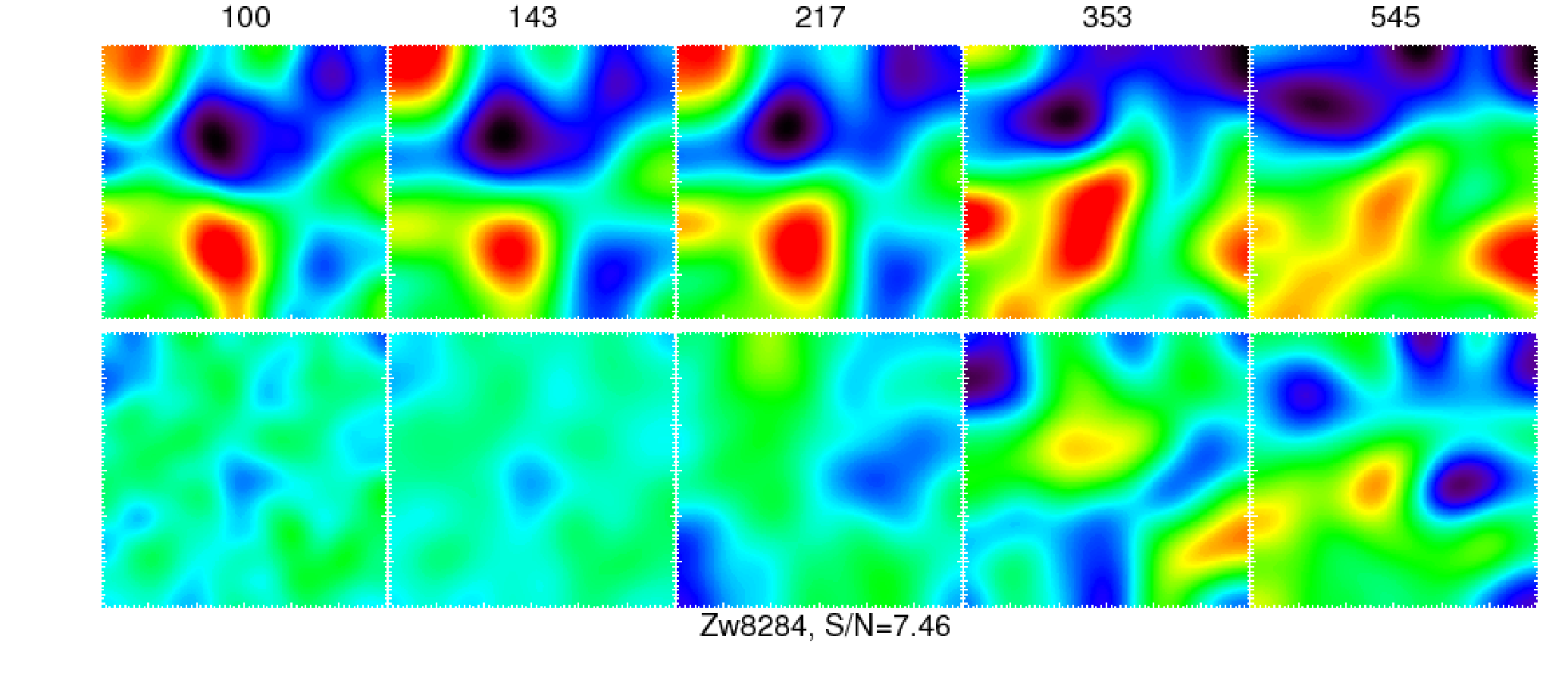}
\includegraphics[width=8cm ]{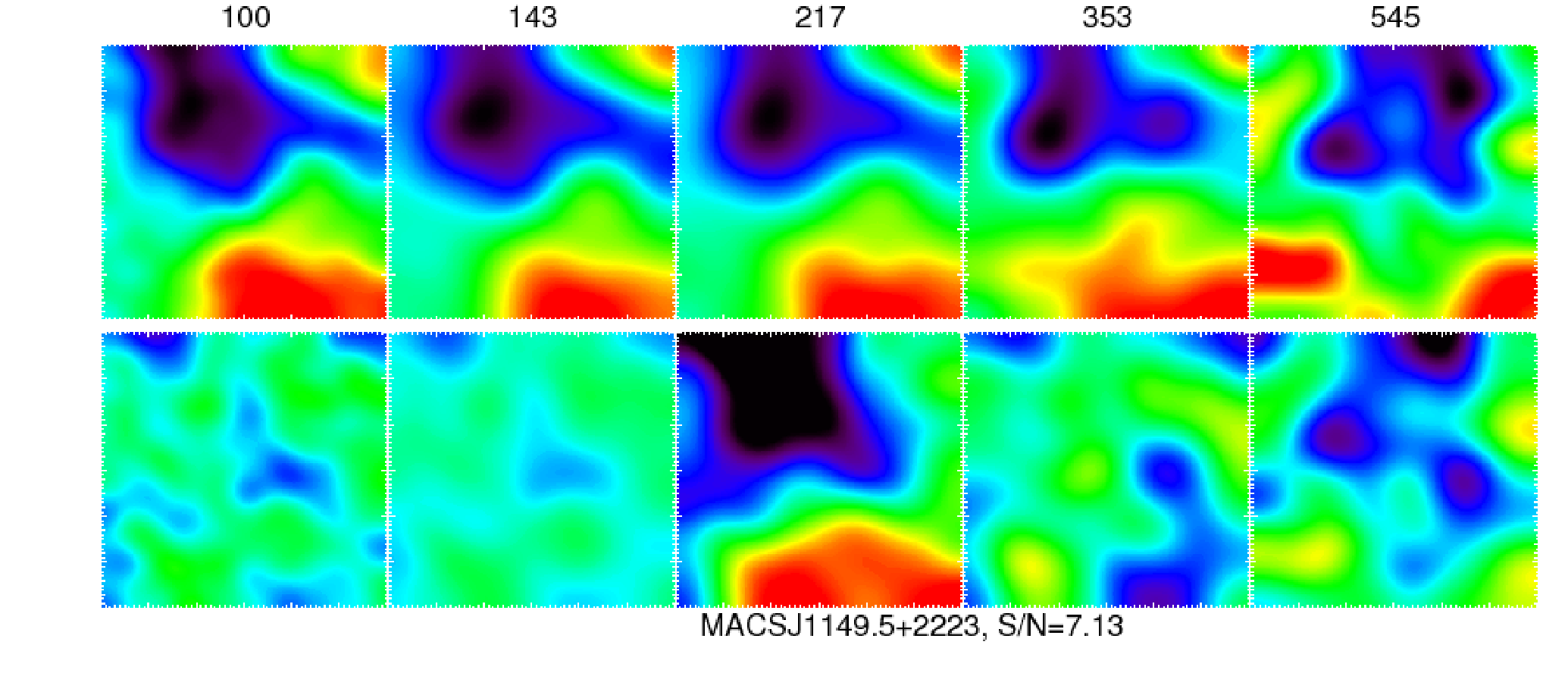}\\
\includegraphics[width=8cm ]{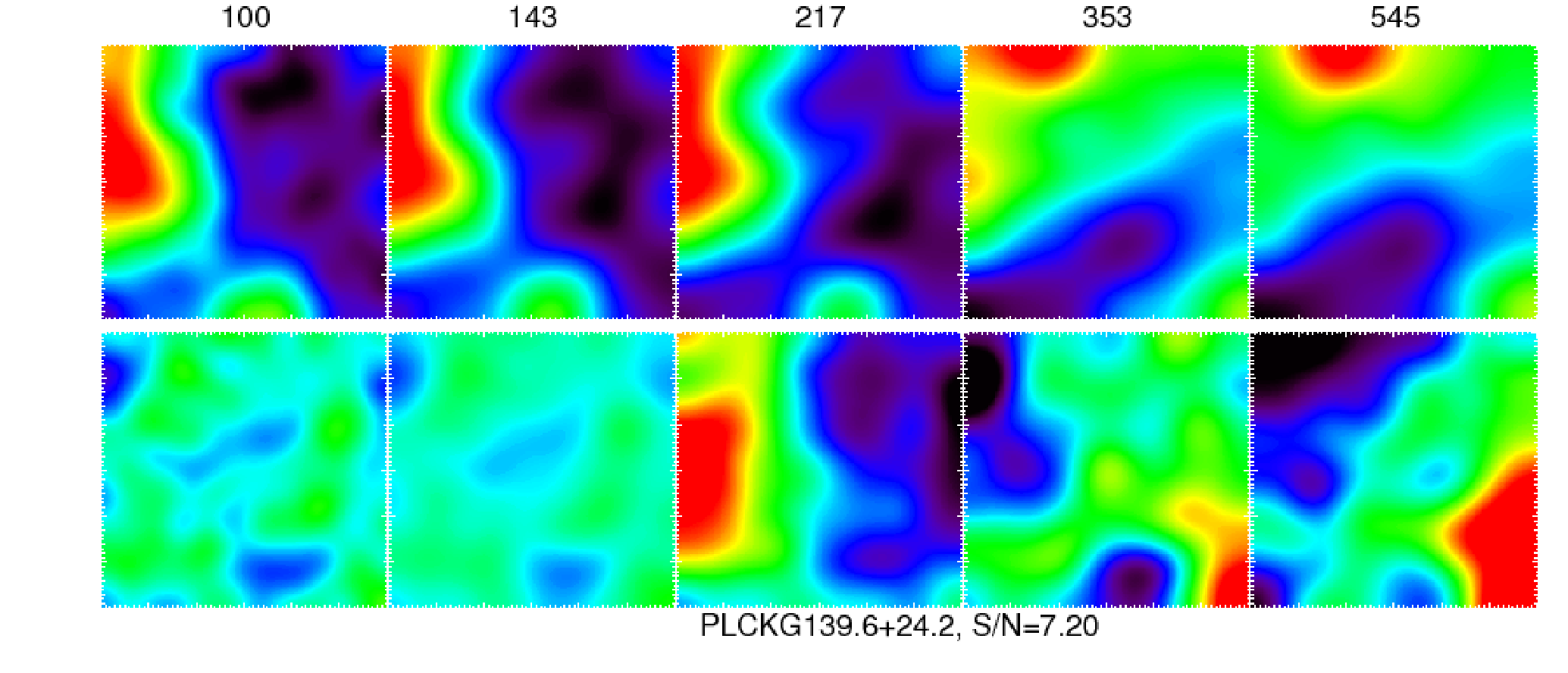}    
\includegraphics[width=8cm ]{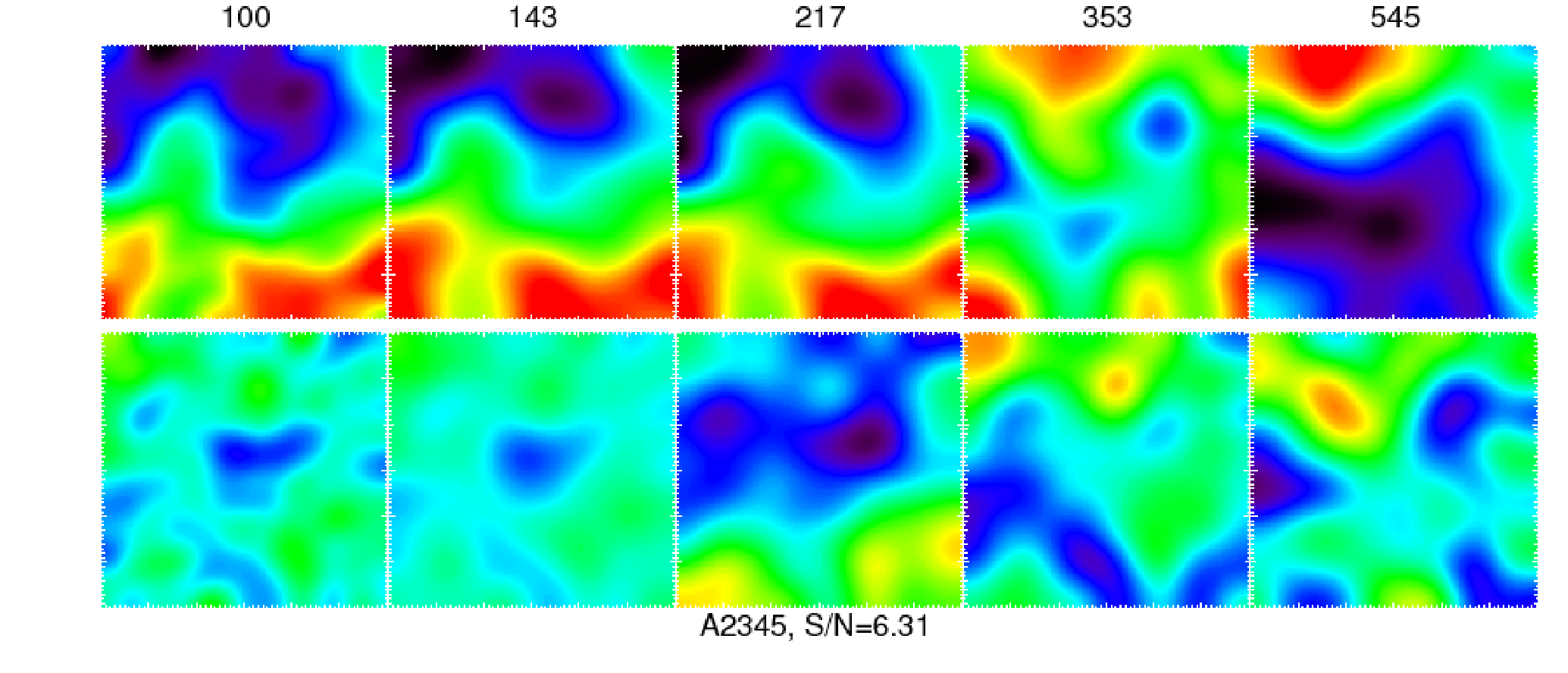}    
\caption{Observations of a few clusters from the ESZ sample. For each
  cluster, the upper panels show the raw (1 square degree) maps at
  100, 143, 217, 353, and 545GHz. The lower panels show the
  corresponding cleaned maps (see Sect.~\ref{sec:cand}). These
  clusters span S/N from 29 to 6 from the upper left to the lower
  right.}
         \label{fig:a2219}
   \end{figure*}

\begin{figure*}
   \centering
\includegraphics[width=5cm]{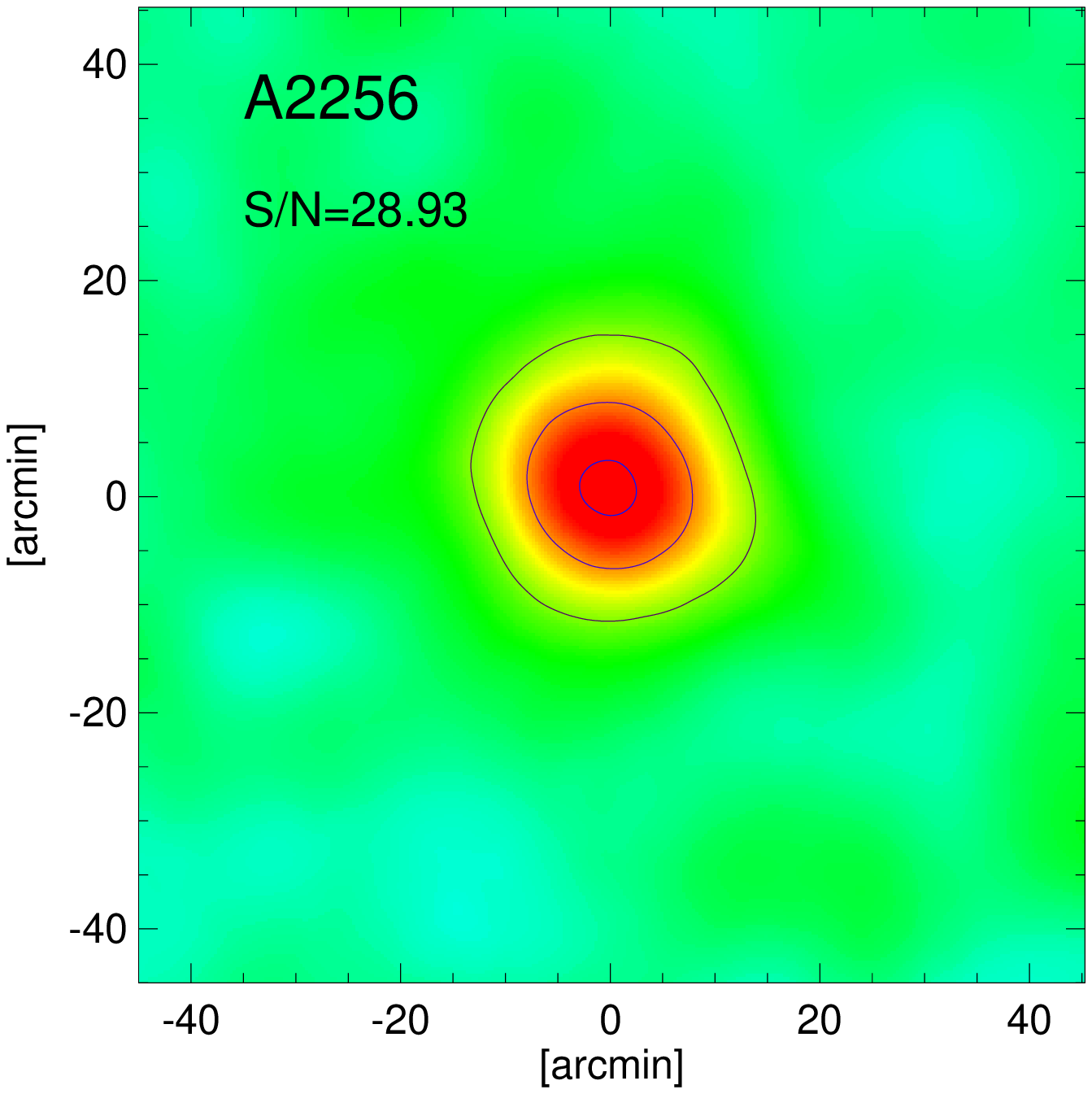}
\includegraphics[width=5cm]{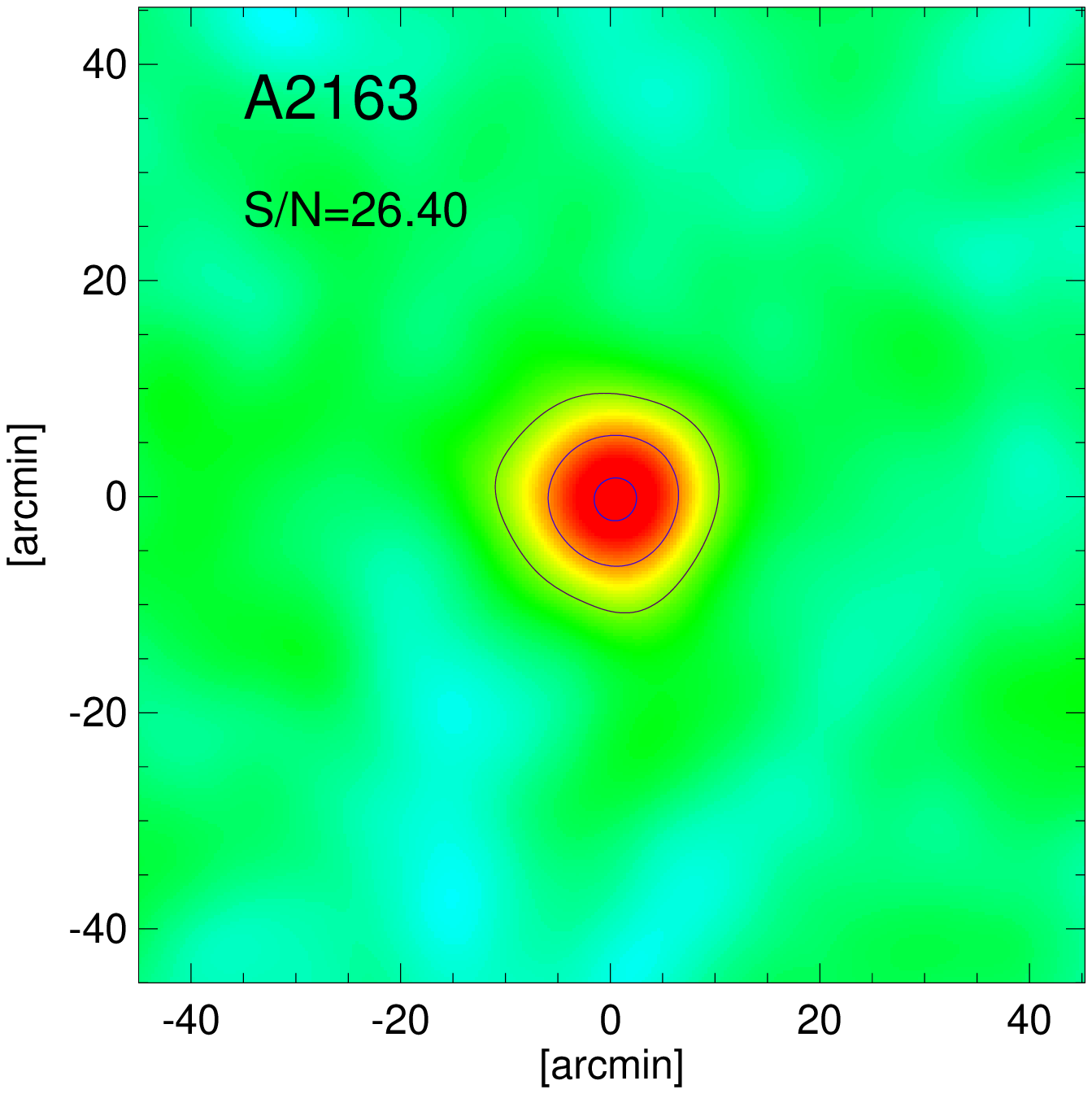}
\includegraphics[width=5cm]{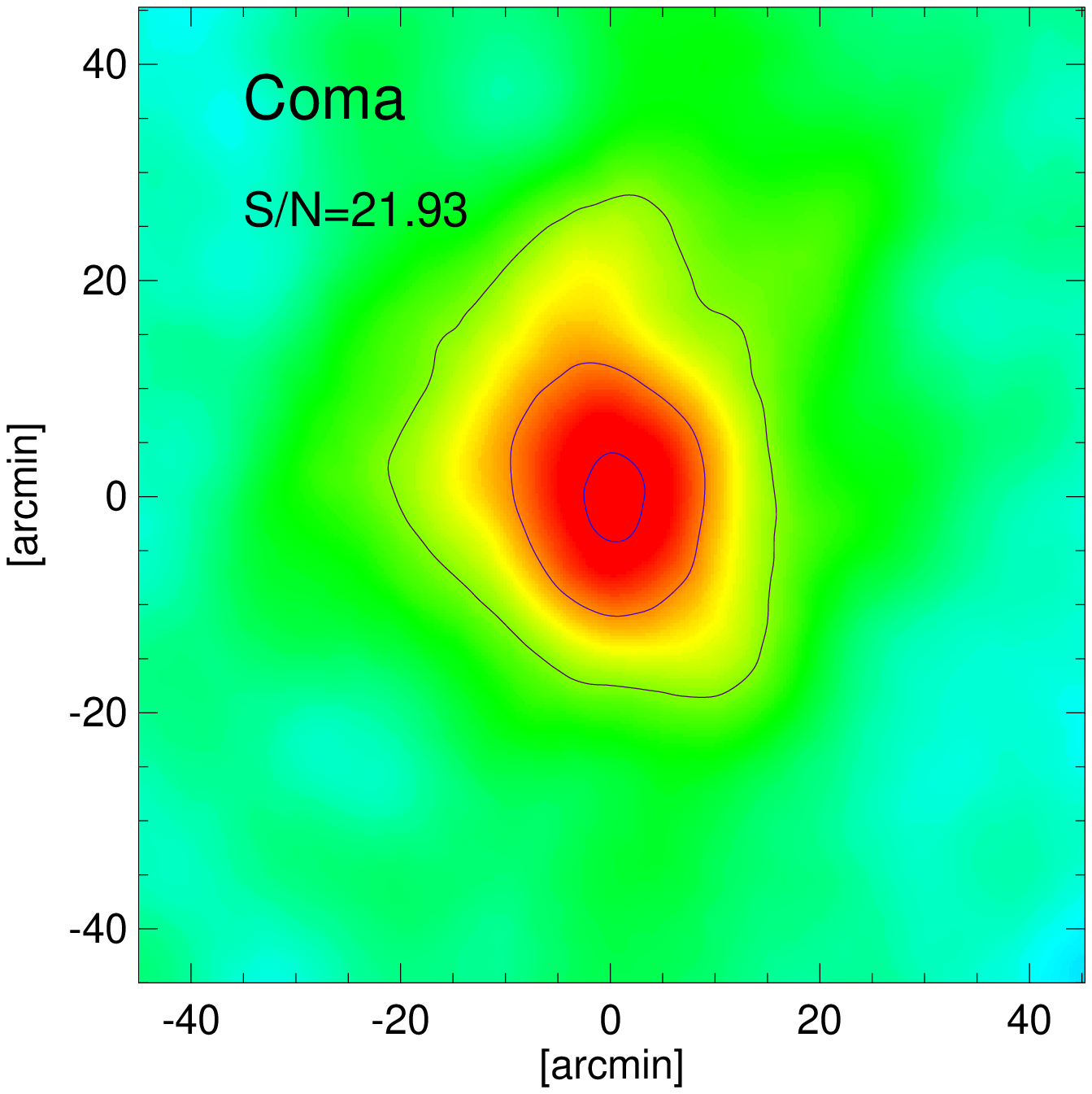}\\
\includegraphics[width=5cm]{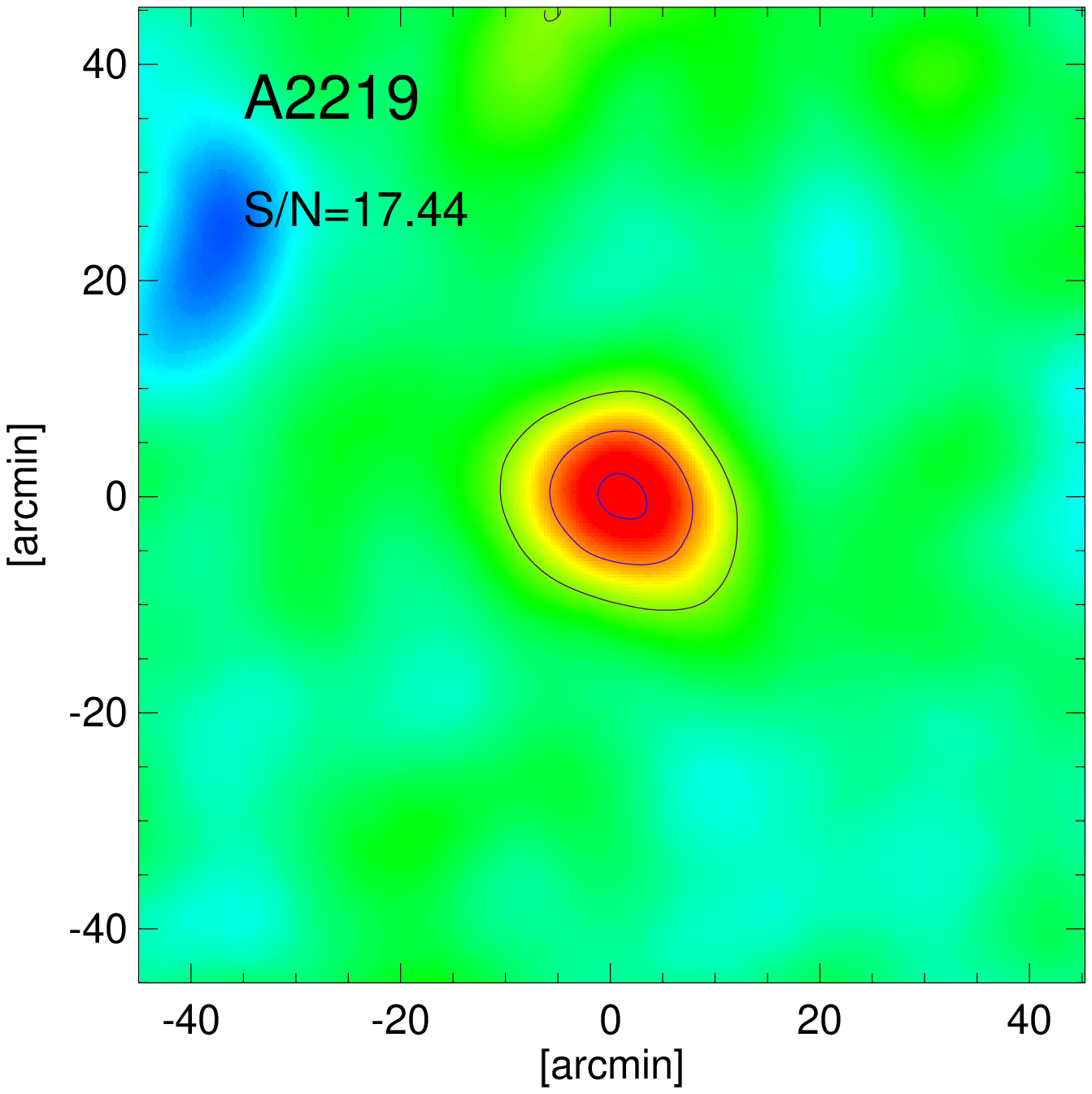}
\includegraphics[width=5cm]{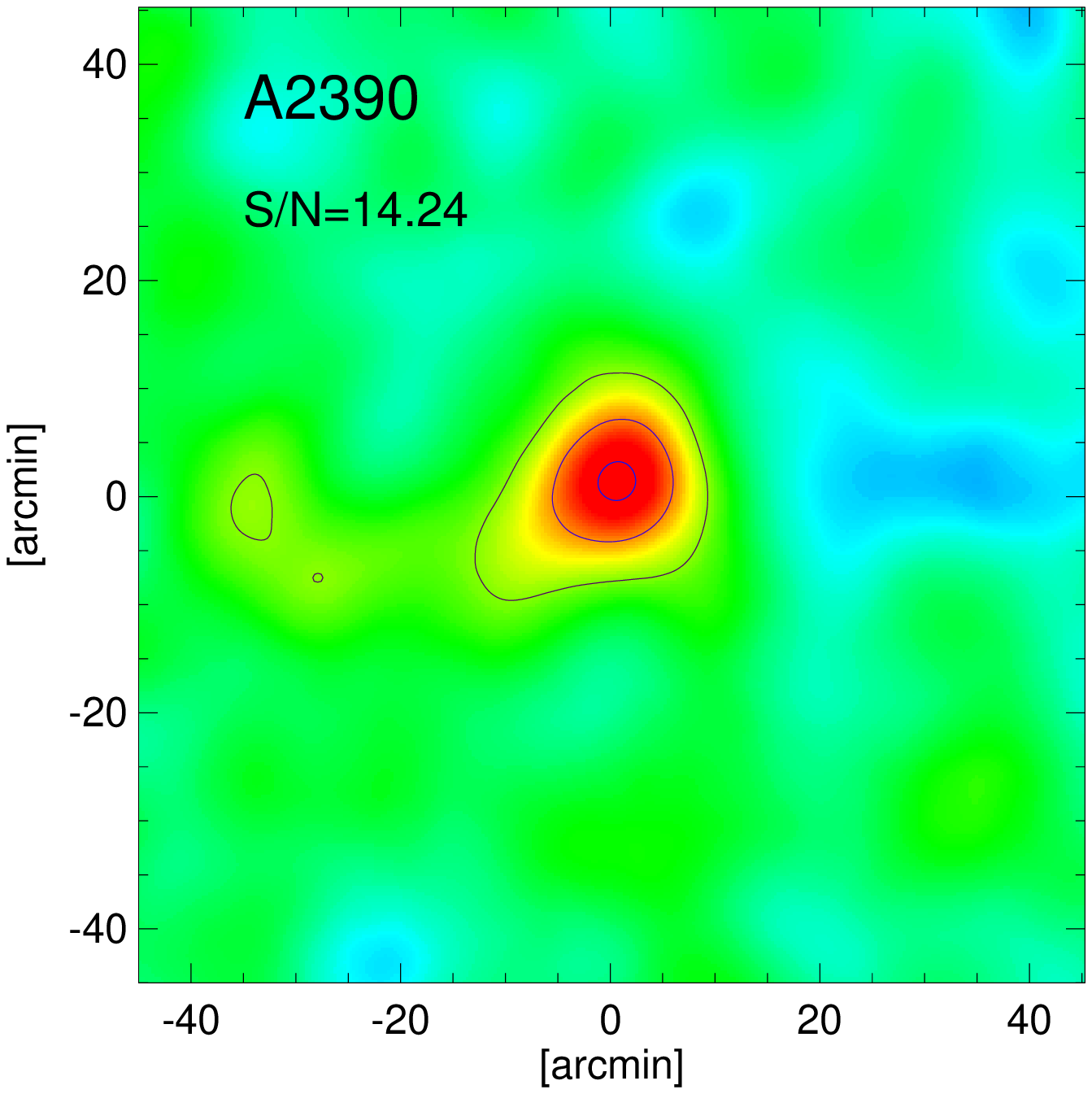}
\includegraphics[width=5cm]{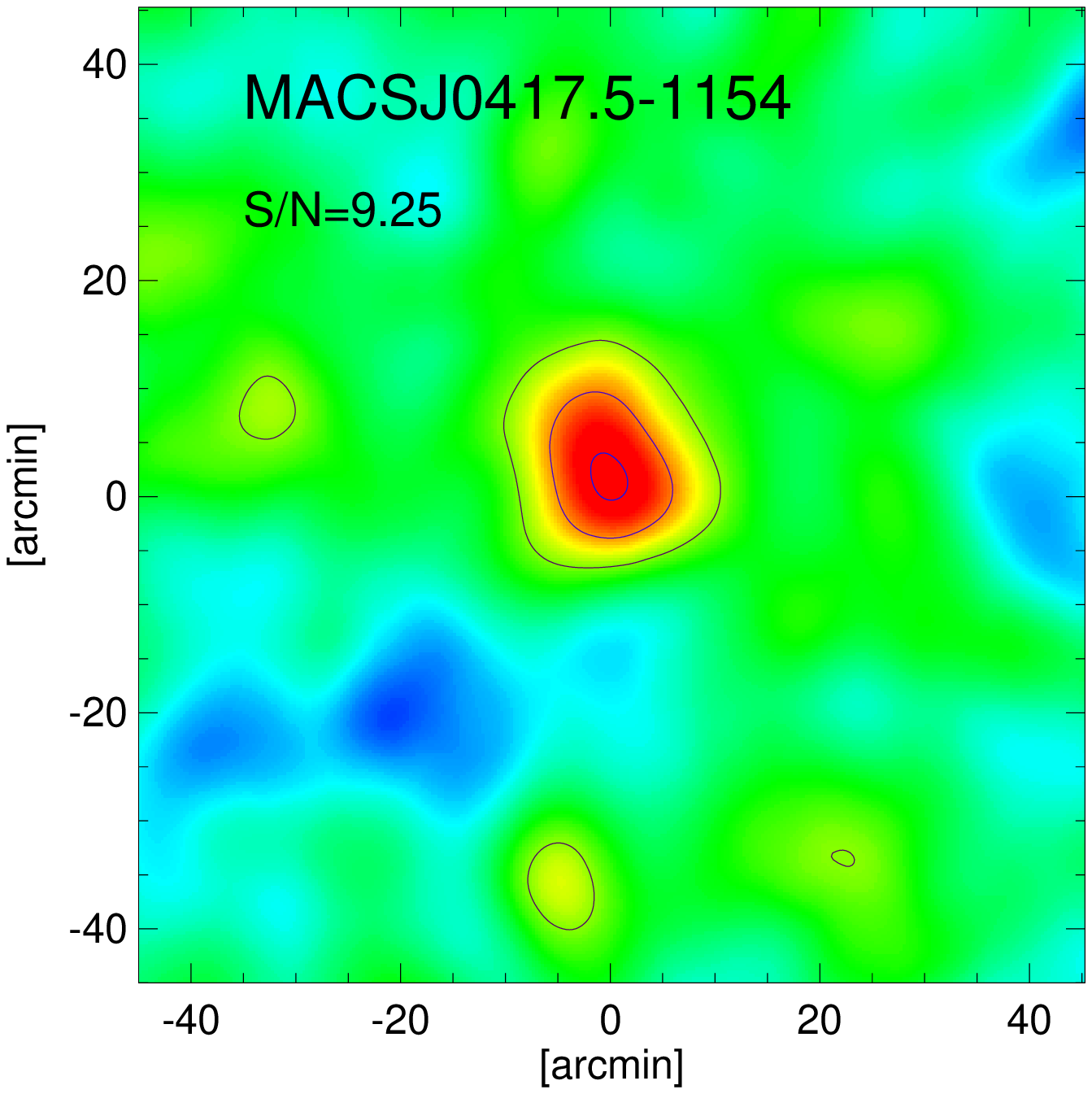}\\
\includegraphics[width=5cm]{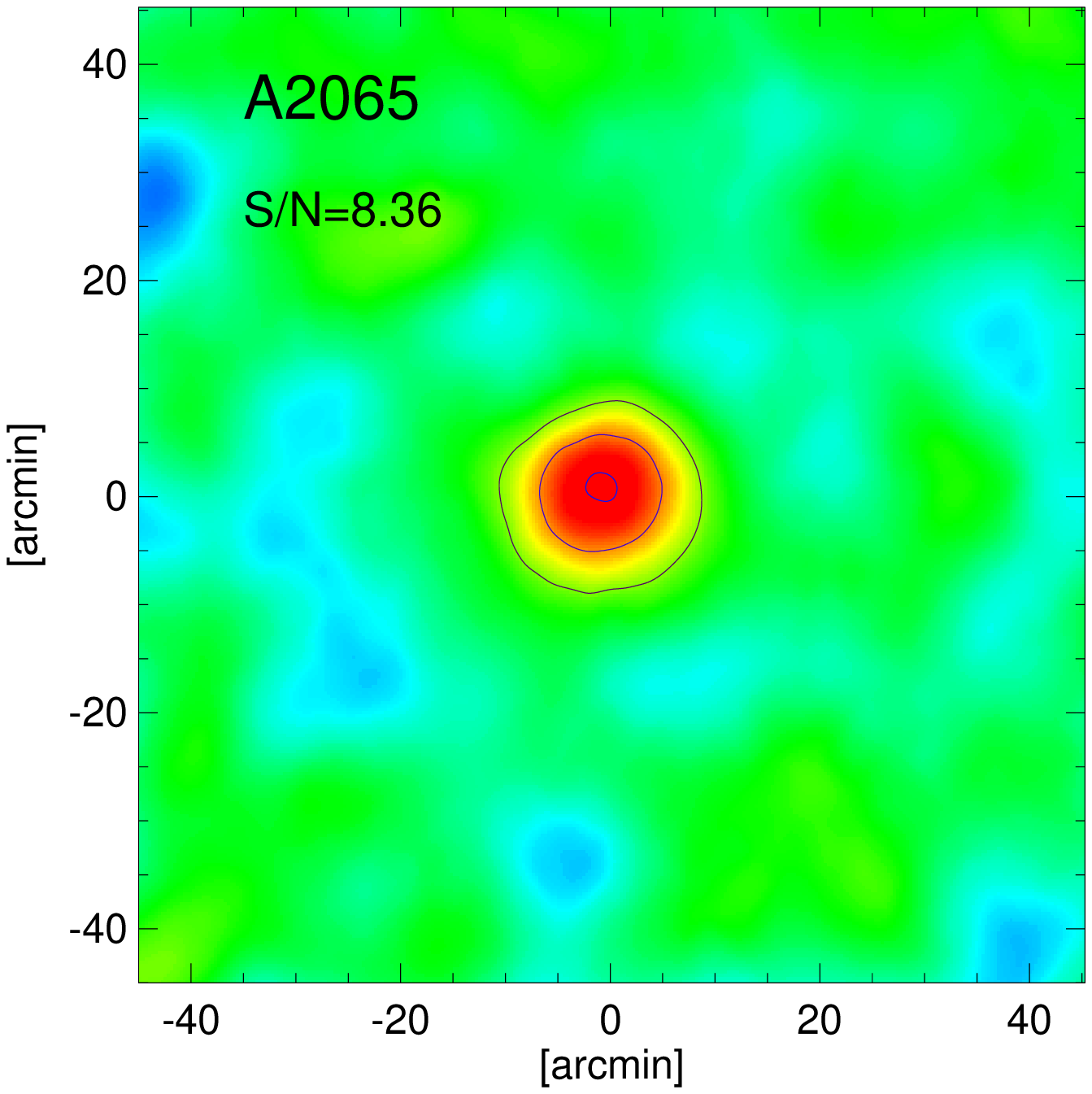}
\includegraphics[width=5cm]{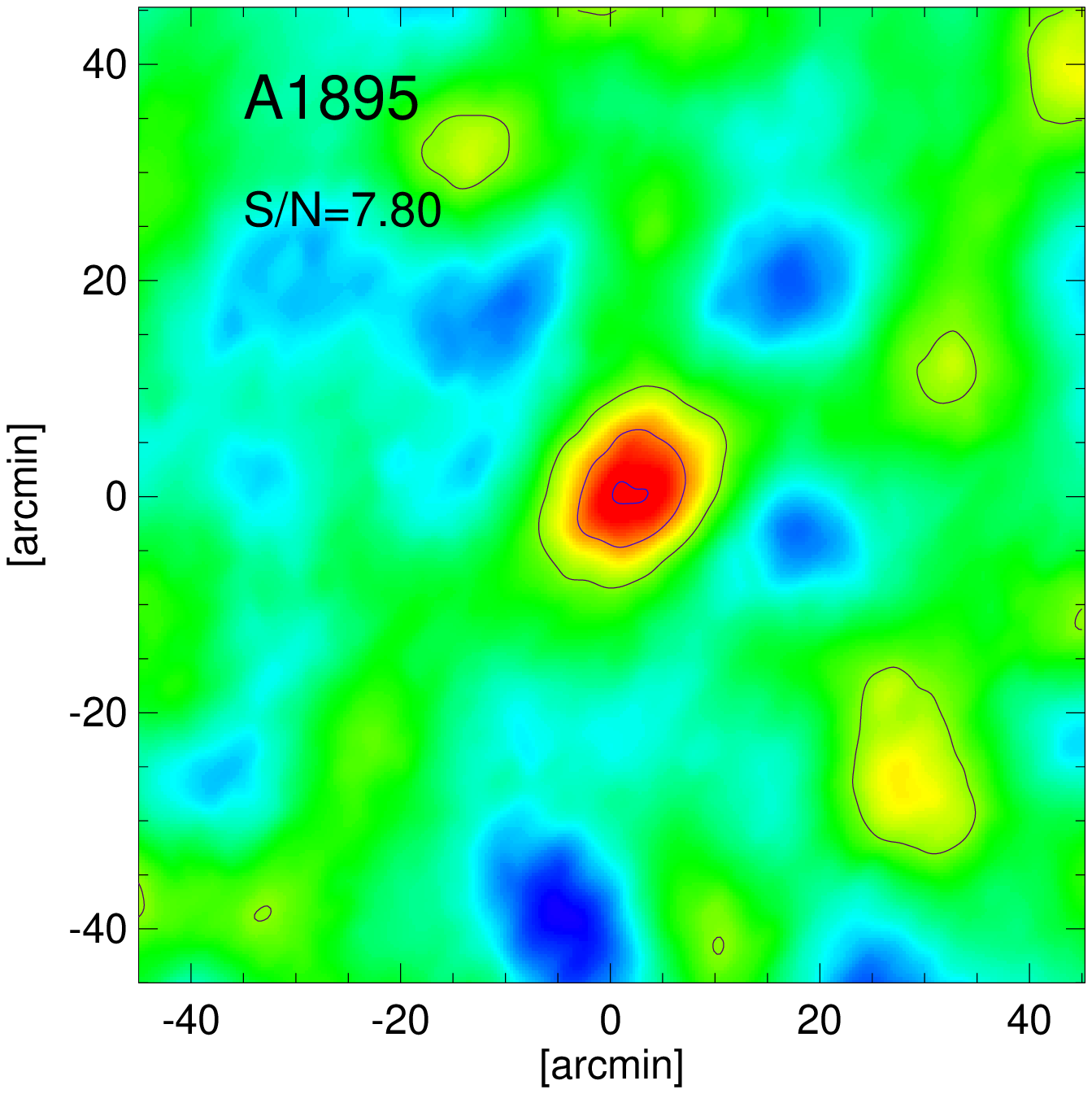}
\includegraphics[width=5cm]{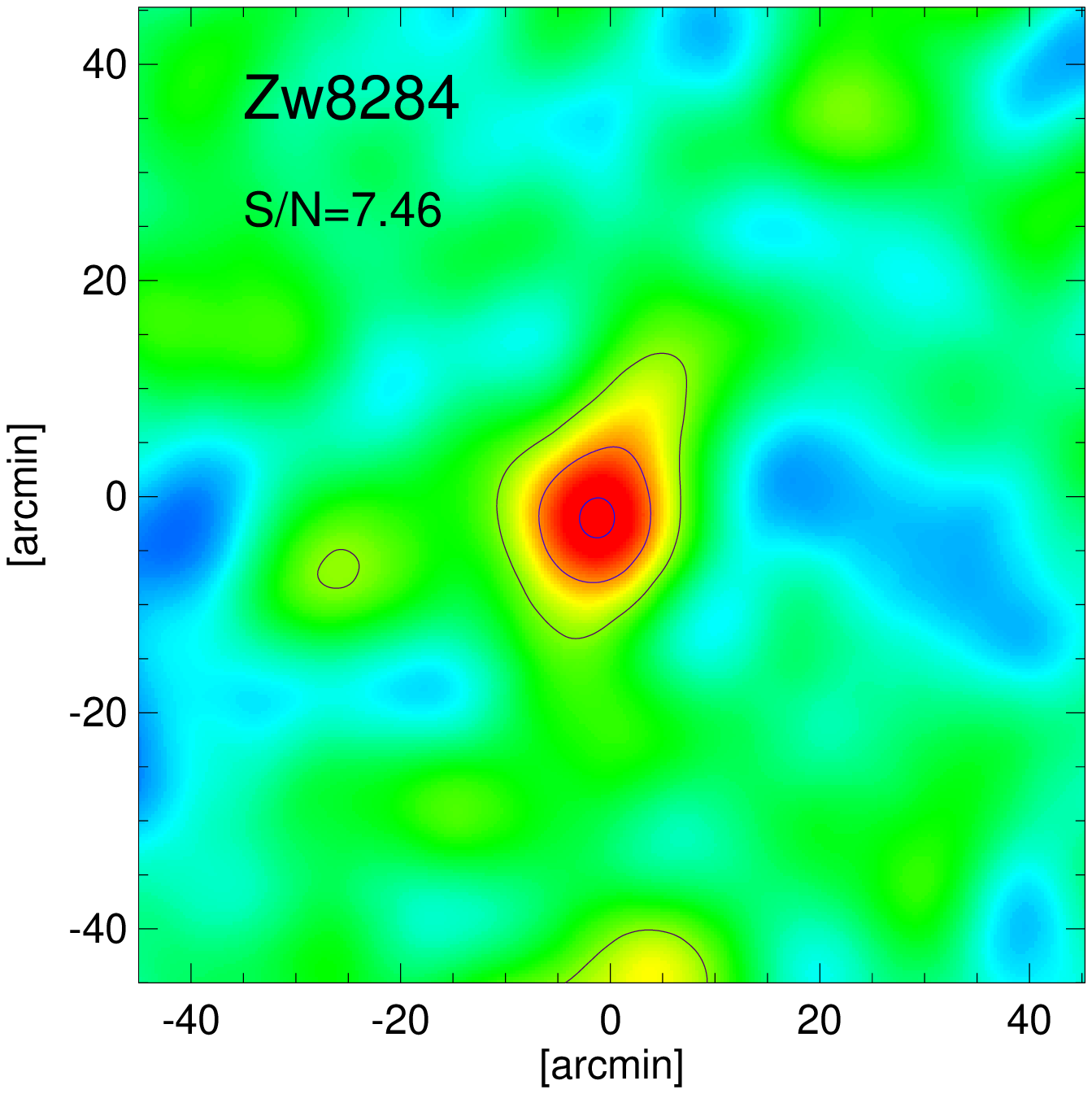}\\
\includegraphics[width=5cm]{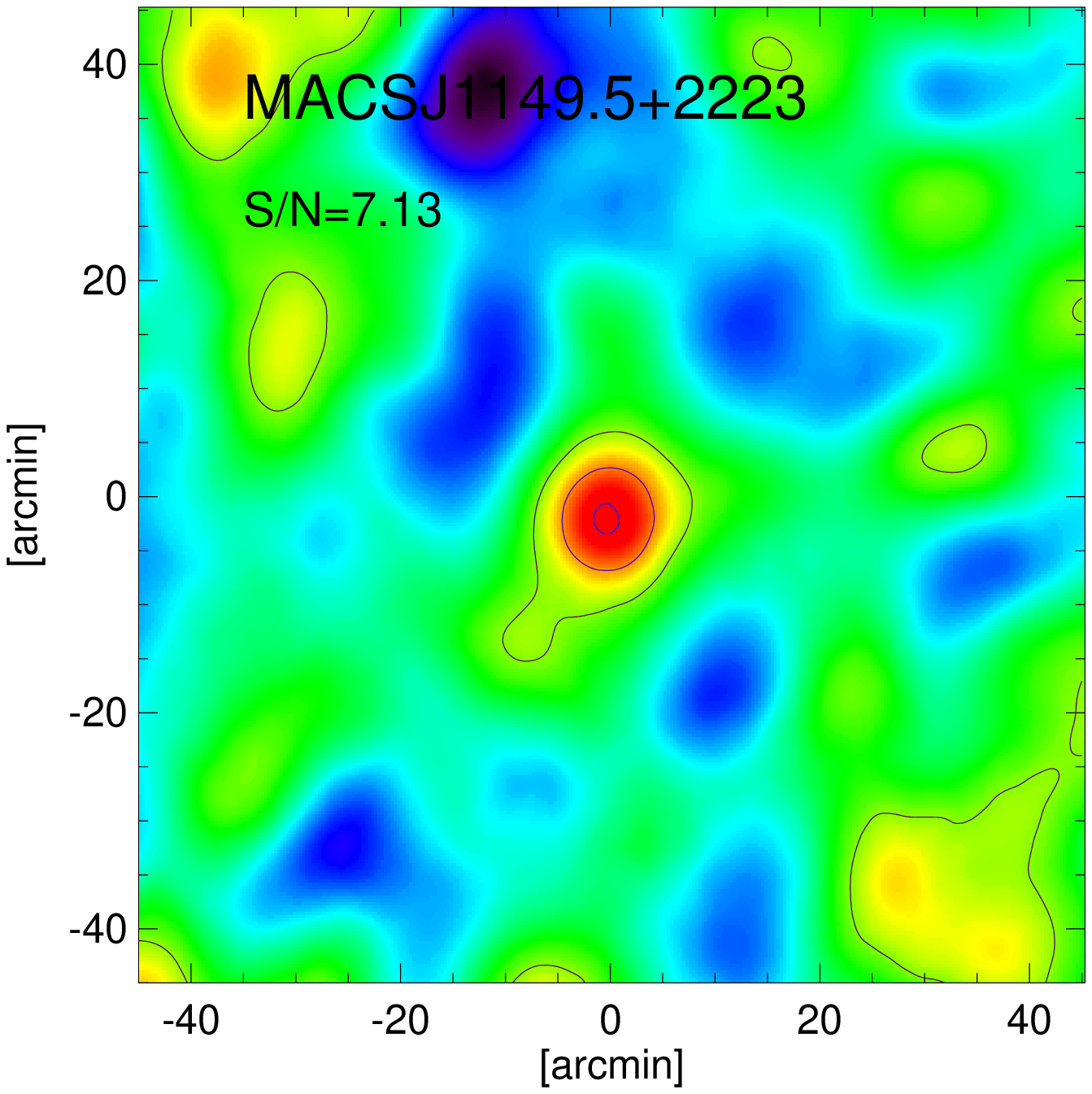}
\includegraphics[width=5cm]{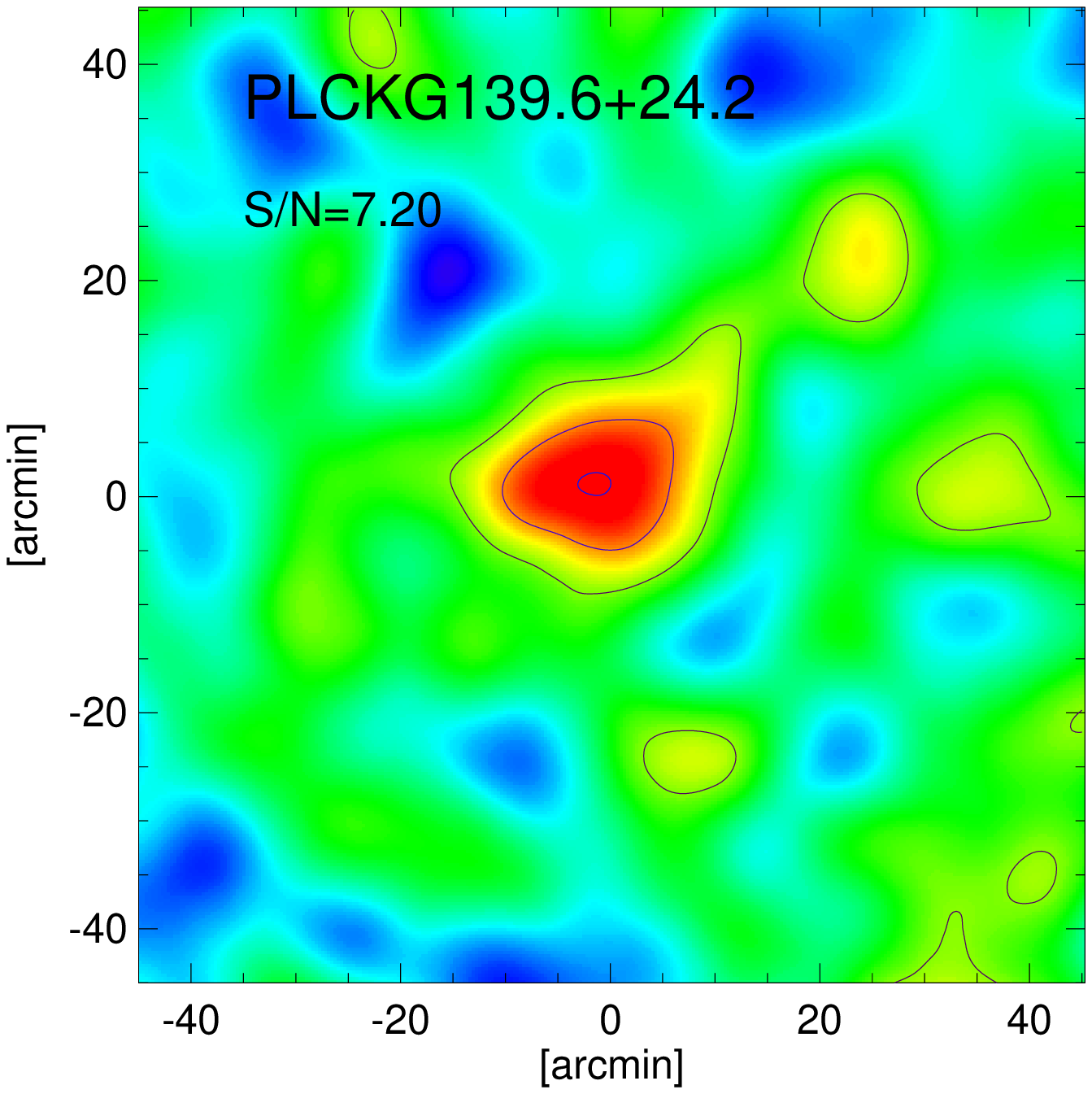}
\includegraphics[width=5cm]{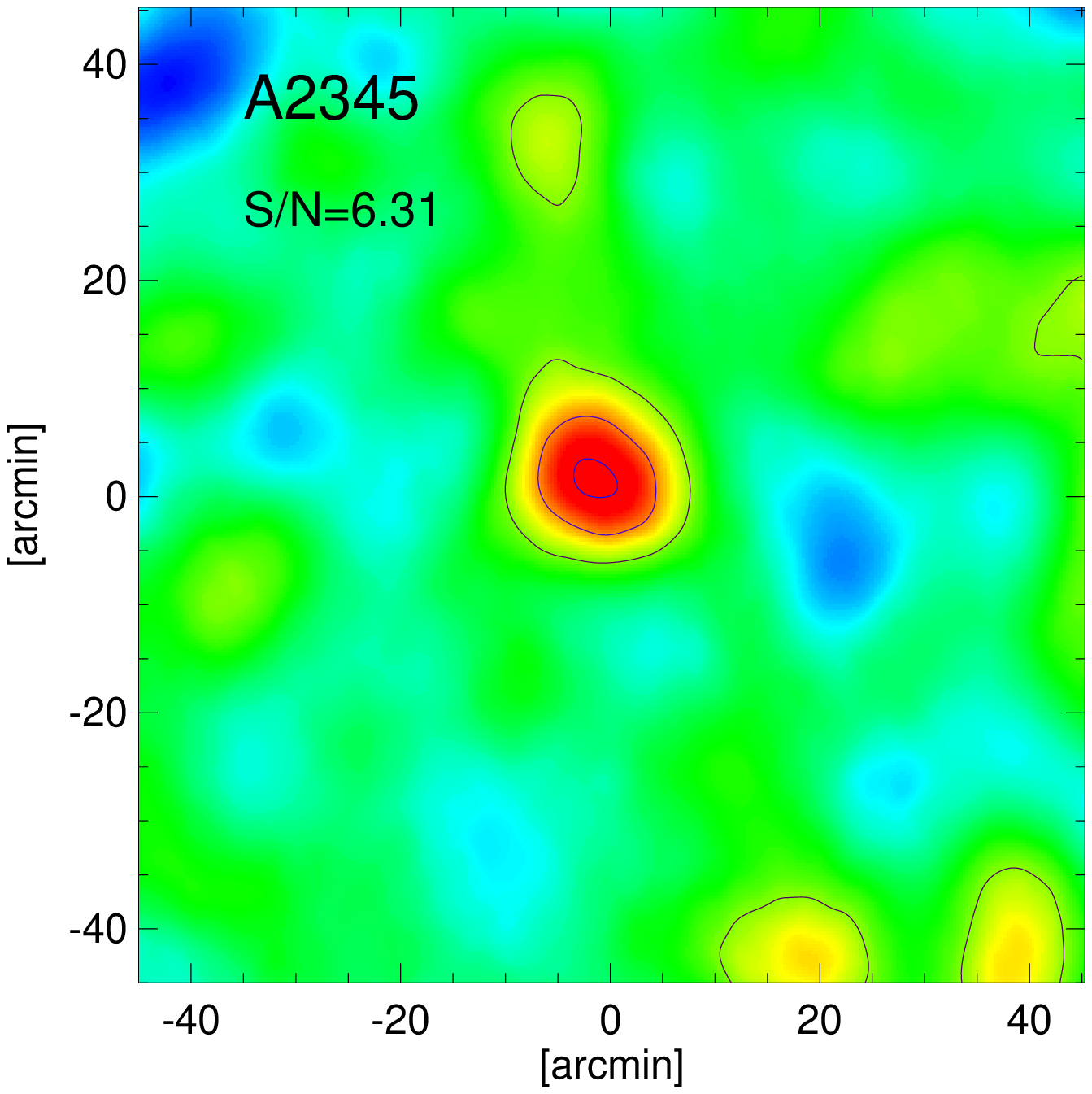}
\caption{Illustrations of reconstructed $y$-maps
  ($1.5\deg\times1.5\deg$, smoothed to 13 arcmin) for clusters
  spanning S/N from 29 to 6 from the upper left to the lower right.}

\label{FigVibStab2}
\end{figure*}

The final released ESZ list thus comprises 189 clusters or candidate
new clusters. The content of the released data\footnote{{\it
    http://www.rssd.esa.int/Planck}} is presented in
Appendix~\ref{tab:eszfits}. Table~\ref{tab:cand} summarises the
different steps of ESZ sample construction and validation detailed in
the previous sections. Figures~\ref{fig:a2219} and \ref{FigVibStab2}
show illustrations of the raw and ``cleaned'' channel maps (from 100
to 545 GHz) as well as corresponding $y$-maps, for a few clusters,
with S/N ranging from the highest ones to more typical ones.

\begin{table*}
\centering
\caption{Summary of the ESZ sample construction and validation steps.
\label{tab:cand}}
\begin{tabular}{l l r r }
\hline
\hline
Selection & & SZ Candidates & Rejected\\
\hline
 & S/N $\geq 6$ and good quality flag on SZ spectrum & 201& \\
 & Detected by one method only & & 11\\
 & Bad quality flag from visual inspection & & 1 \\
\hline\hline
ESZ sample& & 189 & \\
\hline\hline
Known clusters & & 169 & \\
\hline
& X-ray only       & 30& \\
& Optical only     & 5& \\
& NED\/Simbad only & 1& \\
& X-ray + Optical  & 128& \\
& X-ray + SZ       & 1& \\
& SZ + Optical     & 1& \\
& X-ray + Optical +SZ  & 3& \\
\hline
New \Planck\ clusters & & 20& \\
\hline
 & XMM confirmed & 11 & \\
 & AMI confirmed & 1  & \\
 & Candidate new clusters & 8 & \\
\end{tabular}
\end{table*}

\subsection{ESZ candidates identified with known clusters}
\label{sec:known}

The external validation with ancillary data identified 169 clusters in
total out of the 190 candidates detected blindly.  They are known
X-ray or optical clusters and \Planck\ data provide the first measure of
the SZ signal for the majority of them, opening a new observational
window on those already known objects.

Most of the identified SZ candidates, 162 in all, were associated with
known clusters from the MCXC compilation and 158 have known redshifts
(provided in the compilation), X-ray luminosities, X-ray estimated
sizes ($\theta_{500}$), etc. Moreover, as expected, a very large
fraction of them (127 clusters) are at the same time identified in the
optical. They are mostly Abell clusters from the ROSAT X-ray cluster
catalogues.

The remaining seven identified \Planck\ clusters were obtained from
search in SIMBAD (one cluster, RXJ0748.7+5941, observed by ROSAT
  but not part of the NORAS catalogue \citep{app98} and without
published redshift), from logs of observatories (one cluster,
H1821+643 at $z=0.299$ \citep{sch92}) and from optical only, i.e.,
without an X-ray counterpart, identification with Abell or Zwicky
clusters (five clusters). These are ZwCl2120.1+2256, AC114Northern,
A3716S, ZwCl1856.8+6616, and ZwCl0934.8+5216 clusters. The last two
have no published redshifts. For all these clusters, redshifts when
they are available are retrieved from the SIMBAD and NED databases.

The cross-match with known SZ clusters further indicates that one
cluster, AS0520, is common to \Planck, ACT and SPT. Additionally,
five\footnote{One of the candidate new clusters confirmed by
  XMM-{\it Newton} appeared in publication as one of the ACT SZ
  optically-confirmed clusters \citep{men10} to be observed by {\it
    Chandra}, after we scheduled it for observation with 
XMM-{\it Newton}: PLCKESZ
  G262.7-40.9/ACT-CL J0438-5419. We retain it as new candidate in the
  following.}  clusters from ACT are in common with the \Planck\ ESZ
sample, and twelve massive clusters observed by SPT \citep{pla10} are
also observed by \Planck\ and quoted in the ESZ.  Finally by comparing
with the SZ compilation from \cite{dou11}, we find that, in total, 41
clusters from the ESZ sample have already been observed in SZ by
previous experiments. For these clusters \Planck\ provides us with a
homogeneous set of SZ measures. Moreover, out of the full ESZ sample
about 80\% have been observed in SZ for the first time and have a
homogeneous measurement of their Compton parameter from \Planck.

Out of the known clusters in the ESZ sample, a few are given in the
Early Release Compact Source Catalogue (ERCSC)
\citep{planck2011-1.10sup} as they were detected by the source
extraction techniques used to construct the ERCSC. They are 1ES
0657-55.8 (commonly known as the bullet cluster and detected blindly
with an S/N of 19.7), A2218, ACO S0520, CIZA J1938.3+5409, A0119, RXC
J1720.1+2637, A3376, and MACS J2135.2-0102. It is worth noting that
the quoted fluxes in the ERCSC are obtained using aperture photometry
on the channel maps without band merging. They cannot be compared
easily with the obtained integrated Compton parameters in the present
article. Moreover, two of the above-listed clusters, RXC J1720.1+2637
and MACS J2135.2-0102, suffer from astrophysical contamination that
may affect the computed $Y$.
\subsection{New \Planck\ clusters in the ESZ sample}
\label{sec:new}

The ESZ sample contains 20 new clusters or candidates clusters with
S/N ranging from 11.5 to 6. As mentioned above, a follow-up programme
set up to help understand the selection of \Planck\ clusters allowed
us to confirm 12 clusters. Eleven were confirmed with XMM-Newton
snapshot observations, while one cluster was confirmed with AMI
observations and corresponds to an overdensity of galaxies in the WISE
data.

\subsubsection{Confirmed ESZ cluster candidates}  
\label{sec:conf}

The XMM-Newton observations for confirmation of SZ candidates were
based on earlier versions of the channel maps and an earlier version
of the data processing than that used for the ESZ construction. The 25
targets sent for observation were selected in two different campaigns,
a pilot programme (exploring S/N from six down to four) and a higher S/N
programme (above S/N of 5). Among the 21 \Planck\ cluster candidates
confirmed by snapshot observation with XMM-Newton, 11 clusters have a
\Planck\ S/N above six (in the present map version)
and thus meet the ESZ selection criteria. Two of them were found to be
double clusters on the sky. All eleven are published in the ESZ
release. Together with the remaining ten clusters confirmed by
XMM-Newton, all are described in \cite{planck2011-5.1b}. In the
following we just summarise the general properties of the new
confirmed clusters in the ESZ.

The eleven new clusters in the ESZ confirmed by XMM-Newton have S/N
ranging from 11.5 to 6.3. They were found to lie below the REFLEX flux
limit of $3\times 10^{-12}\,\mathrm{erg}\,\mathrm{s}^{-1}$, except for
two confirmed clusters above the limit. These clusters happen to have
associations with BSC sources and to be situated above the MACS limit;
however their redshifts, $z = 0.27$ and $z = 0.09$ are below the
considered redshifts for MACS [see the detailed discussion in
  \citet{planck2011-5.1b}].  The redshifts of the new confirmed
clusters were estimated directly from X-ray observations of iron
emission lines, and range between $z = 0.2$ and 0.44. Only two out of
the eleven confirmed new clusters have optical redshift estimates. For
one new cluster ({PLCKESZ G285.0-23.7}), the agreement between the
X-ray estimated and photometric redshifts is quite good. The second
cluster, {PLCKESZ G262.7-40.9}, was found to be an ACT cluster,
published after the scheduling of XMM-Newton observation, for which
there is a discrepancy between the X-ray-estimated redshift ($z =
0.39$) and the photometric redshift ($z = 0.54$) from
\cite{men10}. The range in temperature spanned by the new confirmed
clusters in the ESZ is from about 4 to 12 keV, and the derived masses
range from about 4 to $15\times 10^{14} M_{\odot}$.  Three new
clusters in the ESZ sample have masses of $10\times 10^{14} M_{\odot}$
or above, including the most massive cluster detected by \Planck\ with
a mass of about $15\times 10^{14} M_{\odot}$. The confirmation of the
\Planck\ new clusters by XMM-Newton provides us with positions and,
most of all, a better estimate of the cluster size that will be
important for the re-extraction of $Y$ values (see Section
\ref{sec:sizeF}).

\begin{figure}
   \centering
 \includegraphics[width=8cm]{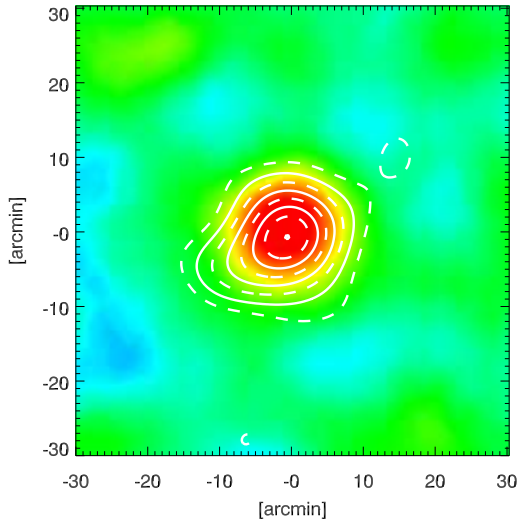}
     \caption{y-map of {PLCKESZ G139.59+24.19} as observed by
       \Planck\ (colour image) and AMI (contours) at a common
       resolution of 13 arcmin.  The contours are from two to nine in
       S/N ratio.  }
         \label{fig:plckami}
   \end{figure}

One additional candidate cluster, {PLCKESZ G139.59+24.19} detected at
S/N = 7.2, was confirmed by a pilot project for confirmation with the
AMI interferometer (see Fig.~\ref{fig:plckami} showing the
\Planck\ $y$ map with the AMI contours, obtained after the subtraction
of bright sources with the large array observations, overlaid). The
\Planck\ cluster was detected at $9\sigma$ by AMI in a long-time
exposure of approximately 30 hours. Preliminary results from AMI give
an integrated Compton parameter of $Y_{5R500} = (17.0 \pm 1.7) \times
10^{-4}$ arcmin$^2$, extracted fixing the cluster size to the
estimated size from \Planck.  The \Planck\ value, $Y_{5R500} = (32 \pm
13)\times 10^{-4}$, is obtained from the blind detection of the
cluster. The error bar takes into account the uncertainty in the
cluster size estimate by the MMF3 algorithm. A detailed comparison is
planned. This same cluster was also confirmed at a S/N level of five
by WISE.

\subsubsection{ESZ candidate new clusters}
\label{sec:cand}

A closer inspection of the ESZ candidate new clusters was performed in
order to ensure the reliability of the retained candidate new
clusters. The same close inspection was also performed, a posteriori,
in order to confirm the rejection of the 11 candidates excluded in the
final steps of the ESZ construction because they were observed solely
by MMF3 (Sect.~\ref{sec:const}). This closer inspection of the
candidates was based on both internal (using \Planck\ alone) and
external data.

For the in-depth inspection of the \Planck\ data, we used cleaned
channel maps, reconstructed $y$-maps and SZ spectra. All these
products are quite sensitive to the procedure used for cleaning the
channel maps, i.e., to the component separation method. We therefore
simultaneously employed different cleaning approaches developed by the
\Planck\ collaboration, briefly described below, in order to ensure
convergence and redundancy in the derived conclusions.  One of the
methods is based on the construction of SZ $y$-maps centred on the ESZ
candidate positions using the Modified Internal Linear Combination
Algorithm (MILCA, \cite{hur10}) applied independently on each
SZ-centred patch. The contribution from other sources of sky emission
such as thermal dust and radio and infrared sources is thus more
accurately reduced. Other approaches based on local component
separation and aperture photometry were also developed in order to
check the $y$-maps and SZ spectra of the candidates. Patches centred
on the SZ candidates are produced from the \Planck\ channel maps and
the IRIS map \citep{mam05}. Local component separation is performed by
decorrelating from the low-frequency channels an extrapolation of the
dust emission computed with the 857 GHz and IRIS maps. The ``dust-free''
217 GHz map is then removed from all channels and visual inspection
can then be performed on these cleaned patches. From this set of
maps we then obtain SZ reconstructed $y$-maps and an SZ spectrum by
applying aperture photometry to each patch. The internal inspection of
the \Planck\ data ($y$-maps, frequency maps and spectra) therefore
provides us with a set of quality flags that were used for the
selection of targets for the follow-up programmes and that are used
for a qualitative assessment of the reliability of the candidates.

Converging negative quality assessments result in the rejection of the
SZ candidates. However in most cases, it is useful to combine and
complement the \Planck-internal quality flags with external
information.  In practice this consists in searching for associations
with FSC (Faint Source Catalogue) and BSC (Bright Source Catalogue)
RASS sources, searching in, and visualising, the RASS maps at the
candidate cluster positions, and finally performing visual checks of
the DSS (Digital Sky Survey) images in the candidate field (within a
five arcminute radius from the \Planck\ position).  Based the lessons
learnt from the XMM-Newton confirmation programme, the association of
candidates with FSC or BSC-RASS sources (in the five arcminute radius
field) was considered as an indication of the reliability of the
candidate.  The presence of an excess in the count-rate RASS images in
the candidate field was also used as a reliability flag. The DSS
images were used simply as an ``empirical'' assessment of the presence
of an overdense region. It is worth noting that the external
information provided in particular by the RASS data never supersede
the \Planck-internal quality flags. As a matter of fact, two of the
confirmed new clusters had neither FSC nor BSC
associations. Conversely, associations with FSC and BSC-RASS sources
were found for SZ candidates that turned out to be false detections
\citep{planck2011-5.1b}.

Using the internal quality flags and the additional external checks,
out of the nine candidate new clusters retained by the ESZ
construction, seven were judged reliable. Two candidate new clusters
had rather poor quality flags and no external associations. One of
them was found to be associated with dust cloud emission. Note that
this source was not flagged by the cross-match with the CC and
Galactic sources, nor identified with a rising spectral distribution at
high frequencies during the internal validation and ESZ
construction. This candidate was rejected from the final ESZ sample,
reducing the total number of clusters and candidate clusters
from 190 to 189. The second cluster candidate with low reliability
({PLCKESZ G189.84-37.24}), was kept in the ESZ list as it was not
associated clearly with any non-SZ source. Table \ref{tab:szcand}
summarises the external information associated with the candidate new
clusters in the ESZ sample\footnote{During the review process 6 of 
the 8 \Planck\ cluster candidates were confirmed by SPT \citep{sto11,wil11} 
and AMI \citep{hur11} experiments independently of the \Planck\ 
collaboration.}.

\begin{table*}
\centering
\caption{For the \Planck\ candidate new clusters not yet confirmed at 
the time of submission, external information from RASS.
\label{tab:szcand}}
\begin{tabular}{l c c c c p{6cm}}
\hline
\hline
Name & RASS& Distance to & S/N of  & S/N of RASS
 & Note \\
 & association & source (arcmin) & RASS source  & in \Planck\ aperture &  \\
\hline
PLCKESZ G115.71+17.52 & BSC & 0.17 & 8.7 & 8.5 &  Possible
      contamination by dust emission\\ 
PLCKESZ G121.11+57.01 & FSC & 1.72 & 2.9 & 4.1 & Possible association with
WHL J125933.4+600409 from \cite{wen09}, $z=0.33$\\
PLCKESZ G189.84-37.24 & None & - & - & 1.3 & Low reliability, high level of 
	 contamination by Galactic emission\\
PLCKESZ G225.92-19.99 & FSC & 1.11 & 2.5 & 6.7 & With XMM-Newton and HST
pointed observations  \\
PLCKESZ G255.62-46.16 & FSC & 0.9 & 2.7 & 3.8 & With ESO and Suzaku
pointed observations\\
PLCKESZ G264.41+19.48 & BSC & 1.22 & 4.6 & 5.7 & \\
PLCKESZ G283.16-22.93 & FSC & 0.54 & 3.6 & 4.2 & \\
PLCKESZ G304.84-41.42 & BSC & 0.55 & 3.6 & 5.1 & With ESO pointed observations 
\\
\hline
\end{tabular}
\end{table*}

\section{Error budget on the cluster parameters}
\label{sec:budg} 
\subsection{Position}

\begin{figure}
\centering
\includegraphics[width=4cm]{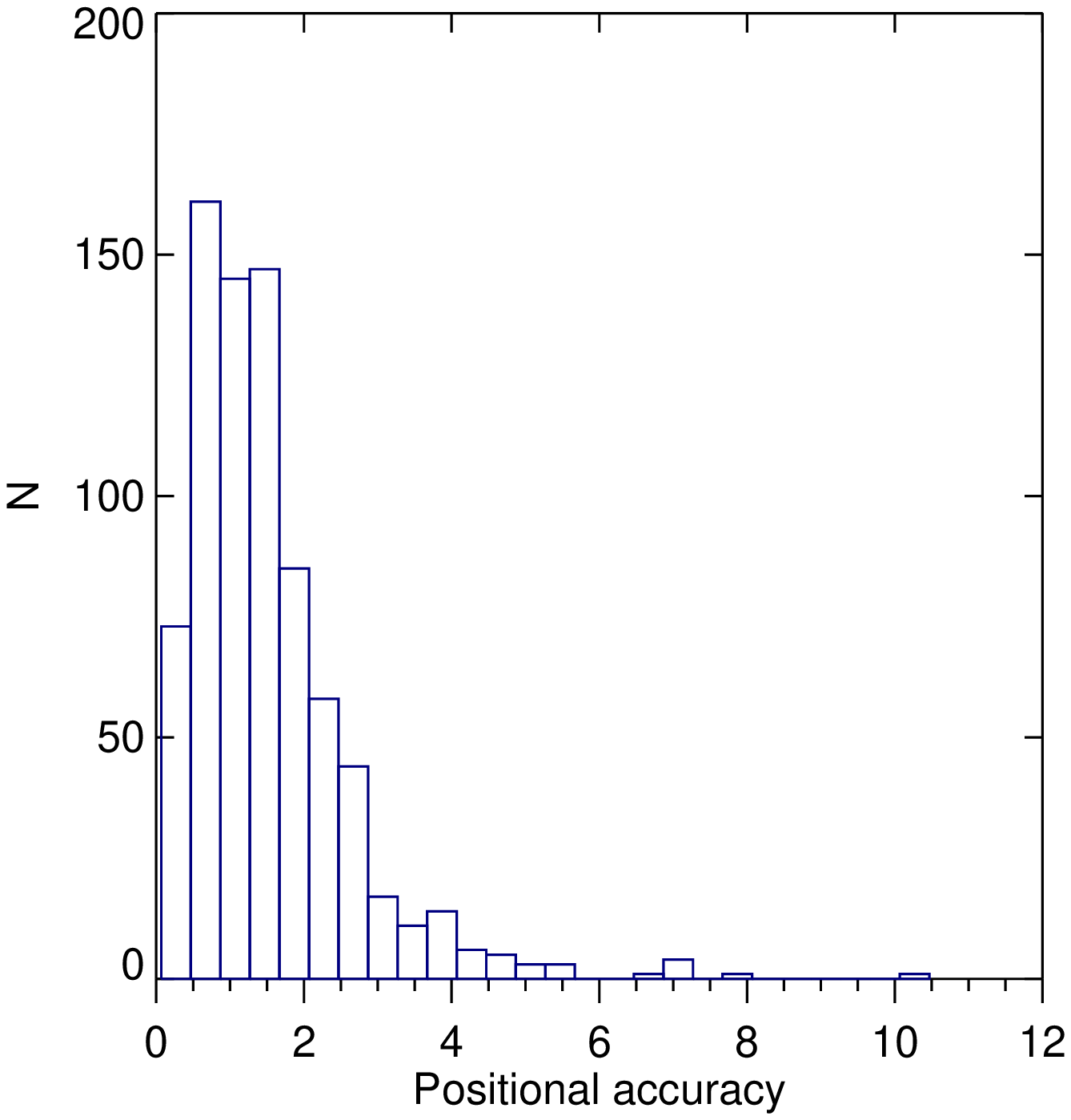}
\includegraphics[width=4.cm]{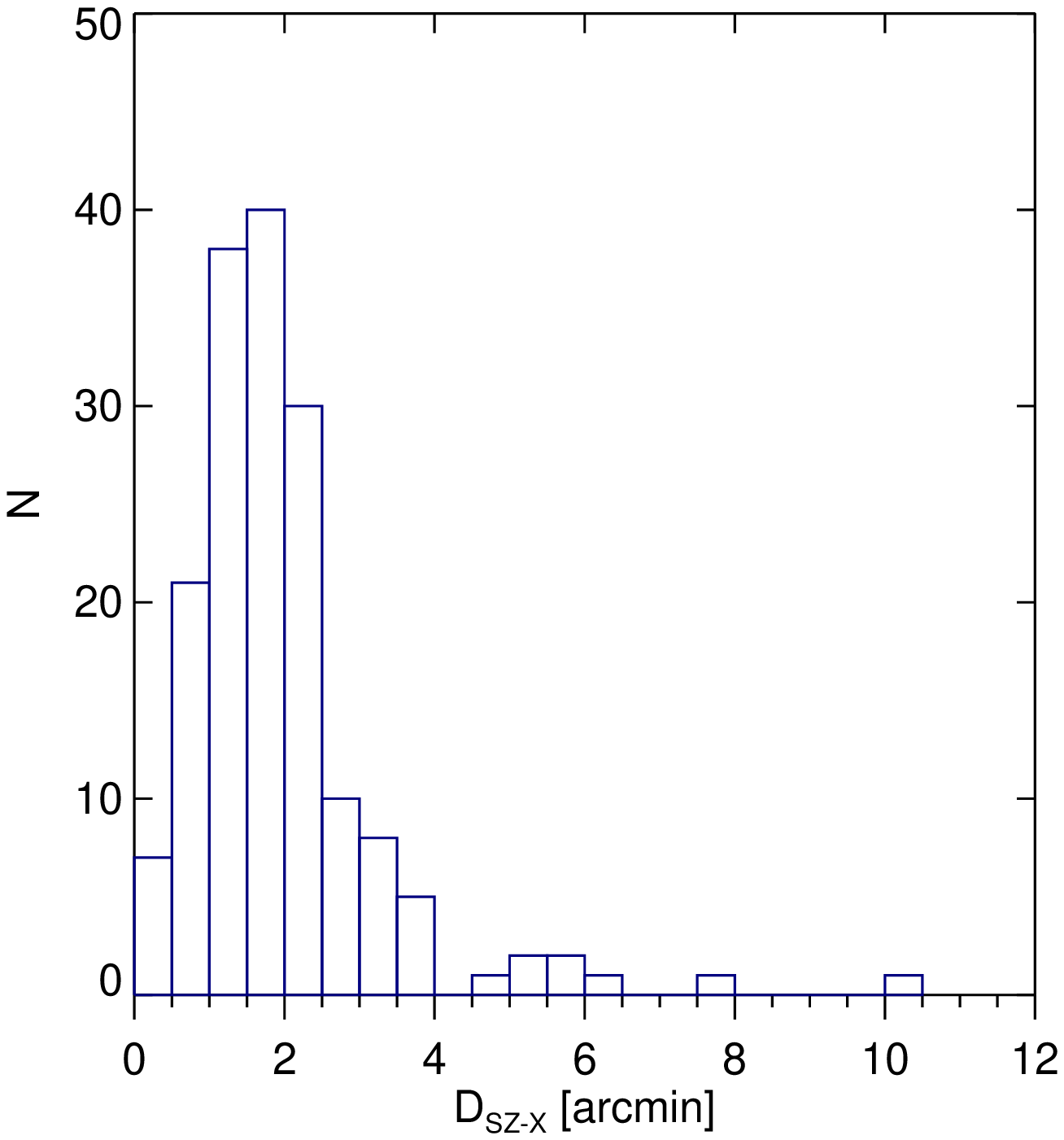}
\includegraphics[width=4.cm]{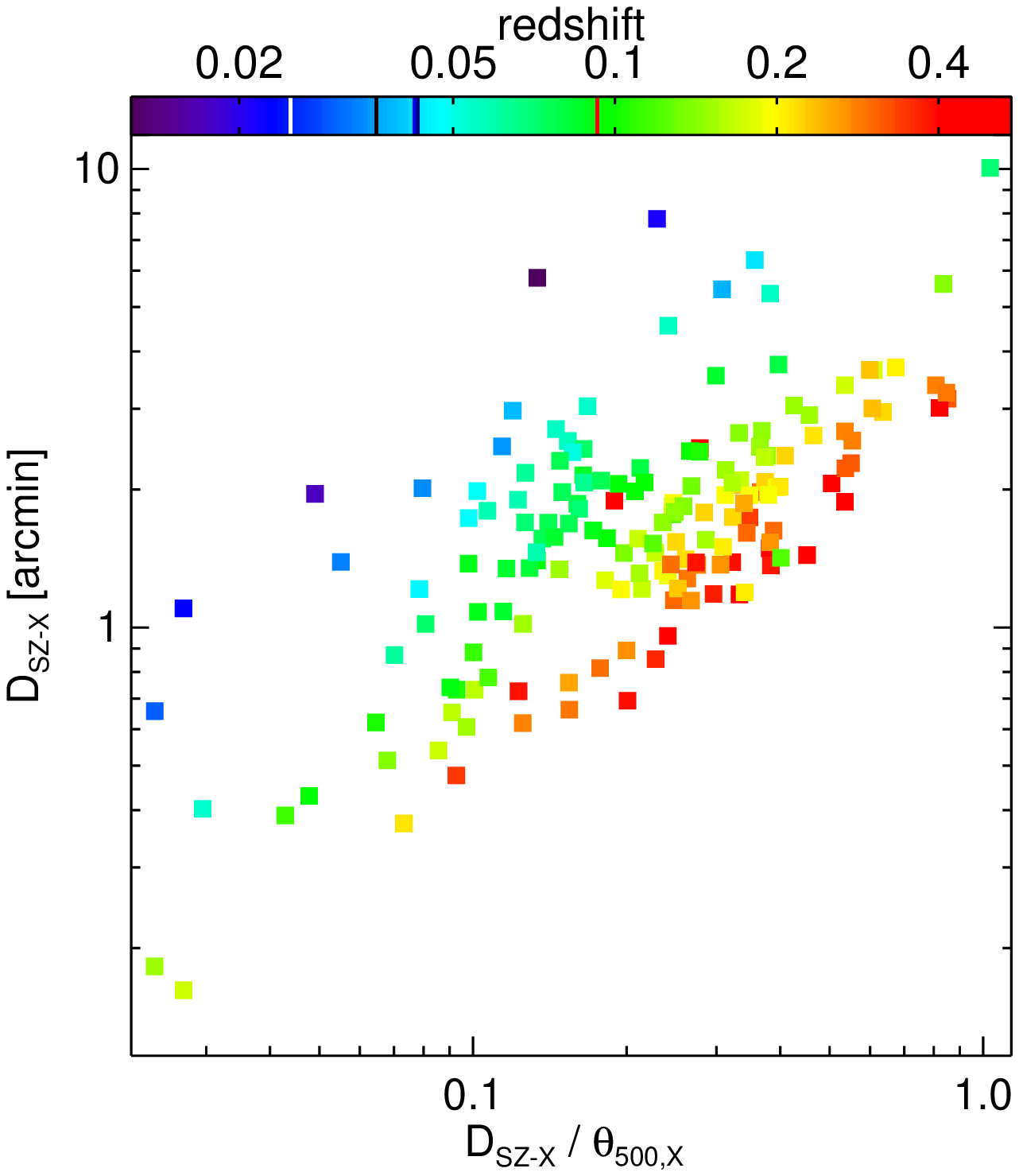}
\caption{Upper left panel: Positional accuracy from MMF3 based on
  simulations for the SZ challenge. Upper right panel: Distribution of
  the distance between the SZ blind position and the X-ray position
  ($D_{\mathrm{SZ-X}}$) for 167 known, or confirmed with XMM-Newton,
  X-ray clusters. Lower panel: Separation of the SZ blind and X-ray
  positions $D_{\mathrm{SZ-X}}$ as a function of $D_{\mathrm{SZ-X}}$
  normalised to the cluster size $\theta_{500, \mathrm{X}}$.}
\label{fig:pos}
\end{figure}

The ESZ sample contains a list of 189 clusters or candidate clusters
distributed over the whole sky with positions obtained from  blind
detection with the MMF3 algorithm.  Based on the simulation used for the
SZ challenge comparison, we find that MMF3 recovers cluster positions
to $\sim 2$~arcmins on average. However, there is a large scatter in
the positional accuracy, as seen in Fig.~\ref{fig:pos}.

For the 158 ESZ candidates identified as X-ray clusters with known
X-ray size, the coordinates of the X-ray counterpart are given by the
MCXC. The X-ray position is also given for the \Planck\ cluster
candidates confirmed by XMM-Newton as single objects. The comparison
of the SZ candidate positions derived from the blind detection with
the X-ray positions of the identified or confirmed clusters for a
total of 167 clusters is shown in Fig.~\ref{fig:pos}, left panel. The
positional offset between \Planck\ blind and X-ray positions,
$D_{\mathrm{SZ-X}}$, is of the order of 2 arcmin on average,
consistent with the estimates obtained from the SZ challenge
simulation. Very few clusters (8 in total over 167) have an offset
$D_{\mathrm{SZ-X}}>4$\,arcmin, and stand out as clear
outliers in the distribution. It is worth noting that such a
positional offset combines both the uncertainty in the position
reconstruction from MMF3 and the possible physical offset between the
centroids of X-ray and SZ signals (e.g., in merging clusters). The
eight clusters with $D_{\mathrm{SZ-X}} > 4$\,arcmin are all nearby
merging clusters or members of larger structures such as A3532 in the
Shapley supercluster, or contaminated by radio source emission. The
cluster A1066 ($z = 0.07$), which has the largest positional offset
($D_{\mathrm{SZ-X}} = 10$ arcmin), is in the Leo Sextans supercluster
\citep{ein01}. In addition, it may suffer from point source
contamination. The cluster Abell 1367 at $z=0.02$ with
$D_{\mathrm{SZ-X}} = 7.8$ arcmin is a young cluster currently forming
at the intersection of two filaments \citep{cor04} with complex gas
density and temperature structures \citep{ghi10}.

As seen from Fig.~\ref{fig:pos}, right panel, large (greater than four
arcmin) offsets are only seen in nearby clusters (seven out of the
eight clusters with $D_{\mathrm{SZ-X}} > 4$ have redshifts lower than
0.08).  They remain smaller than the cluster size, as expected for
offsets dominated by physical effects.  On average, the offsets tend
to decrease with increasing redshift and seem to become independent of
redshift above $z\sim0.3$. This is due to the decreasing contribution
of possible physical offsets, which become unresolved.  The overall
offset, including the absolute reconstruction uncertainty, remains
smaller than the cluster size for most of the clusters in the ESZ
$\theta_{500}$ (Fig.~\ref{fig:pos}, right panel). However, we expect
that it will be of the order of cluster size for clusters at higher
redshifts than the range currently probed. This positional offset is
therefore an additional source of uncertainty in the cluster position
which needs to be taken into account in the follow-up observations for
candidate confirmation.
\subsection{Cluster size-$Y$ degeneracy}
\label{sec:sizeF}

\begin{figure}
\centering
\includegraphics[width=7cm]{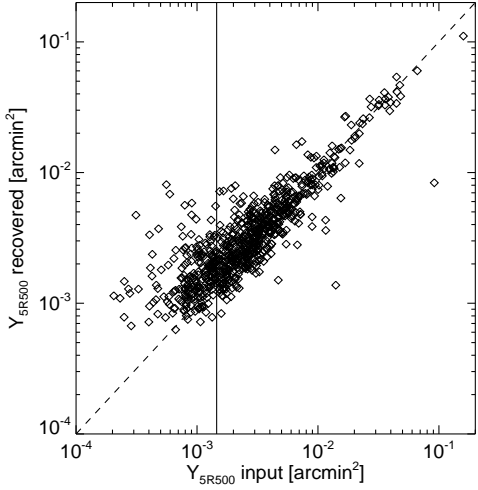}
\caption{Input versus recovered integrated Compton parameter from
  MMF3, based on simulations for the SZ challenge.}
\label{fig:fluxsim}
\end{figure}

The MMF algorithm, and more generally algorithms that are based on the
adjustment of an SZ profile to detect clusters, generally perform
better than algorithms which do not assume an SZ profile. The GNFW
profile used in the present study corresponds to a shape function
characterised by two parameters, the central value and a
characteristic scale $\theta_{\mathrm{s}}$ (with
$\theta_{\mathrm{s}}=\theta_{500}/c_{500}$ and
$c_{500}$ is the concentration parameter).  Simulations
showed that the intrinsic photometric dispersion of the recovered
integrated Compton parameter, with a GNFW profile, could be of order
30\% (see Fig.~\ref{fig:fluxsim}) even with the prior information on
the pressure profile. This is due to the difficulty of estimating the
cluster size, which in turn is degenerate with the SZ $Y$ estimate.

\begin{figure}
\centering
\includegraphics[width=7cm]{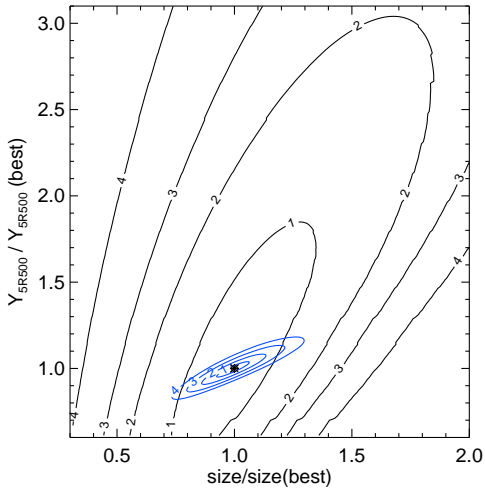}
\caption{Illustration of the {cluster size-$Y$ degeneracy} from
  PwS. Shown are the cases of Coma cluster (high S/N and extended in
  blue), and that of an S/N = 6 unresolved cluster (in
  black). Parameters are plotted with respect to the best fit points
  in each direction.}
\label{fig:deg}
\end{figure}

This {cluster size-$Y$ degeneracy} is illustrated, here using PwS, in
two extreme situations (Fig.~\ref{fig:deg}) showing the likelihood
plots (integrated Compton vs cluster size) of an extended high S/N
cluster such as Coma (blue contours) and an unresolved S/N = 6 cluster
(black contours).  In both cases, the integrated Compton parameter $Y$
is highly correlated with the estimated cluster size.  We find a
correlation coefficient $\rho = 0.91$ and $\rho = 0.75$ for Coma and
the ``unresolved cluster'' respectively.  On average over the ESZ
sample we find a correlation of $\rho = 0.85$.  The degeneracy between
cluster size and $Y$ is extremely detrimental, as it will more than
double the average fractional uncertainty relative to the $Y$ value in
the case where we knew the true value of $\theta_{\mathrm{s}}$
perfectly. As a result, any attempt to constrain the cluster size
(equivalently $\theta_{\mathrm{s}}$), fixing or assuming a prior for
its value, brings a significant reduction on the $Y$ value dispersion.

The issue of the {cluster size-$Y$ degeneracy} is of particular
importance in the case of \Planck, for which a vast majority of clusters
are only marginally resolved. This issue is also crucial when one
wants to use the SZ signal as a mass proxy. Indeed, the dispersion in
$Y$ due to the {cluster size-$Y$ degeneracy} is likely to
dominate the intrinsic scatter of order 10\% of this mass proxy
(\cite{das04}, \cite{arn07}).

\begin{figure*}
\centering

\includegraphics[width=7cm]{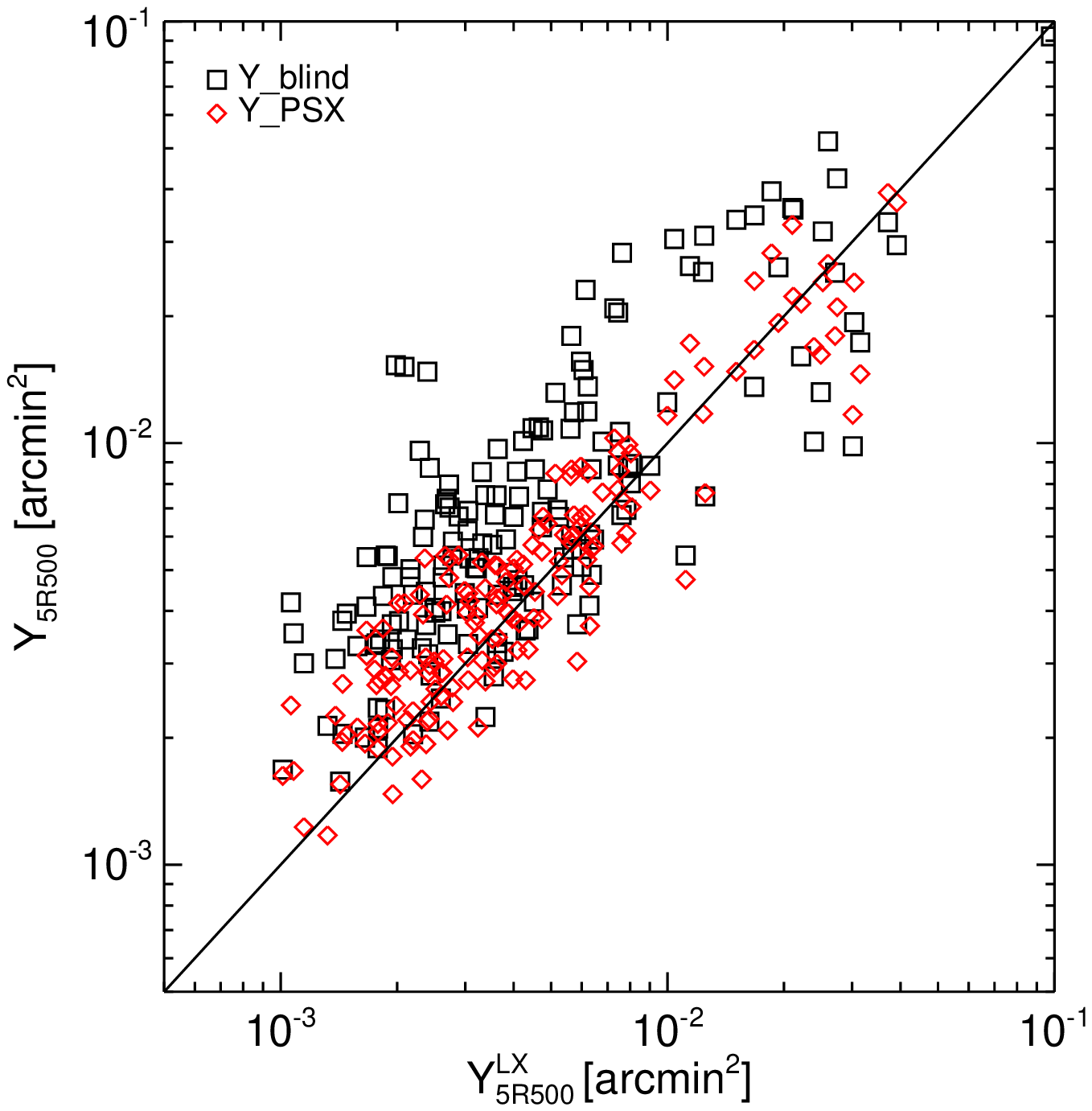}
\includegraphics[width=7cm]{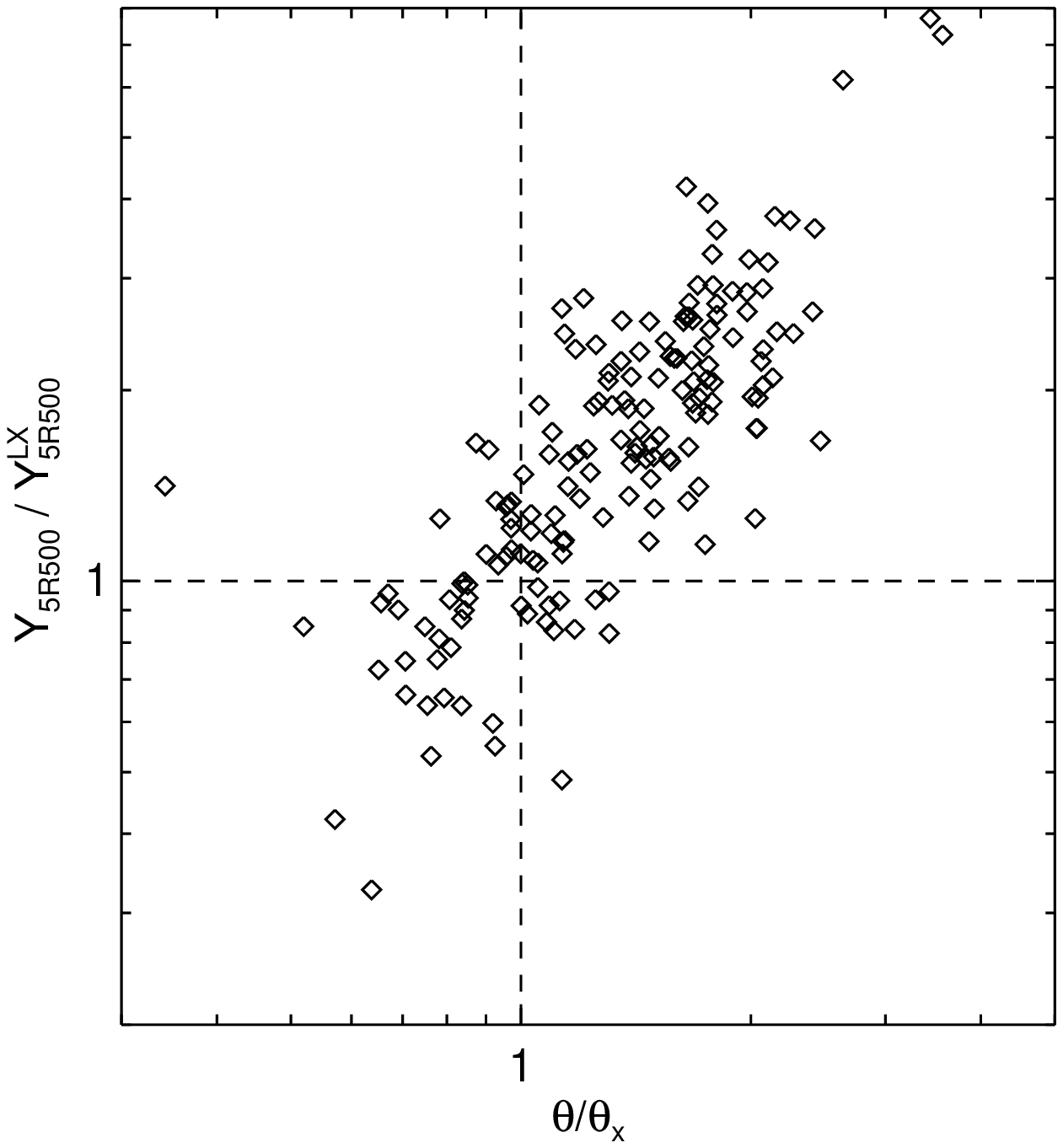}
\caption{Left: The scatterplot of the measured integrated Compton
  parameter $Y_{5R500}$ from the 158 X-ray identified ESZ clusters
  against the predicted $Y^{L_X}$. Black squares: Estimated cluster
  size from blind detections. Red diamonds: Re-computed integrated
  Compton parameter at X-ray positions and with X-ray derived cluster
  size. Right: The ratio between the $Y$ values and the predicted
  $Y^{L_X}$ against the ratio between the estimated cluster size and
  the predicted size ($\theta/\theta_X$).}
\label{fig:recomp_pxcc}
\end{figure*}

\begin{figure}
\centering
\includegraphics[width=7cm]{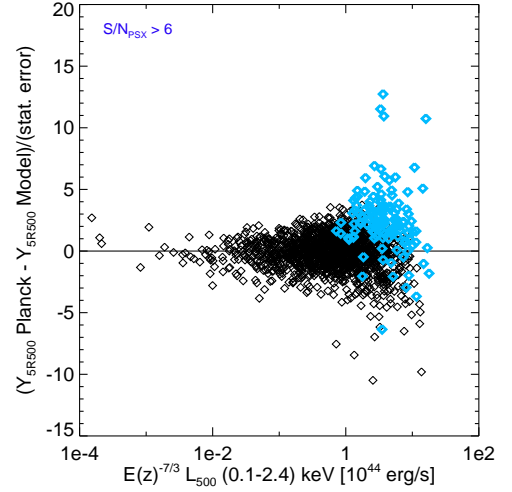}
\caption{Ratio of predicted vs observed $Y_{5R500}$ for the MCXC
  clusters as a function of the X-ray luminosity $L_{500}$ used to
  estimate the cluster properties (radius and integrated Compton
  parameter). The light-blue diamonds indicate a cut of 6 in predicted S/N
  corresponding to the ESZ selection criterion.}
\label{fig:stat_err_pxcc} 
\end{figure}

As a result, we have re-estimated the integrated Compton parameter for
all the ESZ candidates with prior information on their sizes. We have
chosen the X-ray sizes ($\theta_{\mathrm{5R500}}$) derived from the
X-ray luminosities, $L_{500}$, as detailed in \cite{pif10}, as
suitable estimates of the cluster sizes.  Using the MMF3 version
implemented in HFI Core team and SZ validation team, $Y_{5R500}$ were
thus re-computed from the \Planck\ channel maps at fixed X-ray
position and with imposed cluster size equal to the X-ray luminosity based
$\theta_{\mathrm{5R500}}$. The integrated Compton parameter $Y$ was
re-estimated for all the clusters with known X-ray counterparts, being
the 158 ESZ candidates identified with known clusters from the MCXC
and the nine ESZ clusters confirmed by XMM-Newton as single objects. 

Fig.~\ref{fig:recomp_pxcc}, left panel, illustrates the effect of
fixing the position and the cluster size, in the GNFW profile, to
$\theta_{\mathrm{5R500}}$ for the 158 ESZ identified clusters. The
figure displays the measured $Y_{5R500}$ values versus the predicted
$Y^{L_X}_{5R500}$ values using X-ray luminosities. The squares stand
for integrated Compton parameters obtained from the blind detection
whereas the diamonds are integrated Compton parameters re-extracted
from the \Planck\ channel maps for the MCXC-identified clusters. Figure
\ref{fig:recomp_pxcc}, right panel, displays the ratio of blind to
predicted $Y_{5R500}$ versus the ratio of estimated cluster size from
blind detection to X-ray cluster size derived from X-ray
luminosity. This clearly confirms for the 158 identified clusters that
an overestimate of cluster size induces an overestimate of the SZ
signal. As seen in the Fig.~\ref{fig:recomp_pxcc} (left panel), the
scatter is significantly reduced from about 43\% to 34\% by imposing a
cluster size. Likewise the offset changes from 80\% to 14\%.

The dispersion in the predicted integrated Compton parameter is
affected by the intrinsic dispersion in the $L_{500}$-$M$ relation
used to derive the predicted SZ quantities as shown in
Fig.~\ref{fig:stat_err_pxcc}. The selection criterion S/N $\geq 6$
(blue diamonds in the figure) used to construct the ESZ sample
indicates that the high S/N clusters are biased towards larger SZ signals,
showing that the obtained positive offset in
Fig.~\ref{fig:recomp_pxcc} (left panel) is indeed expected.

As emphasised, a prior on the cluster size helps to break the
degeneracy between $Y$ and cluster-size estimates. As a consequence,
the better the cluster size estimate, the more reliable the Compton
$Y$ parameter estimate. From a selected subsample of 62 ESZ clusters
with XMM-Newton archival data \citep{planck2011-5.2b} we have derived
accurate estimates of the X-ray sizes, without using the X-ray
luminosities, and the $Y_{500}$ were re-evaluated on the
\Planck\ channel maps, allowing us to tightly constrain the local SZ
versus X-ray scaling properties. As shown in Appendix A of
\cite{planck2011-5.2b}, the scatter is reduced even more than in
Fig.~\ref{fig:recomp_pxcc} (left panel) and no offset is observed
any more between the predicted and measured $Y_{500}$ values.

\subsection{Systematic effects}

Due to the {cluster size-$Y$ degeneracy} discussed above, beam
uncertainties are likely to have a significant impact on $Y$ estimates
for our candidates because they affect both the original detection and
the estimation of cluster size.  The beams can be characterised by
their shapes and the associated accuracies. The beams for each
frequency channel, used for the detection and $Y$ estimate with all
methods presented in this study, were assumed Gaussian with FWHM given
\cite{planck2011-1.7}.  Uncertainties on the recovered beams have been
estimated in \cite{planck2011-1.7} and found to range between 1 and
7\% (from 100 to 857 GHz).  These uncertainties on the beams have been
propagated to the $Y$ measurements by applying the MMF3 algorithm on
the channel maps varying the beam size within the uncertainties at
$\pm 1\sigma$. In doing so we treat differently the ESZ clusters with
known X-ray cluster size, for which X-ray positions and estimated
$\theta_{5R500}$ are fixed, and the ESZ clusters or candidate clusters
without estimated cluster size for which the $Y$ were re-estimated
without prior information. We find that the uncertainty on the
obtained $Y_{5R500}$ is of the order of 10\% across the ESZ sample.

The \Planck\ HFI maps used for the cluster extraction are calibrated
to better than 2\% for frequencies from 100 to 353 GHz, and to better
than 7\% beyond [see \cite{planck2011-1.7}].  This
uncertainty in the calibration is accounted for again by performing
the SZ-candidate detection with the MMF3 algorithm on the channel
maps. We find that on average, the calibration uncertainty
propagates into an uncertainty on the $Y$ less than 2\%. The
highest \Planck-HFI frequencies, with the largest calibration
uncertainties, have a low impact on the SZ $Y$ measurement and thus
do not impact significantly the overall error budget.

Finally, we have checked that the colour corrections, i.e., the average
SZ signal in the HFI bandpasses, induces less than a 3\%
difference on the estimated $Y_{5R500}$. The SZ-candidate detection
and the $Y$ estimates by the MMF3 algorithm were thus performed
without taking into account the integration of the SZ spectrum in the
\Planck\ bandpasses is negligible.

Table~\ref{tab:error} summarises the effects of beam, calibration, and
colour correction. It shows that the beam effect is the major source
systematic uncertainty in the SZ signal estimate. It is worth noting
that the systematic uncertainties are not included in the
uncertainties quoted in the ESZ table provided at {\it
  http://www.rssd.esa.int/Planck}, \citep{planck2011-1.10sup}.

\begin{table}
\caption{Systematic error budget on the $Y_{5R500}$ values for the ESZ
clusters
\label{tab:error}}
\begin{tabular}{c c c c c c}
\hline \hline
Source & Beam & Calibration & Colour     & Astrophysical \\
       &      &             & correction & contamination \\ \hline
Error  &      &     &      &    \\
contribution  & 8\% &  2\% &  3\% &   3\% \\
\hline
\end{tabular}
\end{table}

\subsection{Contamination by astrophysical sources}
\label{sec:contam}

Galactic and extragalactic sources (both radio and infrared
galaxies) are known to lie in the interior of galaxy clusters and
hence are a possible source of contamination for the SZ clusters and
candidates \citep{rub03,Agh05,Lin09}.

In the course of ESZ validation, we have gone through an
inspection of thirteen known clusters which show some poor quality
flags. All these clusters were annotated and the notes can be found in
\cite{planck2011-1.10sup}. Ten of them are likely to be contaminated
by dust emission from our Galaxy or by IR point sources in their
vicinity. Two of them were found to be contaminated by NVSS [at 1.4
GHz, \cite{nvss}] radio sources that are clearly seen in the LFI
channels.  Combining data from SUMSS [at 0.85 GHz, \cite{sumss}],
NVSS, and \Planck's LFI and HFI frequencies we find that most radio
sources in the ESZ sample have a steep spectrum which makes their
contamination to the SZ signal negligible. Three additional clusters
(beyond the thirteen), have relatively bright ($S_{\mathrm{1.4 GHz}} >
0.2$ Jy) radio sources in their vicinity ($r<15\,$arcmin). NVSS+LFI
data reveal flat spectra (indexes between $\alpha = 0$ and $\alpha =
-0.5$). The flux of the radio sources is thus still significant and
hence the SZ signal could be affected by their presence.
 
A statistical analysis has been performed in order to explore the
astrophysical contamination over the entire ESZ sample, rather than on
an individual cluster basis.

\begin{figure}
\centering
\includegraphics[angle=0,width=8cm]{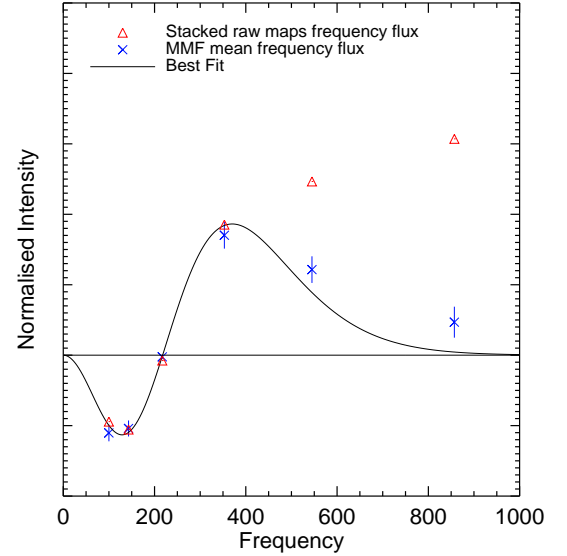}
\caption{Average contamination of the ESZ sample by astrophysical
  sources. Blue crosses: Average $Y$ measurements from MMF3 algorithm
  normalised to the integrated Compton-$y$. Red triangles: $Y$,
  obtained from aperture photometry on the stacked cutouts in the
  channel maps prior filtering by the MMF. Black solid line:
  Normalised theoretical SZ spectrum.}
\label{Fig:color}
\end{figure}

In order to exhibit the initial average level of contamination prior
to the use of the MMF algorithm, we have stacked cutouts 4.5 degrees
on a side from the channel maps centred at the ESZ cluster/candidate
positions from 100 to 857 GHz using a stacking library\footnote{
  http://www.ias.u-psud.fr/irgalaxies/ \citep{bet10}} detailed in
\citet{dol06} and \citet{bet10}. The $Y$ values per frequency,
obtained from aperture photometry on the stacked cutouts, are
displayed in red triangles Fig.~\ref{Fig:color}.  The spectral
signature normalised to the averaged integrated Compton-$y$ over the
whole ESZ sample shows quite good agreement with the theoretical SZ
spectrum at low frequencies (Fig.~\ref{Fig:color}, black solid
line). Above 353 GHz the signal is highly contaminated by IR emission
from Galactic dust and IR point sources.

The $Y$ measurements, per frequency, of the MMF3 algorithm normalised
to the integrated Compton-$y$ are averaged over the ESZ sample and the
resulting spectral energy distribution is compared with the normalised
SZ spectrum (see Fig.~\ref{Fig:color}, blue crosses). The excess of
emission at high frequencies is significantly reduced by the filtering
technique of the MMF algorithm, reinforcing the idea that most of the
excess at the highest frequencies is due to large-scale (larger than
the beam) fluctuations in Galaxy emission. The remaining excess after
the filtering could be due to a combination of small-scale Galactic
fluctuations and/or infrared galaxies. In order to quantify the effect
of this residual IR emission on the integrated Compton-$y$
determination, an SZ spectrum was fitted to the averaged spectrum. The
normalisation was left free. The displayed error bars contain the
dispersion of the measured $Y$ per frequency and, added in quadrature,
the uncertainties due to the beam, the colour correction, and the
calibration ($\sim 10\%$, $\sim 3\%$, $\sim 2\%$ respectively). The
best value for the normalised integrated Compton parameter is
$Y_{\mathrm{fit}}=1.01$, showing an excellent agreement with the
expected spectrum despite the IR excess emission at high
frequencies. The same procedure was applied to the 100, 143, 217, and
353 GHz $Y$ values and led to
$Y_{\mathrm{fit_{100-353}}}=0.97$. This shows that, on average, the
residual IR contamination has a negligible effect ($\sim 3\%$) on the
integrated Compton-$y$ value estimated for the ESZ sample.

%
%_____________________________________________________________________
%
%
\section{Purity and completeness}
\label{sec:pur}

The ESZ sample is characterised by the fact that a significant
fraction of the clusters and candidate clusters lies near a selection
cut. In a catalogue of this sort, the properties of the catalogued
clusters will not be representative of the true underlying cluster
population. For example, if the SZ signal of a cluster is related to a
different cluster property such as mass (collectively referred to as
`scaling relations') the observed integrated Compton-$y$ parameter
values, $Y$, will be biased near the selection cut, an effect known as
Eddington and Malmquist biases [for discussions in a cluster context see
  \cite{man10,and10}].

For the full ESZ sample, we do not always have other cluster
properties to relate the integrated Compton-$y$ to, but we can
nevertheless examine some statistical effects of selection. In order
to do this, we generate large mock cluster catalogues whose properties
are designed to mimic those of the observed sample.  To impose a
selection cut on the mock catalogues, we use the {\em observed}
relation between $Y_{500}$ and S/N from the region significantly above
the selection cut and extrapolate below it, along with an estimate of
scatter again from observations. This is carried out in several
redshift bins, and leads to a predicted  $\mathrm{S/N}$--$Y$
  scaling given by 
$$\mathrm{S/N} = 10^{1.38\pm 0.03}\, (1+z)^{-5.92\pm 0.24}\,
\left[\frac{YE^{-2/3}D_A^2}{10^{-4} \, \mathrm{Mpc}^2}\right] \, , $$
with scatter $\sigma_{\mathrm{log-log}}=0.16$ in log-log scale. We then
construct large mock catalogues of clusters through drawing of Poisson
samples from the \cite{jen10} mass function normalised with
$\sigma_8=0.8$, a value consistent with the latest WMAP constraints.
To each cluster and consistent with \Planck\ observations, we assign 
values of $Y_{5R500}$ by adopting the $Y$--$M$ scaling
relation from \cite{planck2011-5.2b}. An S/N value is then assigned as 
described above, and the cut imposed to create the mock catalogue.

\begin{figure}
 \centering
\includegraphics[width=7cm]{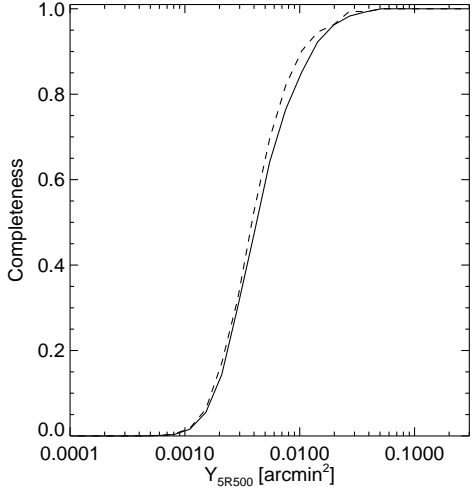}
    \caption{Expected completeness of the ESZ sample as a function of
      $Y_{5R500}$, estimated from mock cluster catalogues.
    }
       \label{fig:compl}
 \end{figure}

We first use these simulations to estimate the completeness of the ESZ
sample as a function of $Y_{5R500}$.  For clusters within a given bin
in $Y_{5R500}$, we extract the fraction of mock clusters which lie
above the selection cut. The result is shown in Fig.~\ref{fig:compl}
(solid line), and indicates that the sample becomes significantly
incomplete (less than 90\% complete) below $Y_{5R500} \simeq 0.013 \,
\mathrm{arcmin}^2$.  This result is fairly insensitive to the assumed
mass function normalisation.  For example, changing to $\sigma_8=0.9$
(dashed line) causes only small variations in the completeness
function. For this case, a completeness of 90\% is obtained at
$Y_{5R500} \simeq 0.010 \, \mathrm{arcmin}^2$. 

\begin{figure}
   \centering
\includegraphics[width=7cm]{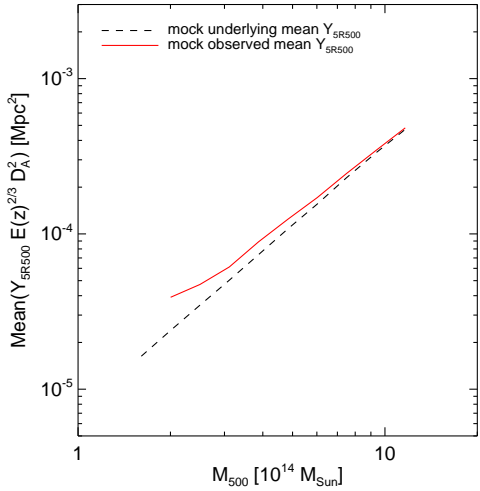}
      \caption{Expected mean $Y_{5R500}$, as a function of cluster
        mass, for the observed sample and for the predicted underlying
        cluster sample.  At low masses, the observed mean rises above
        the true mean due to Malmquist bias.}
         \label{fig:ybias}
   \end{figure}

 We then analyse the extent to which the mean $Y_{5R500}$ of the
observed clusters is biased in relation to the mean $Y_{5R500}$ of the
underlying cluster distribution, through those clusters with low
$Y_{5R500}$ for a given mass being lost via selection. The underlying
mean $Y_{5R500}$ in the mock samples is given by the input
$Y_{500}$--$M_{500}$ scaling relation from \cite{planck2011-5.2b} and
the observed $Y_{5R500}$--$Y_{500}$ scaling; as shown in
Fig.~\ref{fig:ybias} the mean of the observed clusters will be
biased upwards from this, the effect becoming significant for $M_{500}
< 6 \times 10^{14} \msol$. Note that this bias does not imply that the
$Y_{5R500}$ measurements for the ESZ clusters are systematically
wrong; the bias is because the selection cut prevents those clusters
being representative of the true cluster population at those masses.

\begin{figure}
\centering
\includegraphics[width=7cm]{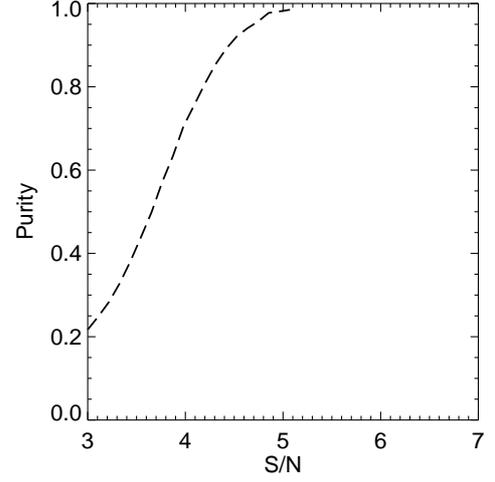}
\caption{Purity as a function of S/N from MMF3, based on simulations
  for the SZ challenge.}
\label{fig:purity}
\end{figure}

Finally, numerical simulations based on the \Planck\ Sky Model were used
to estimate the purity of the \Planck\ SZ catalogue. They showed on a
simulated sky that a cut in S/N of five ensures 100\%
purity of the obtained sample (see Fig.~\ref{fig:purity}). However,
the simulation does not capture the entire complexity of the real sky
and, in particular, the contamination by astrophysical sources
emitting above 217 GHz from IR sources and dust emission or cold
cores was found to be higher than expected. The final ESZ sample
obtained after applying the selection criterion cut in S/N of 6
contained 190 SZ candidates. The validation of the sample showed that
one of them was found to be a spurious source identified with dust
emission and it was rejected. The remaining candidate new clusters are
to be confirmed. The purity of the ESZ sample thus lies between about
95\% and 99\%.

Although an attempt to characterise the completeness and purity is
made, we do not provide a fully characterised selection function along
with the ESZ sample. The {cluster size-$Y$ degeneracy} discussed
above, together with the large scatter in the contamination level of
the SZ detections due mostly to dust emission, makes it difficult to
draw a simple relation between the S/N limit used to construct the
sample and the measured $Y_{5R500}$. It is thus not presently possible
to provide a reliable mass limit to our sample. When the telescope
beam is larger than the cluster size, a survey is limited by SZ
signal. Then, since for the SZ signal the redshift dependence enters through
the angular-diameter distance rather than the luminosity distance, the
mass selection function is more uniform in SZ than in X-ray
surveys. However in our case most of the clusters detected by
\Planck\ are at nearby redshifts ($z_{\mathrm{median}} = 0.15$) and
the majority are resolved, adding even more complexity to the
selection function.

\section{Statistical characterisation of the ESZ sample}

\label{sec:stat}
\begin{figure}
  \centering
\includegraphics[width=7cm]{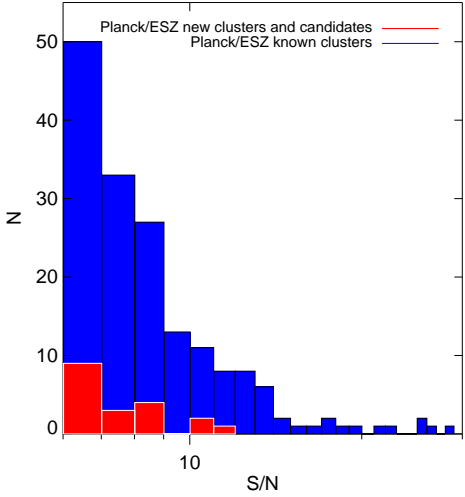}
     \caption{Distribution of S/N (for the full
       ESZ sample: clusters and candidate clusters). }
        \label{fig:snresz}
  \end{figure}

 \begin{figure}
  \centering
\includegraphics[width=7cm]{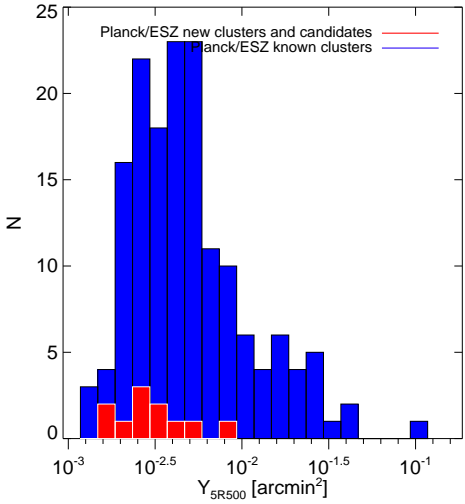}
     \caption{Distribution of ESZ sample in integrated Compton
       parameter $Y$.}
        \label{Fig:yesz}
  \end{figure}

The ESZ sample is the first all-sky sample of high S/N SZ-detected
clusters of galaxies produced by \Planck. Its high reliability is
ensured by the high S/N of the reported detections
and by the subsequent validation process. The S/N of the objects in
the sample, obtained from blind detection using MMF3 on the reference
channel maps, are displayed in Fig.~\ref{fig:snresz}. They range
between 6 and 29 with median S/N of about eight. Six clusters, including
A2163 with S/N = 26 and Coma with S/N = 22, are in the tail of the S/N
distribution with S/N above 20. The twelve confirmed
\Planck\ new SZ clusters, included in the ESZ, have their S/N
distributed between 6.3 and 11.5.  Additional confirmed new clusters
with lower S/N are given in \cite{planck2011-5.1b}. The eight
candidate new clusters yet to be confirmed have S/N ranging from 6 to
8.5.

The ESZ provides us with measures of the integrated Compton parameter
within a $5R500$ sphere, $Y_{5R500}$, for 189 clusters or
candidates. For about 80\% of the known clusters in the ESZ, this is the
very first SZ measure performed in their direction. The integrated
Compton parameter of the whole sample, displayed in
Fig.~\ref{Fig:yesz}, shows that the SZ signal extends over about two
orders of magnitude from about $1.5\times\,10^{-3}$ to
$120\times\,10^{-3}$ arcmin$^2$. Unsurprisingly, the largest value is
that of the Coma cluster. Moreover, the estimated cluster sizes from
the MMF3 algorithm for the ESZ clusters and candidates are all above
$5\theta_{500}=8$ arcmin, indicating that the high S/N clusters under
study can all be  considered as extended sources. We compare the
estimated cluster size (from blind detection) with the X-ray
cluster size obtained from the X-ray observation of the confirmed SZ
clusters, considered as a representative cluster size. We find that
the SZ blind size is generally larger than the X-ray cluster size; it
can be two times larger. As discussed previously, due to the cluster
size-$Y$ degeneracy this affects the integrated Compton parameter
measurement.

Using the MCXC compilation and the XMM-Newton observations of the
confirmed \Planck\ SZ candidates, we obtain masses, $M_{500}$,
estimated from mass proxies (luminosity, $L$--$M$ relation, or
$Y_{\mathrm{X}}$) for 167 clusters out of the 189 of the ESZ
sample. Furthermore, using the redshift information compiled in the
MCXC that we retrieved during the validation process and the
redshift estimates from XMM-Newton follow-up observations, we gather
the redshifts for 173 clusters of the ESZ sample. The distributions of
redshifts and masses are exhibited in Figs. \ref{fig:zdist} and
\ref{fig:mdist}, respectively. The redshifts of the ESZ sample are
distributed in the range of small to moderate redshifts from about
$z=0.01$ to $z=0.55$, with a median redshift of 0.15. The vast
majority of the ESZ clusters, 86\%, are thus nearby ones lying below
$z=0.3$. Most of the newly-discovered clusters confirmed by XMM-Newton
within the ESZ sample have redshifts of the order of 0.4. Among the
new \Planck\ clusters confirmed by XMM-Newton, but with S/N lower than
6, released outside the ESZ we find a cluster with $z=0.54$. As for
the mass distribution of the ESZ clusters, it spans over a decade with
cluster masses ranging from 0.9 to $15\times 10^{14}\, M_{\odot}$
within a surveyed volume of the order of $3.5\times
10^{10}{\mathrm{Mpc}}^3$. It is worth noting that in surveying the
whole sky, \Planck\ has a unique capability to detect rare massive
clusters in the exponential tail of the mass function. Indeed, among
the 21 newly discovered clusters confirmed by XMM-Newton in total
(pilot follow-up programme and high-S/N programme) three have total
masses of $10\times 10^{14}\, M_{\odot}$ or larger and two of them are
high S/N clusters in the ESZ sample.

In order to check the consistency of the cosmological model, we
compare the measured $Y_{5R500}$ with the X-ray predicted
$Y^{L_X}_{5R500}$ that is derived in a given cosmology. To do so, we
use the 158 ESZ clusters with X-ray-based size estimates. We vary the
cosmological parameter  $H_0$, in a range of  30 to 
100 km/s/Mpc assuming a flat
universe ($\Omega_{\mathrm{m}}=0.3$ and $\Omega_\Lambda =
1-\Omega_{\mathrm{m}}$).  The integrated Compton parameters of the 158
clusters were re-estimated from the \Planck\ data with the size
$5R_{500}$ obtained for each explored set of cosmological
parameters. The predicted SZ signals are then compared with the SZ
signal measured by \Planck\, providing us with the best value for
$H_0$. We find that $h$ is barely constrained, with a best estimate of
$H_0=71^{+10}_{-20}\kmsMpc$ (1$\sigma$ uncertainty).

\subsection{Comparison with existing catalogues}
\label{sec:properties}

After the first blind detection of galaxy clusters through their SZ
signature by SPT \citep{sta09} and further discoveries by both SPT
\citep{van10} and ACT \citep{men10}, \Planck\ with its broad frequency
coverage provides the first sample of SZ clusters detected blindly
over the whole sky. For its first and early release, \Planck\ delivers
to the community 189 clusters and candidates with S/N $\geq 6$ in the
ESZ sample, and an additional ten clusters at lower S/N. In total, the
30 new SZ-discovered clusters or candidates by \Planck\ double the
number of new clusters provided by ACT and SPT during the last year
based on their 455 deg$^2$ and 178 deg$^2$ respective
surveys. Moreover, \Planck\ provides the first homogeneous SZ
measurements for many known X-ray or optical clusters.

\begin{figure}
\centering
\includegraphics[width=7cm]{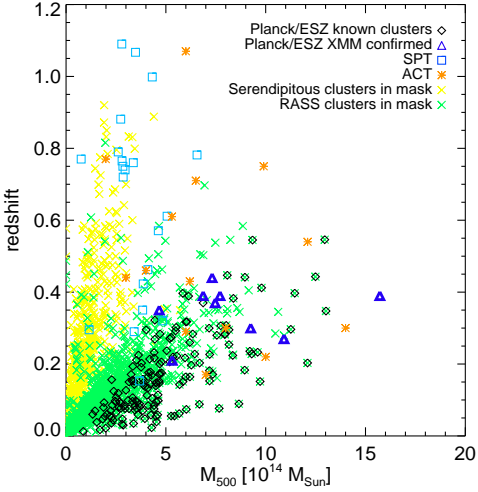}
\caption{The 158 clusters from the \Planck\ ESZ sample identified with
  known X-ray clusters in redshift--mass space, compared with SPT and
  ACT samples from \cite{men10,van10}, as well as serendipitous and
  RASS clusters.}
\label{fig:Mz}
\end{figure}

It is worth examining the distribution of the SZ clusters in the
$M$-$z$ plane (see Fig.~\ref{fig:Mz}).  The range of redshifts
covered by the \Planck\ ESZ sample, from $z=0.01$ to 0.55 with more
than 80\% of the clusters lying below $z=0.3$, is quite complementary
to the high redshift range explored by ACT and SPT experiments, from
$z\simeq 0.15$ to 1.2. The comparison of the estimated masses from the
different experiments is complicated by the fact that they are
obtained using different approaches, from the use of X-ray proxies to
that of mass-significance relations. Overall, we can see from
Fig.~\ref{fig:Mz} that the SPT cluster masses quoted in \cite{van10}
range between 1 and $5\times 10^{14}\, M_{\odot}$. As mentioned
previously, \Planck, being an all-sky survey, spans a broader cluster
mass range from 0.9 to $10\times 10^{14}\, M_{\odot}$ and is
particularly adapted to the detection of very massive clusters in the
tail of the distribution.

\par\bigskip

\begin{figure}
\centering
\includegraphics[width=7cm]{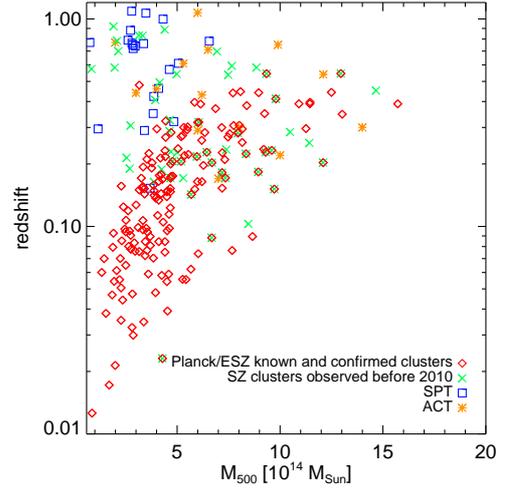}
\caption{The ESZ sample compared to the previously observed SZ
  clusters in redshift--mass space.}
\label{fig:MzSZ}
\end{figure}

The combination of \Planck\ with ACT and SPT experiments already
nicely samples the $M$-$z$ plane (see Fig. \ref{fig:MzSZ}). In
particular the highest redshift clusters are accessible to ACT and SPT
and the most massive clusters to \Planck. Moreover, \Planck\ already
samples the low-mass low-redshift space quite well and will provide us
with a robust reference point in this range. With the deeper
observations of the whole sky, combined with appropriate follow-up
programmes for redshift estimates, \Planck\ will be able to explore
the cluster mass function in its most cosmologically interesting
regimes: high redshifts and high masses. However, the detection of the
highest redshift clusters is likely to be hampered by the dilution by
\Planck\ beam. A combination of the \Planck\, ACT, and SPT carefully
taking into account the selection functions of all three experiments
will thus be needed to fully take advantage of SZ clusters as a
cosmological probe.

Moreover, combining the data from a sample of clusters with different
resolutions (including high-resolution imaging of SZ clusters with
interferometric experiments like SZA and CARMA) will allow us to
perform detailed studies of extended clusters and have a much better
handle on the pressure profile from SZ data directly.

\par\bigskip

Although the ESZ sample is not a catalogue with a fully characterised
selection function, it is worth comparing it to the ROSAT-based
cluster catalogues. To do this we take advantage of the MCXC, which
contains not only NORAS and REFLEX but other survey-based and
serendipitous cluster catalogues. Using the homogenised cluster
properties of the MCXC compilation, we can moreover predict the SZ
signal and the S/N ratio for a measurement of the Compton $Y$
parameter. In order to do this we estimate the \Planck\ noise from
real noise maps at the cluster positions using MMF3. Using this
information, we compared the number of detected clusters in the ESZ at
$\mathrm{S/N} \ge 6$ to the number predicted at that level of
significance. We find very good overall agreement in terms of detected
and predicted clusters, despite the fact that the predictions we use
are based on X-ray-selected clusters from the MCXC compilation and
that the cluster model used for the prediction does not account for
the dispersion in the scaling relations, and despite the noise
properties of channel maps being inhomogeneous across the sky. Only 26
MCXC clusters with predicted S/N $\geq 6$ are not within the ESZ
sample.  For 20 of these clusters information on the presence of a
cool core or peculiar morphology is available in the literature. We
find that 13 of these host cool cores. For these clusters, the X-ray
luminosity is boosted due to the central density peak. The mass
predicted from the luminosity, and hence the predicted SZ signal, is
over-estimated.  For 3 clusters the luminosity measurements adopted in
the MCXC are not reliable because of evidence of AGN contamination
(e.g., A689).  The remaining four clusters are peculiar because they
have very asymmetric morphologies or are located in superclusters
(e.g., A3526 in Centaurus and the A901/A902 system), making the SZ
signal predictions highly uncertain.

 \begin{figure}
   \centering
\includegraphics[width=7cm]{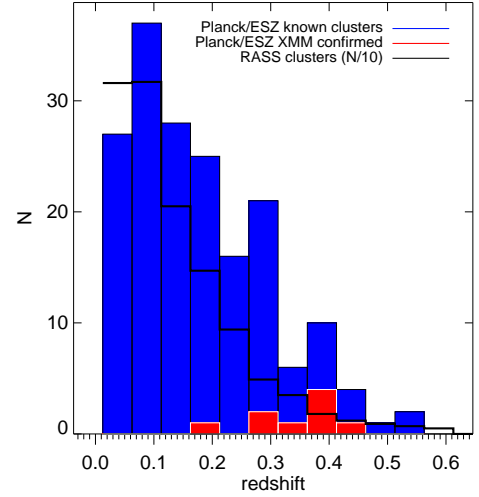}
      \caption{Distribution of ESZ sample in redshift. The 177
        identified ESZ clusters with redshift (from optical or 
	X-ray observations) are in blue, the ESZ
        clusters confirmed with XMM-Newton in red, and the RASS
        clusters (number density divided by 10) in black solid line.
      }
         \label{fig:zdist}
   \end{figure}
\begin{figure}
   \centering
\includegraphics[width=7cm]{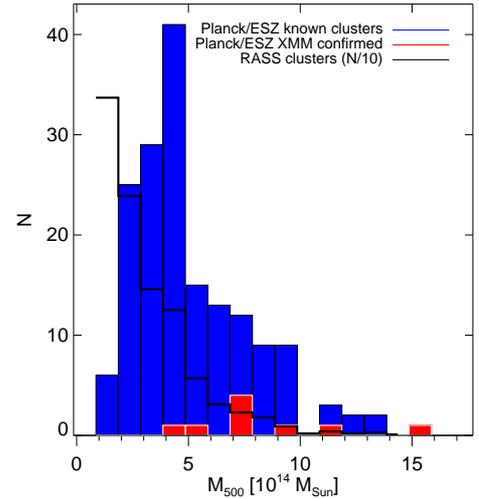}
      \caption{Distribution of ESZ sample in mass. The 167 identified
        ESZ clusters with masses are in blue, the ESZ
        clusters confirmed with XMM-Newton
        in red, and the RASS clusters (number density divided by 10)
        in black solid line.  }
         \label{fig:mdist}
   \end{figure}

\begin{figure}
\centering
\includegraphics[width=7cm]{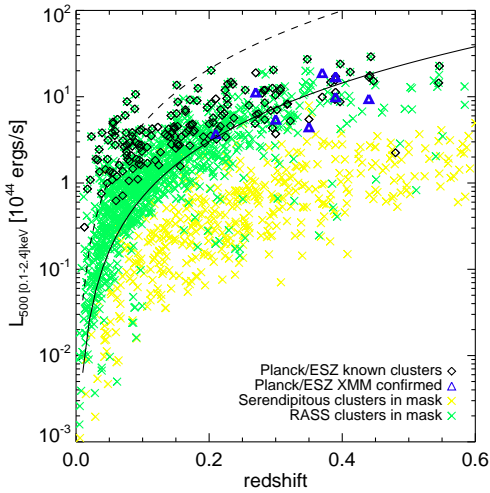}
\caption{The 158 clusters from the \Planck\ ESZ sample identified with
  known X-ray clusters in redshift--luminosity space, compared with
  serendipitous and RASS clusters.}
\label{fig:Lz}
\end{figure}

There is a large overlap between the \Planck\ ESZ sample and the
RASS-based cluster catalogues, in particular REFLEX and NORAS
(Fig. \ref{fig:Lz}). The 162 SZ candidates identified with X-ray
clusters from the MCXC compilation are predominantly clusters from the
REFLEX (74) and NORAS (59) surveys, which corresponds to an overlap of
17\% and 13\% with the REFLEX and NORAS surveys respectively. The
eleven ESZ clusters confirmed by XMM-Newton with $\mathrm{S/N} \ge 6$
were found to lie just around the REFLEX flux limit (only two are
above this limit).

It is thus interesting to compare the ESZ sample mass and redshift
distributions with those of the RASS-based catalogues. This is
illustrated in Figs.~\ref{fig:zdist} and \ref{fig:mdist} in which the
RASS-based mass and redshift distribution divided by ten are
over-plotted on the ESZ histograms in thick solid line. We find that
the ESZ clusters with masses below $4 \times 10^{14}\,M_{\odot}$ represent
only 12\% of the RASS-based clusters in the same mass range; however
they represent 90\% of the RASS-based clusters at higher masses $M
\geq 9 \times 10^{14}\,M_{\odot}$. As for the redshift distribution, the
\Planck\ ESZ clusters represent 14\% of the RASS-based clusters with
redshifts lower than 0.3 and they constitute 31\% of the RASS-based
clusters above $z = 0.3$.

\par\bigskip

The SDSS-MaxBCG cluster catalogue is the basis of the study of
optical-SZ scaling relations \citep{planck2011-5.2c} in \Planck\
data. It is used in particular to measure an integrated Compton
parameter, $Y_{5R500}^\mathrm{MaxBCG}$, from the \Planck\ channel maps
at the MaxBCG position using fixed cluster size according to published
weak-lensing calibrated mass--richness relations for the MaxBCG
catalogue.  Only 20 clusters from the MaxBCG have a measured
S/N larger than six and are thus expected to be
within the ESZ selection\footnote{This number accounts for the
  possible association of a candidate new cluster with a cluster from
  \cite{wen09}.}. Among them, 18 are effectively associated with ESZ
clusters (within a search radius of five arcminutes). One of the two
clusters not in the ESZ sample is Abell 1246 ($z=0.18$). The second is
a fortuitous association with a low-redshift ($z=0.06$) group of the
MaxBCG catalogue near the position of Abell 1795, which is detected in
the ESZ catalogue.

\section{Summary}

Thanks to its all-sky coverage and to its frequency range spanning the
SZ decrement and increment, \Planck\ provides us with the very first
all-sky S/N-selected SZ sample. This early release sample
of high-reliability SZ clusters and candidates (S/N from 6 to 29) was
constructed using a matched multi-filter detection technique. It was
validated using \Planck-internal quality assessment, external X-ray
and optical data, and a multi-frequency follow-up programme for
confirmation relying mostly on XMM-Newton snapshot observations.  The
ESZ sample comprises 189 candidates, of which 20 are candidate new
clusters and 169 have X-ray or optical counterparts. Of these, 162
were observed in X-ray. \Planck\ provides for the first time SZ
observations for about 80\% of the ESZ clusters and hence a
homogeneously measured SZ signal. Twelve candidate clusters in total,
out of the 20, have been confirmed. One candidate was confirmed by
AMI and WISE. Eleven were confirmed with XMM-Newton, including two
candidates found to be double clusters on the sky.

The clusters in the ESZ sample are mostly at moderate redshifts lying
between $z=0.01$ and $z=0.55$, with 86\% of them below $z=0.3$. The
ESZ-cluster masses span over a decade from 0.9 to $15\times 10^{14}\,
M_{\odot}$, i.e. up to the highest masses. The ESZ, constructed using
clear selection criteria, is a nearly complete (90\% above
$E^{-2/3}(z) Y_{5R500}D_{A}^2 \simeq 4\times 10^{-4} \,
\mathrm{Mpc}^2$), high-purity (above 95\%) SZ cluster sample. However,
as mentioned above, it is not possible at the present stage to provide
users with a full selection function.

Thanks to its all-sky coverage, \Planck\ has a unique capability to
detect the rarest and most massive clusters in the exponential tail of
the mass function. Planck is detecting new clusters in a region of
the mass-redshift plane that is sparsely populated by the RASS
catalogues. As a matter of fact, two of the newly-discovered clusters
in the ESZ and confirmed by XMM-Newton have estimated total masses
larger than $10^{15}\, M_{\odot}$. Furthermore, as indicated by
XMM-Newton snapshot observations, most of the new clusters have low
luminosity and a disturbed morphology, suggestive of a complex
dynamical state. \Planck\ may thus have started to reveal a
non-negligible population of massive dynamically-perturbed objects
that is under-represented in X-ray surveys.

A significant fraction of the ESZ clusters have good archival X-ray
and optical data. In addition, the ESZ sample should motivate
follow-up effort by the community. It will hence serve as a
valuable reference for studies of cluster physics at low and moderate
redshifts (e.g., galaxy properties versus intra-cluster gas physics,
metallicities, dynamical state and its evolution, etc). These studies
will require multi-wavelength observations including further SZ
observations at higher spatial resolution and observations in X-rays
(with XMM-Newton, Chandra, and Suzaku), in the optical (imaging and
spectroscopy), and in the radio (e.g., with LOFAR).

The ensemble of early results on the SZ signal in \Planck\, using a
selected local sub-sample of ESZ clusters with high-quality XMM-Newton
archival data \citep{planck2011-5.2b} and using the compilation of
about 1600 MCXC clusters \citep{planck2011-5.2a}, shows excellent
agreement between observed SZ quantities and X-ray-based predictions
underlining the robustness and consistency of our overall view of ICM
properties. These results shed light on long-standing questions
regarding the consistency between the SZ and X-ray view of hot gas in
galaxy clusters. In contrast, the SZ signal-to-optical-richness
relation measured from the SDSS-MaxBCG cluster catalogue
\cite{planck2011-5.2c} has a lower SZ signal than predicted. Extensive
SZ-optical statistical studies of this kind are new. The result, and
the origin of the difference, may be related to the cluster
population, such as the existence of a sub-population of X-ray
under-luminous clusters, or to selection effects in optical cluster
catalogues.

\par\bigskip

In the future, \Planck\ will deliver a larger all-sky SZ cluster
catalogue. The characterisation of the \Planck\ selection function
together with the construction of this legacy catalogue, including its
validation using follow up observations in particular with XMM-Newton,
will be one of the major activities.

The usefulness of the SZ cluster abundance in achieving precise
cosmological constraints relies on several theoretical and
observational requirements. One of them is the ability to obtain
redshift measurements for each confirmed SZ cluster.
Cross-correlation of \Planck\ data with the only available large
optical survey to date, the Sloan Digital Sky Survey (SDSS), can be
used to confirm \Planck\ candidates and provide redshift estimates on
a area restricted to the SDSS coverage area. The XMM-Newton
confirmation observations can provide redshift estimates, but only for
the X-ray brightest clusters.  A significant follow-up effort in the
optical (with ESO, ENO, and NOAO facilities) has thus been put in
place by the \Planck\ collaboration in order to obtain redshifts
(photometric and spectroscopic) for the SZ clusters. Another key
requirement for the cosmological use of the SZ catalogue is the
derivation of the fundamental relation between the integrated Compton
parameter, $Y$, and the cluster mass and its evolution with
redshift. \cite{planck2011-5.2b} have calibrated the local relation
between $Y$ and $Y_X$, the analogue of the SZ signal, measured from
the X-ray gas mass and temperature, to an unprecedented precision and,
for the first time, have demonstrated its remarkably small intrinsic
scatter. We will build an even more robust
and controlled observational proxy of the cluster mass which is
fundamental for cosmological applications. To do this, specific studies
based on the comparison of mass estimates from lensing, X-rays and SZ
observations for a selected representative sample of the SZ catalogue
will be most crucial.

Finally, combining \Planck\ all-sky SZ data with near future and
planned observations of the large-scale structure by large surveys,
e.g., PANSTARRS, LOFAR, Euclid, LSST, and e-ROSITA, will allow us to
understand the physical processes governing large-scale structure
formation and evolution.

\begin{acknowledgements}
The authors thank N. Schartel, ESA XMM-Newton project scientist, for
granting the Director Discretionary Time used for confirmation of SZ
Planck candidates. This research has made use of the following
databases: SIMBAD, operated at CDS, Strasbourg, France; the NED
database, which is operated by the Jet Propulsion Laboratory,
California Institute of Technology, under contract with the National
Aeronautics and Space Administration; BAX, operated by the Laboratoire
d'Astrophysique de Tarbes-Toulouse (LATT), under contract with the
Centre National d'Etudes Spatiales (CNES), SZ repository operated by
IAS Data and Operation Center (IDOC) under contract with CNES. The
authors acknowledge the use of software provided by the US National
Virtual Observatory. A description of the Planck Collaboration and a
list of its members, indicating which technical or scientific
activities they have been involved in, can be found at
http://www.rssd.esa.int/Planck.

\end{acknowledgements}

\bibliographystyle{aa}

\bibliography{esz.bib,Planck_bib.bib}

\begin{thebibliography}{118}
\expandafter\ifx\csname natexlab\endcsname\relax\def\natexlab#1{#1}\fi

\bibitem[{{Abell}(1958)}]{abe58}
{Abell}, G.~O. 1958, \apjs, 3, 211

\bibitem[{{Aghanim} {et~al.}(1997){Aghanim}, {de Luca}, {Bouchet}, {Gispert},
  \& {Puget}}]{agh97}
{Aghanim}, N., {de Luca}, A., {Bouchet}, F.~R., {Gispert}, R., \& {Puget},
  J.~L. 1997, \aap, 325, 9

\bibitem[{{Aghanim} {et~al.}(2005){Aghanim}, {Hansen}, \& {Lagache}}]{Agh05}
{Aghanim}, N., {Hansen}, S.~H., \& {Lagache}, G. 2005, \aap, 439, 901

\bibitem[{{Aghanim} {et~al.}(2008){Aghanim}, {Majumdar}, \& {Silk}}]{agh08}
{Aghanim}, N., {Majumdar}, S., \& {Silk}, J. 2008, Reports on Progress in
  Physics, 71, 066902

\bibitem[{{Ameglio} {et~al.}(2006){Ameglio}, {Borgani}, {Diaferio}, \&
  {Dolag}}]{ame06}
{Ameglio}, S., {Borgani}, S., {Diaferio}, A., \& {Dolag}, K. 2006, \mnras, 369,
  1459

\bibitem[{{Andersson} {et~al.}(2010){Andersson}, {Benson}, {Ade}, {Aird},
  {Armstrong}, {Bautz}, {Bleem}, {Brodwin}, {Carlstrom}, {Chang}, {Crawford},
  {Crites}, {de Haan}, {Desai}, {Dobbs}, {Dudley}, {Foley}, {Forman},
  {Garmire}, {George}, {Gladders}, {Halverson}, {High}, {Holder}, {Holzapfel},
  {Hrubes}, {Jones}, {Joy}, {Keisler}, {Knox}, {Lee}, {Leitch}, {Lueker},
  {Marrone}, {McMahon}, {Mehl}, {Meyer}, {Mohr}, {Montroy}, {Murray}, {Padin},
  {Plagge}, {Pryke}, {Reichardt}, {Rest}, {Ruel}, {Ruhl}, {Schaffer}, {Shaw},
  {Shirokoff}, {Song}, {Spieler}, {Stalder}, {Staniszewski}, {Stark}, {Stubbs},
  {Vanderlinde}, {Vieira}, {Vikhlinin}, {Williamson}, {Yang}, \&
  {Zahn}}]{and10}
{Andersson}, K., {Benson}, B.~A., {Ade}, P.~A.~R., {et~al.} 2010, ArXiv
  e-prints, {\tt arXiv:1006.3068}

\bibitem[{{Appenzeller} {et~al.}(1998){Appenzeller}, {Thiering}, {Zickgraf},
  {Krautter}, {Voges}, {Chavarria}, {Kneer}, {Mujica}, {Pakull}, {Rosso},
  {Ruzicka}, {Serrano}, \& {Ziegler}}]{app98}
{Appenzeller}, I., {Thiering}, I., {Zickgraf}, F., {et~al.} 1998, \apjs, 117,
  319

\bibitem[{Arnaud {et~al.}(2007)Arnaud, Pointecouteau, \& Pratt}]{arn07}
Arnaud, M., Pointecouteau, E., \& Pratt, G.~W. 2007, \aap, 474, L37

\bibitem[{{Arnaud} {et~al.}(2010){Arnaud}, {Pratt}, {Piffaretti},
  {B{\"o}hringer}, {Croston}, \& {Pointecouteau}}]{arn10}
{Arnaud}, M., {Pratt}, G.~W., {Piffaretti}, R., {et~al.} 2010, \aap, 517, A92

\bibitem[{{Battistelli} {et~al.}(2002){Battistelli}, {De Petris}, {Lamagna},
  {Melchiorri}, {Palladino}, {Savini}, {Cooray}, {Melchiorri}, {Rephaeli}, \&
  {Shimon}}]{bat02}
{Battistelli}, E.~S., {De Petris}, M., {Lamagna}, L., {et~al.} 2002, \apjl,
  580, L101

\bibitem[{{Battistelli} {et~al.}(2006){Battistelli}, {De Petris}, {Lamagna},
  {Watson}, {Rebolo}, {Melchiorri}, {G{\'e}nova-Santos}, {Luzzi}, {De Gregori},
  {Rubi{\~n}o-Martin}, {Davies}, {Davis}, {Grainge}, {Hobson}, {Saunders}, \&
  {Scott}}]{bat06}
{Battistelli}, E.~S., {De Petris}, M., {Lamagna}, L., {et~al.} 2006, \apj, 645,
  826

\bibitem[{{Benson} {et~al.}(2003){Benson}, {Church}, {Ade}, {Bock}, {Ganga},
  {Hinderks}, {Mauskopf}, {Philhour}, {Runyan}, \& {Thompson}}]{ben03}
{Benson}, B.~A., {Church}, S.~E., {Ade}, P.~A.~R., {et~al.} 2003, \apj, 592,
  674

\bibitem[{{Bethermin} {et~al.}(2010){Bethermin}, {Dole}, {Beelen}, \&
  {Aussel}}]{bet10}
{Bethermin}, M., {Dole}, H., {Beelen}, A., \& {Aussel}, H. 2010, VizieR Online
  Data Catalog, 351, 29078

\bibitem[{{Birkinshaw}(1999)}]{bir99}
{Birkinshaw}, M. 1999, \physrep, 310, 97

\bibitem[{{Birkinshaw} \& {Gull}(1978)}]{bir78}
{Birkinshaw}, M. \& {Gull}, S.~F. 1978, \nat, 274, 111

\bibitem[{{Birkinshaw} \& {Hughes}(1994)}]{bir94}
{Birkinshaw}, M. \& {Hughes}, J.~P. 1994, \apj, 420, 33

\bibitem[{{Birkinshaw} \& {Lancaster}(2005)}]{bir05}
{Birkinshaw}, M. \& {Lancaster}, K. 2005, in Background Microwave Radiation and
  Intracluster Cosmology, ed. {F.~Melchiorri \& Y.~Rephaeli}, 127--+

\bibitem[{{Bock} {et~al.}(1999){Bock}, {Large}, \& {Sadler}}]{sumss}
{Bock}, D., {Large}, M.~I., \& {Sadler}, E.~M. 1999, \aj, 117, 1578

\bibitem[{{B{\"o}hringer} {et~al.}(2007){B{\"o}hringer}, {Schuecker}, {Pratt},
  {Arnaud}, {Ponman}, {Croston}, {Borgani}, {Bower}, {Briel}, {Collins},
  {Donahue}, {Forman}, {Finoguenov}, {Geller}, {Guzzo}, {Henry}, {Kneissl},
  {Mohr}, {Matsushita}, {Mullis}, {Ohashi}, {Pedersen}, {Pierini}, {Quintana},
  {Raychaudhury}, {Reiprich}, {Romer}, {Rosati}, {Sabirli}, {Temple}, {Viana},
  {Vikhlinin}, {Voit}, \& {Zhang}}]{boe07}
{B{\"o}hringer}, H., {Schuecker}, P., {Pratt}, G.~W., {et~al.} 2007, \aap, 469,
  363

\bibitem[{{Bonamente} {et~al.}(2006){Bonamente}, {Joy}, {LaRoque}, {Carlstrom},
  {Reese}, \& {Dawson}}]{bon06}
{Bonamente}, M., {Joy}, M.~K., {LaRoque}, S.~J., {et~al.} 2006, \apj, 647, 25

\bibitem[{{Borgani}(2006)}]{bor06}
{Borgani}, S. 2006, ArXiv Astrophysics e-prints {{\tt astro-ph/0605575}},

\bibitem[{{Borgani} {et~al.}(2004){Borgani}, {Murante}, {Springel}, {Diaferio},
  {Dolag}, {Moscardini}, {Tormen}, {Tornatore}, \& {Tozzi}}]{bor04}
{Borgani}, S., {Murante}, G., {Springel}, V., {et~al.} 2004, \mnras, 348, 1078

\bibitem[{{Budav{\'a}ri} {et~al.}(2007){Budav{\'a}ri}, {Dobos}, {Szalay},
  {Greene}, {Gray}, \& {Rots}}]{vo-footprint-paper}
{Budav{\'a}ri}, T., {Dobos}, L., {Szalay}, A.~S., {et~al.} 2007, in
  Astronomical Society of the Pacific Conference Series, Vol. 376, Astronomical
  Data Analysis Software and Systems XVI, ed. {R.~A.~Shaw, F.~Hill, \&
  D.~J.~Bell}, 559--+

\bibitem[{{Carlstrom} {et~al.}(2009){Carlstrom}, {Ade}, {Aird}, {Benson},
  {Bleem}, {Busetti}, {Chang}, {Chauvin}, {Cho}, {Crawford}, {Crites}, {Dobbs},
  {Halverson}, {Heimsath}, {Holzapfel}, {Hrubes}, {Joy}, {Keisler}, {Lanting},
  {Lee}, {Leitch}, {Leong}, {Lu}, {Lueker}, {McMahon}, {Mehl}, {Meyer}, {Mohr},
  {Montroy}, {Padin}, {Plagge}, {Pryke}, {Ruhl}, {Schaffer}, {Schwan},
  {Shirokoff}, {Spieler}, {Staniszewski}, {Stark}, \& {Vieira}}]{spt}
{Carlstrom}, J.~E., {Ade}, P.~A.~R., {Aird}, K.~A., {et~al.} 2009, ArXiv
  e-prints, {\tt arXiv:0907.4445}

\bibitem[{{Carlstrom} {et~al.}(2002){Carlstrom}, {Holder}, \& {Reese}}]{car02}
{Carlstrom}, J.~E., {Holder}, G.~P., \& {Reese}, E.~D. 2002, \araa, 40, 643

\bibitem[{{Carvalho} {et~al.}(2009){Carvalho}, {Rocha}, \& {Hobson}}]{car09}
{Carvalho}, P., {Rocha}, G., \& {Hobson}, M.~P. 2009, \mnras, 393, 681

\bibitem[{{Cavaliere} \& {Fusco-Femiano}(1978)}]{cav78}
{Cavaliere}, A. \& {Fusco-Femiano}, R. 1978, \aap, 70, 677

\bibitem[{{Colafrancesco} {et~al.}(2003){Colafrancesco}, {Marchegiani}, \&
  {Palladino}}]{col03}
{Colafrancesco}, S., {Marchegiani}, P., \& {Palladino}, E. 2003, \aap, 397, 27

\bibitem[{{Condon} {et~al.}(1998){Condon}, {Cotton}, {Greisen}, {Yin},
  {Perley}, {Taylor}, \& {Broderick}}]{nvss}
{Condon}, J.~J., {Cotton}, W.~D., {Greisen}, E.~W., {et~al.} 1998, \aj, 115,
  1693

\bibitem[{{Cortese} {et~al.}(2004){Cortese}, {Gavazzi}, {Boselli},
  {Iglesias-Paramo}, \& {Carrasco}}]{cor04}
{Cortese}, L., {Gavazzi}, G., {Boselli}, A., {Iglesias-Paramo}, J., \&
  {Carrasco}, L. 2004, \aap, 425, 429

\bibitem[{{da Silva} {et~al.}(2001){da Silva}, {Barbosa}, {Liddle}, \&
  {Thomas}}]{das01}
{da Silva}, A.~C., {Barbosa}, D., {Liddle}, A.~R., \& {Thomas}, P.~A. 2001,
  \mnras, 326, 155

\bibitem[{{da Silva} {et~al.}(2004){da Silva}, {Kay}, {Liddle}, \&
  {Thomas}}]{das04}
{da Silva}, A.~C., {Kay}, S.~T., {Liddle}, A.~R., \& {Thomas}, P.~A. 2004,
  \mnras, 348, 1401

\bibitem[{{Dawson} {et~al.}(2001){Dawson}, {Holzapfel}, {Carlstrom}, {Joy},
  {LaRoque}, \& {Reese}}]{daw01}
{Dawson}, K.~S., {Holzapfel}, W.~L., {Carlstrom}, J.~E., {et~al.} 2001, \apjl,
  553, L1

\bibitem[{{De Petris} {et~al.}(1999){De Petris}, {Mainella}, {Nerozzi}, {de
  Bernardis}, {Garavini}, {Granata}, {Guarini}, {Masi}, {Melchiorri},
  {Melchiorri}, {Nobili}, {Orlando}, {Palummo}, {Pisano}, \&
  {Terracina}}]{dep99}
{De Petris}, M., {Mainella}, G., {Nerozzi}, A., {et~al.} 1999, \nar, 43, 297

\bibitem[{{Dobbs} {et~al.}(2006){Dobbs}, {Halverson}, {Ade}, {Basu}, {Beelen},
  {Bertoldi}, {Cohalan}, {Cho}, {G{\"u}sten}, {Holzapfel}, {Kermish},
  {Kneissl}, {Kov{\'a}cs}, {Kreysa}, {Lanting}, {Lee}, {Lueker}, {Mehl},
  {Menten}, {Muders}, {Nord}, {Plagge}, {Richards}, {Schilke}, {Schwan},
  {Spieler}, {Weiss}, \& {White}}]{dob06}
{Dobbs}, M., {Halverson}, N.~W., {Ade}, P.~A.~R., {et~al.} 2006, \nar, 50, 960

\bibitem[{{Dole} {et~al.}(2006){Dole}, {Lagache}, {Puget}, {Caputi},
  {Fern{\'a}ndez-Conde}, {Le Floc'h}, {Papovich}, {P{\'e}rez-Gonz{\'a}lez},
  {Rieke}, \& {Blaylock}}]{dol06}
{Dole}, H., {Lagache}, G., {Puget}, J., {et~al.} 2006, \aap, 451, 417

\bibitem[{{Douspis} {et~al.}(2011){Douspis}, {Aghanim}, {Evrard}, \&
  {Langer}}]{dou11}
{Douspis}, M., {Aghanim}, N., {Evrard}, L., \& {Langer}, M. 2011, in
  preparation

\bibitem[{{Douspis} {et~al.}(2006){Douspis}, {Aghanim}, \& {Langer}}]{dou06}
{Douspis}, M., {Aghanim}, N., \& {Langer}, M. 2006, \aap, 456, 819

\bibitem[{{Einasto} {et~al.}(2001){Einasto}, {Einasto}, {Tago}, {M{\"u}ller},
  \& {Andernach}}]{ein01}
{Einasto}, M., {Einasto}, J., {Tago}, E., {M{\"u}ller}, V., \& {Andernach}, H.
  2001, \aj, 122, 2222

\bibitem[{{Fixsen} {et~al.}(1994){Fixsen}, {Cheng}, {Cottingham}, {Eplee},
  {Hewagama}, {Isaacman}, {Jensen}, {Mather}, {Massa}, {Meyer}, {Noerdlinger},
  {Read}, {Rosen}, {Shafer}, {Trenholme}, {Weiss}, {Bennett}, {Boggess},
  {Wilkinson}, \& {Wright}}]{fix94}
{Fixsen}, D.~J., {Cheng}, E.~S., {Cottingham}, D.~A., {et~al.} 1994, \apj, 420,
  457

\bibitem[{{G{\'e}nova-Santos} {et~al.}(2005){G{\'e}nova-Santos},
  {Rubi{\~n}o-Mart{\'{\i}}n}, {Rebolo}, {Cleary}, {Davies}, {Davis},
  {Dickinson}, {Falc{\'o}n}, {Grainge}, {Guti{\'e}rrez}, {Hobson}, {Jones},
  {Kneissl}, {Lancaster}, {Padilla-Torres}, {Saunders}, {Scott}, {Taylor}, \&
  {Watson}}]{gen05}
{G{\'e}nova-Santos}, R., {Rubi{\~n}o-Mart{\'{\i}}n}, J.~A., {Rebolo}, R.,
  {et~al.} 2005, \mnras, 363, 79

\bibitem[{{Ghizzardi} {et~al.}(2010){Ghizzardi}, {Rossetti}, \&
  {Molendi}}]{ghi10}
{Ghizzardi}, S., {Rossetti}, M., \& {Molendi}, S. 2010, \aap, 516, A32+

\bibitem[{{Glenn} {et~al.}(1998){Glenn}, {Bock}, {Chattopadhyay}, {Edgington},
  {Lange}, {Zmuidzinas}, {Mauskopf}, {Rownd}, {Yuen}, \& {Ade}}]{gle98}
{Glenn}, J., {Bock}, J.~J., {Chattopadhyay}, G., {et~al.} 1998, in Presented at
  the Society of Photo-Optical Instrumentation Engineers (SPIE) Conference,
  Vol. 3357, Society of Photo-Optical Instrumentation Engineers (SPIE)
  Conference Series, ed. {T.~G.~Phillips}, 326--334

\bibitem[{{G{\'o}mez} {et~al.}(2003){G{\'o}mez}, {Romer}, {Peterson},
  {Cantalupo}, {Holzapfel}, {Kuo}, {Newcomb}, {Ruhl}, {Goldstein}, {Torbet}, \&
  {Runyan}}]{gom03}
{G{\'o}mez}, P.~L., {Romer}, K.~A., {Peterson}, J., {et~al.} 2003, in
  Astronomical Society of the Pacific Conference Series, Vol. 301, Astronomical
  Society of the Pacific Conference Series, ed. {S.~Bowyer \& C.-Y.~Hwang},
  495--+

\bibitem[{{G{\'o}rski} {et~al.}(2005){G{\'o}rski}, {Hivon}, {Banday},
  {Wandelt}, {Hansen}, {Reinecke}, \& {Bartelmann}}]{gor05}
{G{\'o}rski}, K.~M., {Hivon}, E., {Banday}, A.~J., {et~al.} 2005, \apj, 622,
  759

\bibitem[{{Grego} {et~al.}(2001){Grego}, {Carlstrom}, {Reese}, {Holder},
  {Holzapfel}, {Joy}, {Mohr}, \& {Patel}}]{gre01}
{Grego}, L., {Carlstrom}, J.~E., {Reese}, E.~D., {et~al.} 2001, \apj, 552, 2

\bibitem[{{Haiman} {et~al.}(2001){Haiman}, {Mohr}, \& {Holder}}]{hai01}
{Haiman}, Z., {Mohr}, J.~J., \& {Holder}, G.~P. 2001, \apj, 553, 545

\bibitem[{{Herranz} {et~al.}(2002){Herranz}, {Sanz}, {Hobson}, {Barreiro},
  {Diego}, {Mart{\'i}nez-Gonz{\'a}lez}, \& {Lasenby}}]{her02}
{Herranz}, D., {Sanz}, J.~L., {Hobson}, M.~P., {et~al.} 2002, \mnras, 336, 1057

\bibitem[{{Hinshaw} {et~al.}(2009){Hinshaw}, {Weiland}, {Hill}, {Odegard},
  {Larson}, {Bennett}, {Dunkley}, {Gold}, {Greason}, {Jarosik}, {Komatsu},
  {Nolta}, {Page}, {Spergel}, {Wollack}, {Halpern}, {Kogut}, {Limon}, {Meyer},
  {Tucker}, \& {Wright}}]{hin09}
{Hinshaw}, G., {Weiland}, J.~L., {Hill}, R.~S., {et~al.} 2009, \apjs, 180, 225

\bibitem[{{Holzapfel} {et~al.}(1997){Holzapfel}, {Wilbanks}, {Ade}, {Church},
  {Fischer}, {Mauskopf}, {Osgood}, \& {Lange}}]{hol97}
{Holzapfel}, W.~L., {Wilbanks}, T.~M., {Ade}, P.~A.~R., {et~al.} 1997, \apj,
  479, 17

\bibitem[{{Horellou} {et~al.}(2005){Horellou}, {Nord}, {Johansson}, \&
  {L{\'e}vy}}]{hor05}
{Horellou}, C., {Nord}, M., {Johansson}, D., \& {L{\'e}vy}, A. 2005, \aap, 441,
  435

\bibitem[{{Hurier} {et~al.}(2010){Hurier}, {Hildebrandt}, \&
  {Macias-Perez}}]{hur10}
{Hurier}, G., {Hildebrandt}, S.~R., \& {Macias-Perez}, J.~F. 2010, ArXiv
  e-prints, {\tt arXiv:1007.1149}

\bibitem[{{Hurley-Walker} {et~al.}(2011){Hurley-Walker}, {Brown}, {Davies},
  {Feroz}, {Franzen}, {Grainge}, {Hobson}, {Lasenby}, {Olamaie}, {Pooley},
  {Rodriguez-Gonzalvez}, {Saunders}, {Schammel}, {Scaife}, {Scott}, {Shimwell},
  {Titterington}, \& {Waldram}}]{hur11}
{Hurley-Walker}, T.~A.~C.~N., {Brown}, M.~L., {Davies}, M.~L., {et~al.} 2011,
  ArXiv e-prints, {\tt arXiv:1103.0947}

\bibitem[{{Itoh} {et~al.}(1998){Itoh}, {Kohyama}, \& {Nozawa}}]{ito98}
{Itoh}, N., {Kohyama}, Y., \& {Nozawa}, S. 1998, \apj, 502, 7

\bibitem[{{Jenkins} {et~al.}(2001){Jenkins}, {Frenk}, {White}, {Colberg},
  {Cole}, {Evrard}, {Couchman}, \& {Yoshida}}]{jen10}
{Jenkins}, A., {Frenk}, C.~S., {White}, S.~D.~M., {et~al.} 2001, \mnras, 321,
  372

\bibitem[{{Jones} {et~al.}(1993){Jones}, {Saunders}, {Alexander}, {Birkinshaw},
  {Dilon}, {Grainge}, {Hancock}, {Lasenby}, {Lefebvre}, \& {Pooley}}]{jon93}
{Jones}, M., {Saunders}, R., {Alexander}, P., {et~al.} 1993, \nat, 365, 320

\bibitem[{{Kashlinsky} {et~al.}(2008){Kashlinsky}, {Atrio-Barandela},
  {Kocevski}, \& {Ebeling}}]{kas08}
{Kashlinsky}, A., {Atrio-Barandela}, F., {Kocevski}, D., \& {Ebeling}, H. 2008,
  \apjl, 686, L49

\bibitem[{{Kobayashi} {et~al.}(1996){Kobayashi}, {Sasaki}, \& {Suto}}]{kob96}
{Kobayashi}, S., {Sasaki}, S., \& {Suto}, Y. 1996, \pasj, 48, L107

\bibitem[{{Koester} {et~al.}(2007){Koester}, {McKay}, {Annis}, {Wechsler},
  {Evrard}, {Bleem}, {Becker}, {Johnston}, {Sheldon}, {Nichol}, {Miller},
  {Scranton}, {Bahcall}, {Barentine}, {Brewington}, {Brinkmann}, {Harvanek},
  {Kleinman}, {Krzesinski}, {Long}, {Nitta}, {Schneider}, {Sneddin}, {Voges},
  \& {York}}]{koe07}
{Koester}, B.~P., {McKay}, T.~A., {Annis}, J., {et~al.} 2007, \apj, 660, 239

\bibitem[{{Komatsu} {et~al.}(1999){Komatsu}, {Kitayama}, {Suto}, {Hattori},
  {Kawabe}, {Matsuo}, {Schindler}, \& {Yoshikawa}}]{kom99}
{Komatsu}, E., {Kitayama}, T., {Suto}, Y., {et~al.} 1999, \apjl, 516, L1

\bibitem[{{Kravtsov} {et~al.}(2006){Kravtsov}, {Vikhlinin}, \& {Nagai}}]{kra06}
{Kravtsov}, A.~V., {Vikhlinin}, A., \& {Nagai}, D. 2006, \apj, 650, 128

\bibitem[{{Lamarre} {et~al.}(2010){Lamarre}, {Puget}, {Ade}, {Bouchet},
  {Guyot}, {Lange}, {Pajot}, {Arondel}, {Benabed}, {Beney}, {Beno{\^i}t},
  {Bernard}, {Bhatia}, {Blanc}, {Bock}, {Br{\'e}elle}, {Bradshaw}, {Camus},
  {Catalano}, {Charra}, {Charra}, {Church}, {Couchot}, {Coulais}, {Crill},
  {Crook}, {Dassas}, {de Bernardis}, {Delabrouille}, {de Marcillac}, {Delouis},
  {D{\'e}sert}, {Dumesnil}, {Dupac}, {Efstathiou}, {Eng}, {Evesque},
  {Fourmond}, {Ganga}, {Giard}, {Gispert}, {Guglielmi}, {Haissinski},
  {Henrot-Versill{\'e}}, {Hivon}, {Holmes}, {Jones}, {Koch}, {Lagard{\`e}re},
  {Lami}, {Land{\'e}}, {Leriche}, {Leroy}, {Longval},
  {Mac{\'{\i}}as-P{\'e}rez}, {Maciaszek}, {Maffei}, {Mansoux}, {Marty}, {Masi},
  {Mercier}, {Miville-Desch{\^e}nes}, {Moneti}, {Montier}, {Murphy},
  {Narbonne}, {Nexon}, {Paine}, {Pahn}, {Perdereau}, {Piacentini}, {Piat},
  {Plaszczynski}, {Pointecouteau}, {Pons}, {Ponthieu}, {Prunet}, {Rambaud},
  {Recouvreur}, {Renault}, {Ristorcelli}, {Rosset}, {Santos}, {Savini},
  {Serra}, {Stassi}, {Sudiwala}, {Sygnet}, {Tauber}, {Torre}, {Tristram},
  {Vibert}, {Woodcraft}, {Yurchenko}, \& {Yvon}}]{Lamarre2010}
{Lamarre}, J., {Puget}, J., {Ade}, P.~A.~R., {et~al.} 2010, \aap, 520, A9+

\bibitem[{{Lamarre} {et~al.}(1998){Lamarre}, {Giard}, {Pointecouteau},
  {Bernard}, {Serra}, {Pajot}, {D{\'e}sert}, {Ristorcelli}, {Torre}, {Church},
  {Coron}, {Puget}, \& {Bock}}]{lam98}
{Lamarre}, J.~M., {Giard}, M., {Pointecouteau}, E., {et~al.} 1998, \apjl, 507,
  L5

\bibitem[{{Lancaster} {et~al.}(2005){Lancaster}, {Genova-Santos}, {Falc{\`o}n},
  {Grainge}, {Guti{\`e}rrez}, {Kneissl}, {Marshall}, {Pooley}, {Rebolo},
  {Rubi{\~n}o-Martin}, {Saunders}, {Waldram}, \& {Watson}}]{lan05}
{Lancaster}, K., {Genova-Santos}, R., {Falc{\`o}n}, N., {et~al.} 2005, \mnras,
  359, 16

\bibitem[{{Levine} {et~al.}(2002){Levine}, {Schulz}, \& {White}}]{lev02}
{Levine}, E.~S., {Schulz}, A.~E., \& {White}, M. 2002, \apj, 577, 569

\bibitem[{{Lin} {et~al.}(2009){Lin}, {Partridge}, {Pober}, {Bouchefry},
  {Burke}, {Klein}, {Coish}, \& {Huffenberger}}]{Lin09}
{Lin}, Y., {Partridge}, B., {Pober}, J.~C., {et~al.} 2009, \apj, 694, 992

\bibitem[{{Luzzi} {et~al.}(2009){Luzzi}, {Shimon}, {Lamagna}, {Rephaeli}, {De
  Petris}, {Conte}, {De Gregori}, \& {Battistelli}}]{luz09}
{Luzzi}, G., {Shimon}, M., {Lamagna}, L., {et~al.} 2009, \apj, 705, 1122

\bibitem[{{Majumdar} \& {Mohr}(2004)}]{maj04}
{Majumdar}, S. \& {Mohr}, J.~J. 2004, \apj, 613, 41

\bibitem[{{Mantz} {et~al.}(2010){Mantz}, {Allen}, {Ebeling}, {Rapetti}, \&
  {Drlica-Wagner}}]{man10}
{Mantz}, A., {Allen}, S.~W., {Ebeling}, H., {Rapetti}, D., \& {Drlica-Wagner},
  A. 2010, \mnras, 406, 1773

\bibitem[{{Marriage} {et~al.}(2010){Marriage}, {Acquaviva}, {Ade}, {Aguirre},
  {Amiri}, {Appel}, {Barrientos}, {Battistelli}, {Bond}, {Brown}, {Burger},
  {Chervenak}, {Das}, {Devlin}, {Dicker}, {Doriese}, {Dunkley}, {Dunner},
  {Essinger-Hileman}, {Fisher}, {Fowler}, {Hajian}, {Halpern}, {Hasselfield},
  {Hern'andez-Monteagudo}, {Hilton}, {Hilton}, {Hincks}, {Hlozek},
  {Huffenberger}, {Hughes}, {Hughes}, {Infante}, {Irwin}, {Juin}, {Kaul},
  {Klein}, {Kosowsky}, {Lau}, {Limon}, {Lin}, {Lupton}, {Marsden}, {Martocci},
  {Mauskopf}, {Menanteau}, {Moodley}, {Moseley}, {Netterfield}, {Niemack},
  {Nolta}, {Page}, {Parker}, {Partridge}, {Quintana}, {Reese}, {Reid},
  {Sehgal}, {Sherwin}, {Sievers}, {Spergel}, {Staggs}, {Swetz}, {Switzer},
  {Thornton}, {Trac}, {Tucker}, {Warne}, {Wilson}, {Wollack}, \&
  {Zhao}}]{mar10}
{Marriage}, T.~A., {Acquaviva}, V., {Ade}, P.~A.~R., {et~al.} 2010, ArXiv
  e-prints, {\tt arXiv:1010.1065}

\bibitem[{{Melin} {et~al.}(2006){Melin}, {Bartlett}, \& {Delabrouille}}]{mel06}
{Melin}, J., {Bartlett}, J.~G., \& {Delabrouille}, J. 2006, \aap, 459, 341

\bibitem[{{Melin} {et~al.}(2011){Melin}, {Aghanim}, {Bartelmann}, {Bartlett},
  {Betoule}, {Bobin}, {Carvalho}, {Chon}, {Delabrouille}, {Diego}, {Harrison},
  {Herranz}, {Hobson}, {Kneissl}, {Lasenby}, {Le Jeune}, {Lopez-Caniego},
  {Mazzotta}, {Rocha}, {Schaefer}, {Starck}, {Waizmann}, \& {Yvon}}]{szchal}
{Melin}, J.-B., {Aghanim}, N., {Bartelmann}, M., {et~al.} 2011

\bibitem[{{Menanteau} {et~al.}(2010){Menanteau}, {Gonz{\'a}lez}, {Juin},
  {Marriage}, {Reese}, {Acquaviva}, {Aguirre}, {Appel}, {Baker}, {Barrientos},
  {Battistelli}, {Bond}, {Das}, {Deshpande}, {Devlin}, {Dicker}, {Dunkley},
  {D{\"u}nner}, {Essinger-Hileman}, {Fowler}, {Hajian}, {Halpern},
  {Hasselfield}, {Hern{\'a}ndez-Monteagudo}, {Hilton}, {Hincks}, {Hlozek},
  {Huffenberger}, {Hughes}, {Infante}, {Irwin}, {Klein}, {Kosowsky}, {Lin},
  {Marsden}, {Moodley}, {Niemack}, {Nolta}, {Page}, {Parker}, {Partridge},
  {Sehgal}, {Sievers}, {Spergel}, {Staggs}, {Swetz}, {Switzer}, {Thornton},
  {Trac}, {Warne}, \& {Wollack}}]{men10}
{Menanteau}, F., {Gonz{\'a}lez}, J., {Juin}, J., {et~al.} 2010, \apj, 723, 1523

\bibitem[{{Miville-Desch{\^e}nes} \& {Lagache}(2005)}]{mam05}
{Miville-Desch{\^e}nes}, M. \& {Lagache}, G. 2005, \apjs, 157, 302

\bibitem[{{Montier} {et~al.}(2010){Montier}, {Pelkonen}, {Juvela},
  {Ristorcelli}, \& {Marshall}}]{mon10}
{Montier}, L.~A., {Pelkonen}, V., {Juvela}, M., {Ristorcelli}, I., \&
  {Marshall}, D.~J. 2010, \aap, 522, A83+

\bibitem[{{Motl} {et~al.}(2005){Motl}, {Hallman}, {Burns}, \& {Norman}}]{mot05}
{Motl}, P.~M., {Hallman}, E.~J., {Burns}, J.~O., \& {Norman}, M.~L. 2005,
  \apjl, 623, L63

\bibitem[{{Muchovej} {et~al.}(2007){Muchovej}, {Mroczkowski}, {Carlstrom},
  {Cartwright}, {Greer}, {Hennessy}, {Loh}, {Pryke}, {Reddall}, {Runyan},
  {Sharp}, {Hawkins}, {Lamb}, {Woody}, {Joy}, {Leitch}, \& {Miller}}]{muc07}
{Muchovej}, S., {Mroczkowski}, T., {Carlstrom}, J.~E., {et~al.} 2007, \apj,
  663, 708

\bibitem[{{Nagai} {et~al.}(2007){Nagai}, {Kravtsov}, \& {Vikhlinin}}]{nag07}
{Nagai}, D., {Kravtsov}, A.~V., \& {Vikhlinin}, A. 2007, \apj, 668, 1

\bibitem[{{Pfrommer} {et~al.}(2007){Pfrommer}, {En{\ss}lin}, {Springel},
  {Jubelgas}, \& {Dolag}}]{pfr07}
{Pfrommer}, C., {En{\ss}lin}, T.~A., {Springel}, V., {Jubelgas}, M., \&
  {Dolag}, K. 2007, \mnras, 378, 385

\bibitem[{{Piffaretti} {et~al.}(2010){Piffaretti}, {Arnaud}, {Pratt},
  {Pointecouteau}, \& {Melin}}]{pif10}
{Piffaretti}, R., {Arnaud}, M., {Pratt}, G.~W., {Pointecouteau}, E., \&
  {Melin}, J. 2010, arXiv e-print, {\tt arXiv:1007.1916}

\bibitem[{{Piffaretti} \& {Valdarnini}(2008)}]{pif08}
{Piffaretti}, R. \& {Valdarnini}, R. 2008, \aap, 491, 71

\bibitem[{{Plagge} {et~al.}(2010){Plagge}, {Benson}, {Ade}, {Aird}, {Bleem},
  {Carlstrom}, {Chang}, {Cho}, {Crawford}, {Crites}, {de Haan}, {Dobbs},
  {George}, {Hall}, {Halverson}, {Holder}, {Holzapfel}, {Hrubes}, {Joy},
  {Keisler}, {Knox}, {Lee}, {Leitch}, {Lueker}, {Marrone}, {McMahon}, {Mehl},
  {Meyer}, {Mohr}, {Montroy}, {Padin}, {Pryke}, {Reichardt}, {Ruhl},
  {Schaffer}, {Shaw}, {Shirokoff}, {Spieler}, {Stalder}, {Staniszewski},
  {Stark}, {Vanderlinde}, {Vieira}, {Williamson}, \& {Zahn}}]{pla10}
{Plagge}, T., {Benson}, B.~A., {Ade}, P.~A.~R., {et~al.} 2010, \apj, 716, 1118

\bibitem[{{Planck Collaboration}(2011{\natexlab{a}})}]{planck2011-1.10}
{Planck Collaboration}. 2011{\natexlab{a}}, {Planck early results 07: The Early
  Release Compact Source Catalogue} ({Submitted to \aap,
  [arXiv:astro-ph/1101.2041]})

\bibitem[{{Planck Collaboration}(2011{\natexlab{b}})}]{planck2011-5.1b}
{Planck Collaboration}. 2011{\natexlab{b}}, {Planck early results 09:
  XMM-Newton follow-up for validation of Planck cluster candidates} ({Submitted
  to \aap, [arXiv:astro-ph/1101.2025]})

\bibitem[{{Planck Collaboration}(2011{\natexlab{c}})}]{planck2011-5.2a}
{Planck Collaboration}. 2011{\natexlab{c}}, {Planck early results 10:
  Statistical analysis of Sunyaev-Zeldovich scaling relations for X-ray galaxy
  clusters} ({Submitted to \aap, [arXiv:astro-ph/1101.2043]})

\bibitem[{{Planck Collaboration}(2011{\natexlab{d}})}]{planck2011-5.2b}
{Planck Collaboration}. 2011{\natexlab{d}}, {Planck early results 11:
  Calibration of the local galaxy cluster Sunyaev-Zeldovich scaling relations}
  ({Submitted to \aap, [arXiv:astro-ph/1101.2026]})

\bibitem[{{Planck Collaboration}(2011{\natexlab{e}})}]{planck2011-5.2c}
{Planck Collaboration}. 2011{\natexlab{e}}, {Planck early results 12: Cluster
  Sunyaev-Zeldovich optical Scaling relations} ({Submitted to \aap,
  [arXiv:astro-ph/1101.2027]})

\bibitem[{{Planck Collaboration}(2011{\natexlab{f}})}]{planck2011-7.7b}
{Planck Collaboration}. 2011{\natexlab{f}}, {Planck early results 23: The
  Galactic cold core population revealed by the first all-sky survey}
  ({Submitted to \aap, [arXiv:astro-ph/1101.2035]})

\bibitem[{{Planck Collaboration}(2011{\natexlab{g}})}]{planck2011-1.10sup}
{Planck Collaboration}. 2011{\natexlab{g}}, {The Explanatory Supplement to the
  Planck Early Release Compact Source Catalogue} ({ESA})

\bibitem[{{Planck HFI Core Team}(2011{\natexlab{a}})}]{planck2011-1.5}
{Planck HFI Core Team}. 2011{\natexlab{a}}, {Planck early results 04: First
  assessment of the High Frequency Instrument in-flight performance}
  ({Submitted to \aap, [arXiv:astro-ph/1101.2039]})

\bibitem[{{Planck HFI Core Team}(2011{\natexlab{b}})}]{planck2011-1.7}
{Planck HFI Core Team}. 2011{\natexlab{b}}, {Planck early results 06: The High
  Frequency Instrument data processing} ({Submitted to \aap,
  [arXiv:astro-ph/1101.2048]})

\bibitem[{{Pointecouteau} {et~al.}(1998){Pointecouteau}, {Giard}, \&
  {Barret}}]{poi98}
{Pointecouteau}, E., {Giard}, M., \& {Barret}, D. 1998, \aap, 336, 44

\bibitem[{{Pointecouteau} {et~al.}(1999){Pointecouteau}, {Giard}, {Benoit},
  {D{\'e}sert}, {Aghanim}, {Coron}, {Lamarre}, \& {Delabrouille}}]{poi99}
{Pointecouteau}, E., {Giard}, M., {Benoit}, A., {et~al.} 1999, \apjl, 519, L115

\bibitem[{{Pratt} {et~al.}(2009){Pratt}, {Croston}, {Arnaud}, \&
  {B{\"o}hringer}}]{pra09}
{Pratt}, G.~W., {Croston}, J.~H., {Arnaud}, M., \& {B{\"o}hringer}, H. 2009,
  \aap, 498, 361

\bibitem[{{Reese} {et~al.}(2002){Reese}, {Carlstrom}, {Joy}, {Mohr}, {Grego},
  \& {Holzapfel}}]{ree02}
{Reese}, E.~D., {Carlstrom}, J.~E., {Joy}, M., {et~al.} 2002, \apj, 581, 53

\bibitem[{{Rephaeli}(1995)}]{rep95}
{Rephaeli}, Y. 1995, \araa, 33, 541

\bibitem[{{Rubi{\~n}o-Mart{\'{\i}}n} \& {Sunyaev}(2003)}]{rub03}
{Rubi{\~n}o-Mart{\'{\i}}n}, J.~A. \& {Sunyaev}, R.~A. 2003, \mnras, 344, 1155

\bibitem[{{Sadat} {et~al.}(2004){Sadat}, {Blanchard}, {Kneib}, {Mathez},
  {Madore}, \& {Mazzarella}}]{sad04}
{Sadat}, R., {Blanchard}, A., {Kneib}, J., {et~al.} 2004, \aap, 424, 1097

\bibitem[{{Schneider} {et~al.}(1992){Schneider}, {Bahcall}, {Gunn}, \&
  {Dressler}}]{sch92}
{Schneider}, D.~P., {Bahcall}, J.~N., {Gunn}, J.~E., \& {Dressler}, A. 1992,
  \aj, 103, 1047

\bibitem[{{Shimon} \& {Rephaeli}(2004)}]{shi04}
{Shimon}, M. \& {Rephaeli}, Y. 2004, \na, 9, 69

\bibitem[{{Silk} \& {White}(1978)}]{sil78}
{Silk}, J. \& {White}, S.~D.~M. 1978, \apjl, 226, L103

\bibitem[{{Staniszewski} {et~al.}(2009){Staniszewski}, {Ade}, {Aird}, {Benson},
  {Bleem}, {Carlstrom}, {Chang}, {Cho}, {Crawford}, {Crites}, {de Haan},
  {Dobbs}, {Halverson}, {Holder}, {Holzapfel}, {Hrubes}, {Joy}, {Keisler},
  {Lanting}, {Lee}, {Leitch}, {Loehr}, {Lueker}, {McMahon}, {Mehl}, {Meyer},
  {Mohr}, {Montroy}, {Ngeow}, {Padin}, {Plagge}, {Pryke}, {Reichardt}, {Ruhl},
  {Schaffer}, {Shaw}, {Shirokoff}, {Spieler}, {Stalder}, {Stark},
  {Vanderlinde}, {Vieira}, {Zahn}, \& {Zenteno}}]{sta09}
{Staniszewski}, Z., {Ade}, P.~A.~R., {Aird}, K.~A., {et~al.} 2009, \apj, 701,
  32

\bibitem[{{Story} {et~al.}(2011){Story}, {Aird}, {Andersson}, {Armstrong},
  {Bazin}, {Benson}, {Bleem}, {Bonamente}, {Brodwin}, {Carlstrom}, {Chang},
  {Clocchiatti}, {Crawford}, {Crites}, {de Haan}, {Desai}, {Dobbs}, {Dudley},
  {Foley}, {George}, {Gladders}, {Gonzalez}, {Halverson}, {High}, {Holder},
  {Holzapfel}, {Hoover}, {Hrubes}, {Joy}, {Keisler}, {Knox}, {Lee}, {Leitch},
  {Lueker}, {Luong-Van}, {Marrone}, {McMahon}, {Mehl}, {Meyer}, {Mohr},
  {Montroy}, {Padin}, {Plagge}, {Pryke}, {Reichardt}, {Rest}, {Ruel}, {Ruhl},
  {Saliwanchik}, {Saro}, {Schaffer}, {Shaw}, {Shirokoff}, {Song}, {Spieler},
  {Stalder}, {Staniszewski}, {Stark}, {Stubbs}, {Vanderlinde}, {Vieira},
  {Williamson}, \& {Zenteno}}]{sto11}
{Story}, K., {Aird}, K.~A., {Andersson}, K., {et~al.} 2011, ArXiv e-prints,
  {\tt arXiv:1102.2189}

\bibitem[{{Sunyaev} \& {Zeldovich}(1980)}]{sun80}
{Sunyaev}, R.~A. \& {Zeldovich}, I.~B. 1980, \araa, 18, 537

\bibitem[{Sunyaev \& Zeldovich(1972)}]{sun72}
Sunyaev, R.~A. \& Zeldovich, Y.~B. 1972, Comments on Astrophysics and Space
  Physics, 4, 173

\bibitem[{{Udomprasert} {et~al.}(2004){Udomprasert}, {Mason}, {Readhead}, \&
  {Pearson}}]{udo04}
{Udomprasert}, P.~S., {Mason}, B.~S., {Readhead}, A.~C.~S., \& {Pearson}, T.~J.
  2004, \apj, 615, 63

\bibitem[{{Uzan} {et~al.}(2004){Uzan}, {Aghanim}, \& {Mellier}}]{uza04}
{Uzan}, J., {Aghanim}, N., \& {Mellier}, Y. 2004, \prd, 70, 083533

\bibitem[{{Vanderlinde} {et~al.}(2010){Vanderlinde}, {Crawford}, {de Haan},
  {Dudley}, {Shaw}, {Ade}, {Aird}, {Benson}, {Bleem}, {Brodwin}, {Carlstrom},
  {Chang}, {Crites}, {Desai}, {Dobbs}, {Foley}, {George}, {Gladders}, {Hall},
  {Halverson}, {High}, {Holder}, {Holzapfel}, {Hrubes}, {Joy}, {Keisler},
  {Knox}, {Lee}, {Leitch}, {Loehr}, {Lueker}, {Marrone}, {McMahon}, {Mehl},
  {Meyer}, {Mohr}, {Montroy}, {Ngeow}, {Padin}, {Plagge}, {Pryke}, {Reichardt},
  {Rest}, {Ruel}, {Ruhl}, {Schaffer}, {Shirokoff}, {Song}, {Spieler},
  {Stalder}, {Staniszewski}, {Stark}, {Stubbs}, {van Engelen}, {Vieira},
  {Williamson}, {Yang}, {Zahn}, \& {Zenteno}}]{van10}
{Vanderlinde}, K., {Crawford}, T.~M., {de Haan}, T., {et~al.} 2010, \apj, 722,
  1180

\bibitem[{{Vikhlinin} {et~al.}(2009){Vikhlinin}, {Burenin}, {Ebeling},
  {Forman}, {Hornstrup}, {Jones}, {Kravtsov}, {Murray}, {Nagai}, {Quintana}, \&
  {Voevodkin}}]{vik09}
{Vikhlinin}, A., {Burenin}, R.~A., {Ebeling}, H., {et~al.} 2009, \apj, 692,
  1033

\bibitem[{{Weller} {et~al.}(2002){Weller}, {Battye}, \& {Kneissl}}]{wel02}
{Weller}, J., {Battye}, R.~A., \& {Kneissl}, R. 2002, Physical Review Letters,
  88, 231301

\bibitem[{{Wen} {et~al.}(2009){Wen}, {Han}, \& {Liu}}]{wen09}
{Wen}, Z.~L., {Han}, J.~L., \& {Liu}, F.~S. 2009, \apjs, 183, 197

\bibitem[{{Williamson} {et~al.}(2011){Williamson}, {Benson}, {High},
  {Vanderlinde}, {Ade}, {Aird}, {Andersson}, {Armstrong}, {Ashby}, {Bautz},
  {Bazin}, {Bertin}, {Bleem}, {Bonamente}, {Brodwin}, {Carlstrom}, {Chang},
  {Clocchiatti}, {Crawford}, {Crites}, {de Haan}, {Desai}, {Dobbs}, {Dudley},
  {Fazio}, {Foley}, {Forman}, {Garmire}, {George}, {Gladders}, {Gonzalez},
  {Halverson}, {Holder}, {Holzapfel}, {Hoover}, {Hrubes}, {Jones}, {Joy},
  {Keisler}, {Knox}, {Lee}, {Leitch}, {Lueker}, {Luong-Van}, {Marrone},
  {McMahon}, {Mehl}, {Meyer}, {Mohr}, {Montroy}, {Murray}, {Padin}, {Plagge},
  {Pryke}, {Reichardt}, {Rest}, {Ruel}, {Ruhl}, {Saliwanchik}, {Saro},
  {Schaffer}, {Shaw}, {Shirokoff}, {Song}, {Spieler}, {Stalder}, {Stanford},
  {Staniszewski}, {Stark}, {Story}, {Stubbs}, {Vieira}, {Vikhlinin}, \&
  {Zenteno}}]{wil11}
{Williamson}, R., {Benson}, B.~A., {High}, F.~W., {et~al.} 2011, ArXiv
  e-prints, {\tt arXiv:1101.1290}

\bibitem[{{Wright} {et~al.}(2010){Wright}, {Eisenhardt}, {Mainzer}, {Ressler},
  {Cutri}, {Jarrett}, {Kirkpatrick}, {Padgett}, {McMillan}, {Skrutskie},
  {Stanford}, {Cohen}, {Walker}, {Mather}, {Leisawitz}, {Gautier}, {McLean},
  {Benford}, {Lonsdale}, {Blain}, {Mendez}, {Irace}, {Duval}, {Liu}, {Royer},
  {Heinrichsen}, {Howard}, {Shannon}, {Kendall}, {Walsh}, {Larsen}, {Cardon},
  {Schick}, {Schwalm}, {Abid}, {Fabinsky}, {Naes}, \& {Tsai}}]{wri10}
{Wright}, E.~L., {Eisenhardt}, P.~R.~M., {Mainzer}, A.~K., {et~al.} 2010, \aj,
  140, 1868

\bibitem[{{Wu} {et~al.}(2008){Wu}, {Chiueh}, {Huang}, {Liao}, {Wang},
  {Altimirano}, {Chang}, {Chang}, {Chang}, {Chen}, {Chereau}, {Han}, {Ho},
  {Huang}, {Hwang}, {Jiang}, {Koch}, {Kubo}, {Li}, {Lin}, {Liu},
  {Martin-Cocher}, {Molnar}, {Nishioka}, {Raffin}, {Umetsu}, {Kesteven},
  {Wilson}, {Birkinshaw}, \& {Lancaster}}]{wu08}
{Wu}, J., {Chiueh}, T., {Huang}, C., {et~al.} 2008, Modern Physics Letters A,
  23, 1675

\bibitem[{{Zemcov} {et~al.}(2007){Zemcov}, {Borys}, {Halpern}, {Mauskopf}, \&
  {Scott}}]{zem07}
{Zemcov}, M., {Borys}, C., {Halpern}, M., {Mauskopf}, P., \& {Scott}, D. 2007,
  \mnras, 376, 1073

\bibitem[{{Zemcov} {et~al.}(2010){Zemcov}, {Rex}, {Rawle}, {Bock}, {Egami},
  {Altieri}, {Blain}, {Boone}, {Bridge}, {Clement}, {Combes}, {Dowell},
  {Dessauges-Zavadsky}, {Fadda}, {Ilbert}, {Ivison}, {Jauzac}, {Kneib}, {Lutz},
  {Pell{\'o}}, {Pereira}, {P{\'e}rez-Gonz{\'a}lez}, {Richard}, {Rieke},
  {Rodighiero}, {Schaerer}, {Smith}, {Valtchanov}, {Walth}, {van der Werf}, \&
  {Werner}}]{zem10}
{Zemcov}, M., {Rex}, M., {Rawle}, T.~D., {et~al.} 2010, \aap, 518, L16+

\bibitem[{{Zwart} {et~al.}(2008){Zwart}, {Barker}, {Biddulph}, {Bly}, {Boysen},
  {Brown}, {Clementson}, {Crofts}, {Culverhouse}, {Czeres}, {Dace}, {Davies},
  {D'Alessandro}, {Doherty}, {Duggan}, {Ely}, {Felvus}, {Feroz}, {Flynn},
  {Franzen}, {Geisb{\"u}sch}, {G{\'e}nova-Santos}, {Grainge}, {Grainger},
  {Hammett}, {Hills}, {Hobson}, {Holler}, {Hurley-Walker}, {Jilley}, {Jones},
  {Kaneko}, {Kneissl}, {Lancaster}, {Lasenby}, {Marshall}, {Newton}, {Norris},
  {Northrop}, {Odell}, {Petencin}, {Pober}, {Pooley}, {Pospieszalski}, {Quy},
  {Rodr{\'{\i}}guez-Gonz{\'a}lvez}, {Saunders}, {Scaife}, {Schofield}, {Scott},
  {Shaw}, {Shimwell}, {Smith}, {Taylor}, {Titterington}, {Veli{\'c}},
  {Waldram}, {West}, {Wood}, {Yassin}, \& {AMI Consortium}}]{ami}
{Zwart}, J.~T.~L., {Barker}, R.~W., {Biddulph}, P., {et~al.} 2008, \mnras, 391,
  1545

\bibitem[{{Zwicky} {et~al.}(1961){Zwicky}, {Herzog}, \& {Wild}}]{zwi61}
{Zwicky}, F., {Herzog}, E., \& {Wild}, P. 1961, {Catalogue of galaxies and of
  clusters of galaxies, Vol. I}, ed. {Zwicky, F., Herzog, E., \& Wild, P.}

\end{thebibliography}

\begin{appendix}

\section{ESZ sample extract}

Table \ref{tab:eszfits} is an extract from the \Planck\ ESZ sample
available at {\it www.rssd.esa.int/Planck} aiming at presenting the
content of the released product. Four entries are given as examples
for each category (\Planck\ ESZ known clusters, \Planck\ ESZ new
confirmed clusters, \Planck\ ESZ clusters candidates). In the present
extract, only Galactic longitudes and latitudes are given. The ESZ
sample contains, in addition, the right ascensions and declinations
for all the entries.

For each entry the following fields are provided:
\begin{itemize}
  \item Name: \Planck\ Name of Cluster Candidate
  \item GLON: Galactic Longitude from \Planck\
  \item GLAT: Galactic latitude from \Planck\
  \item S/N: Signal-to-noise ratio returned by the matched
    multi-Filter algorithm (MMF3)
  \item ID: External Identifier of \Planck\ Clusters e.g. Coma, Abell
    2163 etc 
  \item z: Redshift of Cluster from the MCXC X-ray cluster compilation
    unless otherwise stated in the individual notes
  \item $\Theta_X$: Angular size at $5R500$ from X-ray data
  \item $Y_{PSX}$: Integrated Compton parameter at X-ray position and within
    $5R500$ ($\Theta_X$) in arcmin$^2$,
  \item $Y_{PSX}^{ERR}$: Uncertainty in Integrated Compton parameter at X-ray
    position and within $5R500$ ($\Theta_X$) in arcmin$^2$  
  \item $\Theta$: Estimated angular size from matched multi-Filter (MMF3),
  \item $Y$: Integrated Compton parameter at \Planck\ position and
    within $\Theta$, from matched multi-Filter (MMF3)  in arcmin$^2$
  \item $Y^{ERR}$: Uncertainty in Integrated Compton parameter at
    \Planck\ position and within $\Theta$ from matched multi-Filter
    (MMF3) in arcmin$^2$
\end{itemize}

 \begin{table*}[!h]
\caption{ESZ sample. }\label{tab:eszfits}
\begin{tabular}{|l|r|r|r|r|r|r|r|r|r|r|r|r|r|r|r|r|r|r}
\hline
\tiny
Name & GLON & GLAT & S/N & ID & z & $\Theta_X$ & $Y_{PSX}$ & $Y_{PSX}^{ERR}$ & $\Theta$ & $Y$ & $Y^{ERR}$ \\ \hline
PLCKG111.0+31.7 & 110.98 & 31.73 & 28.93 & A2256 & 0.06 & 83.14 & 0.0242 & 0.0009 & NaN & NaN & NaN \\
PLCKG57.3+88.0 & 57.34 & 88.01 & 21.94 & Coma & 0.02 & 203.37 & 0.1173 & 0.0054 & NaN & NaN & NaN \\
PLCKG239.3+24.8 & 239.28 & 24.77 & 25.67 & A0754 & 0.05 & 90.56 & 0.0330 & 0.0012 & NaN & NaN & NaN \\
PLCKG272.1-40.2 & 272.11 & -40.15 & 25.90 & A3266 & 0.06 & 84.19 & 0.0282 & 0.0012 & NaN & NaN & NaN \\
PLCKG6.8+30.5 & 6.78 & 30.47 & 26.40 & A2163 & 0.20 & 37.81 & 0.0173 & 0.0007 & NaN & NaN & NaN \\
PLCKG340.9-33.3 & 340.89 & -33.35 & 22.02 & A3667 & 0.06 & 93.11 & 0.0266 & 0.0014 & NaN & NaN & NaN \\
PLCKG266.0-21.3 & 266.04 & -21.25 & 19.75 & 1ES 0657-55.8 & 0.30 & 26.96 & 0.0067 & 0.0003 & NaN & NaN & NaN \\
PLCKG44.2+48.7 & 44.23 & 48.68 & 18.46 & A2142 & 0.09 & 70.33 & 0.0241 & 0.0013 & NaN & NaN & NaN \\
PLCKG93.9+34.9 & 93.92 & 34.91 & 17.31 & A2255 & 0.08 & 58.11 & 0.0103 & 0.0006 & NaN & NaN & NaN \\
PLCKG164.2-38.9 & 164.19 & -38.89 & 13.79 & A0401 & 0.07 & 74.42 & 0.0193 & 0.0016 & NaN & NaN & NaN \\
PLCKG72.6+41.5 & 72.63 & 41.46 & 17.44 & A2219 & 0.23 & 31.40 & 0.0085 & 0.0005 & NaN & NaN & NaN \\
PLCKG263.7-22.5 & 263.67 & -22.54 & 16.70 & A3404 & 0.16 & 35.89 & 0.0064 & 0.0004 & NaN & NaN & NaN \\
PLCKG97.7+38.1 & 97.74 & 38.12 & 14.65 & A2218 & 0.17 & 31.96 & 0.0044 & 0.0003 & NaN & NaN & NaN \\
PLCKG263.2-25.2 & 263.21 & -25.21 & 11.24 & A3395 & 0.05 & 76.98 & 0.0073 & 0.0009 & NaN & NaN & NaN \\
PLCKG262.3-35.4 & 262.25 & -35.37 & 15.19 & ACO S0520 & 0.30 & 24.30 & 0.0034 & 0.0003 & NaN & NaN & NaN \\
PLCKG74.0-27.8 & 73.97 & -27.82 & 14.25 & A2390 & 0.23 & 31.17 & 0.0056 & 0.0005 & NaN & NaN & NaN \\
PLCKG332.2-46.4 & 332.23 & -46.37 & 13.89 & A3827 & 0.10 & 52.41 & 0.0086 & 0.0007 & NaN & NaN & NaN \\
PLCKG265.0-48.9 & 265.01 & -48.95 & 13.95 & A3158 & 0.06 & 77.52 & 0.0117 & 0.0010 & NaN & NaN & NaN \\
PLCKG115.2-72.1 & 115.16 & -72.09 & 13.14 & A0085 & 0.06 & 94.22 & 0.0210 & 0.0018 & NaN & NaN & NaN \\
PLCKG316.3+28.5 & 316.35 & 28.54 & 12.85 & A3571 & 0.04 & 124.14 & 0.0372 & 0.0031 & NaN & NaN & NaN \\
PLCKG86.5+15.3 & 86.46 & 15.30 & 12.33 & CIZA J1938.3+5409 & 0.26 & 24.59 & 0.0031 & 0.0003 & NaN & NaN & NaN \\
PLCKG33.8+77.2 & 33.78 & 77.16 & 12.39 & A1795 & 0.06 & 85.79 & 0.0169 & 0.0014 & NaN & NaN & NaN \\
PLCKG6.5+50.5 & 6.48 & 50.55 & 13.36 & A2029 & 0.08 & 77.99 & 0.0180 & 0.0015 & NaN & NaN & NaN \\
PLCKG349.5-59.9 & 349.46 & -59.95 & 13.93 & ACO S1063 & 0.35 & 24.94 & 0.0046 & 0.0003 & NaN & NaN & NaN \\
PLCKG186.4+37.3 & 186.39 & 37.26 & 12.61 & A0697 & 0.28 & 25.00 & 0.0051 & 0.0005 & NaN & NaN & NaN \\
PLCKG229.9+15.3 & 229.94 & 15.30 & 12.46 & A0644 & 0.07 & 67.06 & 0.0116 & 0.0010 & NaN & NaN & NaN \\
PLCKG149.7+34.7 & 149.73 & 34.70 & 11.57 & A0665 & 0.18 & 34.94 & 0.0060 & 0.0005 & NaN & NaN & NaN \\
PLCKG3.9-59.4 & 3.91 & -59.42 & 12.06 & A3888 & 0.15 & 38.39 & 0.0061 & 0.0005 & NaN & NaN & NaN \\
PLCKG312.0+30.7 & 312.00 & 30.72 & 9.04 & A3558 & 0.05 & 97.34 & 0.0223 & 0.0024 & NaN & NaN & NaN \\
PLCKG313.9-17.1 & 313.87 & -17.11 & 11.57 & CIZA J1601.7-7544 & 0.15 & 40.63 & 0.0078 & 0.0007 & NaN & NaN & NaN \\
PLCKG335.6-46.5 & 335.59 & -46.46 & 10.17 & A3822 & 0.08 & 57.18 & 0.0084 & 0.0008 & NaN & NaN & NaN \\
PLCKG288.6-37.7 & 288.62 & -37.66 & 9.86 & A3186 & 0.13 & 35.86 & 0.0053 & 0.0006 & NaN & NaN & NaN \\
PLCKG315.7-18.0 & 315.71 & -18.04 & 11.44 & A3628 & 0.10 & 48.10 & 0.0088 & 0.0008 & NaN & NaN & NaN \\
PLCKG263.2-23.4 & 263.16 & -23.41 & 10.08 & ACO S0592 & 0.23 & 28.97 & 0.0032 & 0.0003 & NaN & NaN & NaN \\
PLCKG149.2+54.2 & 149.24 & 54.19 & 11.58 & A1132 & 0.14 & 36.78 & 0.0052 & 0.0005 & NaN & NaN & NaN \\
PLCKG21.1+33.3 & 21.09 & 33.26 & 10.61 & A2204 & 0.15 & 45.35 & 0.0076 & 0.0007 & NaN & NaN & NaN \\
PLCKG322.0-48.0 & 321.96 & -47.98 & 11.27 & A3921 & 0.09 & 49.92 & 0.0053 & 0.0006 & NaN & NaN & NaN \\
PLCKG182.4-28.3 & 182.44 & -28.30 & 12.77 & A0478 & 0.09 & 65.31 & 0.0167 & 0.0014 & NaN & NaN & NaN \\
PLCKG242.0+14.9 & 241.97 & 14.86 & 10.49 & A3411 & 0.17 & 31.37 & 0.0041 & 0.0005 & NaN & NaN & NaN \\
PLCKG29.0+44.6 & 29.01 & 44.56 & 10.25 & A2147 & 0.04 & 108.90 & 0.0148 & 0.0021 & NaN & NaN & NaN \\
PLCKG228.5+53.1 & 228.50 & 53.13 & 12.20 & Zw 3179 & 0.14 & 33.59 & 0.0022 & 0.0005 & NaN & NaN & NaN \\
PLCKG62.9+43.7 & 62.93 & 43.71 & 10.03 & A2199 & 0.03 & 138.17 & 0.0241 & 0.0023 & NaN & NaN & NaN \\
PLCKG206.0-39.5 & 205.96 & -39.48 & 9.26 & MACS J0417.5-1154 & 0.44 & 20.43 & 0.0038 & 0.0004 & NaN & NaN & NaN \\
PLCKG336.6-55.4 & 336.59 & -55.45 & 10.29 & A3911 & 0.10 & 47.65 & 0.0057 & 0.0006 & NaN & NaN & NaN \\
PLCKG67.2+67.5 & 67.23 & 67.46 & 11.03 & A1914 & 0.17 & 37.08 & 0.0057 & 0.0005 & NaN & NaN & NaN \\
PLCKG92.7+73.5 & 92.73 & 73.46 & 11.26 & A1763 & 0.23 & 27.79 & 0.0045 & 0.0004 & NaN & NaN & NaN \\
PLCKG146.3-15.6 & 146.33 & -15.59 & 7.10 & CIZA J0254.4+4134 & 0.02 & 199.50 & 0.0392 & 0.0060 & NaN & NaN & NaN \\
PLCKG112.5+57.0 & 112.46 & 57.04 & 9.81 & A1767 & 0.07 & 56.20 & 0.0053 & 0.0006 & NaN & NaN & NaN \\
PLCKG55.6+31.9 & 55.60 & 31.86 & 9.27 & A2261 & 0.22 & 30.74 & 0.0049 & 0.0005 & NaN & NaN & NaN \\
PLCKG58.3+18.6 & 58.28 & 18.59 & 9.19 & CIZA J1825.3+3026 & 0.06 & 62.65 & 0.0087 & 0.0009 & NaN & NaN & NaN \\
PLCKG159.9-73.5 & 159.86 & -73.47 & 10.63 & A0209 & 0.21 & 28.17 & 0.0053 & 0.0005 & NaN & NaN & NaN \\
PLCKG282.5+65.2 & 282.49 & 65.17 & 8.49 & ZwCl 1215.1+0400 & 0.08 & 60.24 & 0.0095 & 0.0012 & NaN & NaN & NaN \\
PLCKG313.4+61.1 & 313.36 & 61.12 & 10.12 & A1689 & 0.18 & 37.34 & 0.0071 & 0.0008 & NaN & NaN & NaN \\
PLCKG53.5+59.5 & 53.52 & 59.54 & 8.50 & A2034 & 0.11 & 44.04 & 0.0055 & 0.0008 & NaN & NaN & NaN \\
PLCKG244.3-32.1 & 244.34 & -32.14 & 8.39 & RBS0653 & 0.28 & 24.88 & 0.0029 & 0.0004 & NaN & NaN & NaN \\
PLCKG46.9+56.5 & 46.88 & 56.50 & 9.07 & A2069 & 0.11 & 45.41 & 0.0067 & 0.0008 & NaN & NaN & NaN \\
PLCKG294.7-37.0 & 294.67 & -37.03 & 8.64 & RXCJ0303.7-7752 & 0.27 & 22.29 & 0.0028 & 0.0004 & NaN & NaN & NaN \\
PLCKG346.6+35.0 & 346.60 & 35.05 & 9.38 & RXCJ1514.9-1523 & 0.22 & 26.97 & 0.0048 & 0.0006 & NaN & NaN & NaN \\
PLCKG243.6+67.8 & 243.57 & 67.76 & 8.57 & A1307 & 0.08 & 52.20 & 0.0062 & 0.0007 & NaN & NaN & NaN \\
PLCKG166.1+43.4 & 166.13 & 43.39 & 9.23 & A0773 & 0.22 & 28.20 & 0.0038 & 0.0004 & NaN & NaN & NaN \\
PLCKG226.2+76.8 & 226.25 & 76.77 & 9.18 & A1413 & 0.14 & 39.97 & 0.0058 & 0.0006 & NaN & NaN & NaN \\
PLCKG107.1+65.3 & 107.11 & 65.31 & 8.85 & A1758A & 0.28 & 24.73 & 0.0031 & 0.0004 & NaN & NaN & NaN \\
PLCKG42.8+56.6 & 42.83 & 56.62 & 8.36 & A2065 & 0.07 & 63.04 & 0.0099 & 0.0011 & NaN & NaN & NaN \\
PLCKG125.6-64.1 & 125.59 & -64.14 & 10.47 & A0119 & 0.04 & 88.65 & 0.0141 & 0.0017 & NaN & NaN & NaN \\
PLCKG57.3-45.4 & 57.27 & -45.36 & 8.11 & MACS J2211.7-0349 & 0.40 & 21.56 & 0.0032 & 0.0004 & NaN & NaN & NaN\\
PLCKG33.5-48.4 & 33.46 & -48.43 & 9.24 & A2384A & 0.09 & 44.21 & 0.0054 & 0.0006 & NaN & NaN & NaN \\
\hline
\hline
\end{tabular}
\end{table*}

 \begin{table*}[!h]
\caption{The ESZ sample (continued) }\label{tab:eszfits}
\begin{tabular}{|l|r|r|r|r|r|r|r|r|r|r|r|r|r|r|r|r|}
\hline
\tiny
Name & GLON & GLAT & S/N & ID & z & $\Theta_X$ & $Y_{PSX}$ & $Y_{PSX}^{ERR}$ & $\Theta$ & $Y$ & $Y^{ERR}$ \\ \hline
PLCKG241.8-24.0 & 241.78 & -24.00 & 8.94 & A3378 & 0.14 & 37.79 & 0.0038 & 0.0005 & NaN & NaN & NaN \\
PLCKG46.5-49.4 & 46.50 & -49.44 & 8.55 & A2420 & 0.08 & 54.47 & 0.0064 & 0.0008 & NaN & NaN & NaN \\
PLCKG304.9+45.5 & 304.90 & 45.45 & 8.99 & A1644 & 0.05 & 88.28 & 0.0152 & 0.0018 & NaN & NaN & NaN \\
PLCKG209.6-36.5 & 209.56 & -36.49 & 7.96 & A0496 & 0.03 & 126.18 & 0.0162 & 0.0021 & NaN & NaN & NaN \\
PLCKG57.0-55.1 & 56.97 & -55.08 & 8.16 & MACS J2243.3-0935 & 0.45 & 17.55 & 0.0029 & 0.0004 & NaN & NaN & NaN \\
PLCKG56.8+36.3 & 56.81 & 36.32 & 9.15 & A2244 & 0.10 & 53.36 & 0.0058 & 0.0007 & NaN & NaN & NaN \\
PLCKG57.6+34.9 & 57.61 & 34.94 & 9.54 & A2249 & 0.08 & 52.39 & 0.0052 & 0.0007 & NaN & NaN & NaN \\
PLCKG49.2+30.9 & 49.20 & 30.86 & 8.33 & RXC J1720.1+2637 & 0.16 & 36.32 & 0.0043 & 0.0005 & NaN & NaN & NaN \\
PLCKG6.7-35.5 & 6.70 & -35.54 & 8.45 & A3695 & 0.09 & 52.90 & 0.0059 & 0.0008 & NaN & NaN & NaN \\
PLCKG77.9-26.6 & 77.91 & -26.65 & 8.36 & A2409 & 0.15 & 36.46 & 0.0040 & 0.0005 & NaN & NaN & NaN \\
PLCKG8.9-81.2 & 8.94 & -81.24 & 8.39 & A2744 & 0.31 & 23.17 & 0.0042 & 0.0005 & NaN & NaN & NaN \\
PLCKG106.7-83.2 & 106.73 & -83.23 & 8.55 & A2813 & 0.29 & 21.45 & 0.0036 & 0.0004 & NaN & NaN & NaN \\
PLCKG269.5+26.4 & 269.52 & 26.42 & 8.40 & A1060 & 0.01 & 216.51 & 0.0215 & 0.0029 & NaN & NaN & NaN \\
PLCKG180.2+21.0 & 180.24 & 21.05 & 8.36 & MACS J0717.5+3745 & 0.55 & 17.74 & 0.0028 & 0.0004 & NaN & NaN & NaN \\
PLCKG241.7-30.9 & 241.74 & -30.89 & 7.42 & RXCJ0532.9-3701 & 0.27 & 22.43 & 0.0028 & 0.0004 & NaN & NaN & NaN \\
PLCKG332.9-19.3 & 332.89 & -19.28 & 7.72 & CIZA J1813.3-6127 & 0.15 & 35.99 & 0.0043 & 0.0006 & NaN & NaN & NaN \\
PLCKG48.1+57.2 & 48.05 & 57.18 & 7.14 & A2061 & 0.08 & 54.65 & 0.0067 & 0.0010 & NaN & NaN & NaN \\
PLCKG139.2+56.4 & 139.20 & 56.36 & 7.65 & A1351 & 0.32 & 18.52 & 0.0012 & 0.0003 & NaN & NaN & NaN \\
PLCKG306.7+61.1 & 306.68 & 61.06 & 8.02 & A1650 & 0.08 & 57.80 & 0.0095 & 0.0012 & NaN & NaN & NaN \\
PLCKG167.7+17.6 & 167.66 & 17.65 & 8.11 & ZwCl 0634.1+4750 & 0.17 & 31.51 & 0.0045 & 0.0005 & NaN & NaN & NaN \\
PLCKG49.3+44.4 & 49.34 & 44.38 & 7.40 & A2175 & 0.10 & 42.77 & 0.0054 & 0.0009 & NaN & NaN & NaN \\
PLCKG226.2-21.9 & 226.18 & -21.91 & 7.28 & A0550 & 0.10 & 45.53 & 0.0047 & 0.0007 & NaN & NaN & NaN \\
PLCKG195.8-24.3 & 195.77 & -24.31 & 7.23 & A0520 & 0.20 & 31.04 & 0.0046 & 0.0006 & NaN & NaN & NaN \\
PLCKG253.5-33.7 & 253.48 & -33.72 & 6.73 & A3343 & 0.19 & 26.95 & 0.0022 & 0.0004 & NaN & NaN & NaN \\
PLCKG250.9-36.3 & 250.91 & -36.26 & 8.62 & A3322 & 0.20 & 27.92 & 0.0028 & 0.0004 & NaN & NaN & NaN \\
PLCKG256.5-65.7 & 256.45 & -65.71 & 7.77 & A3016 & 0.22 & 26.97 & 0.0029 & 0.0004 & NaN & NaN & NaN \\
PLCKG324.5-45.0 & 324.50 & -44.97 & 6.22 & RBS1847 & 0.10 & 44.97 & 0.0039 & 0.0005 & NaN & NaN & NaN \\
PLCKG113.8+44.4 & 113.82 & 44.35 & 7.80 & A1895 & 0.22 & 23.20 & 0.0012 & 0.0002 & NaN & NaN & NaN \\
PLCKG125.7+53.9 & 125.71 & 53.86 & 7.36 & A1576 & 0.30 & 20.94 & 0.0019 & 0.0003 & NaN & NaN & NaN \\
PLCKG266.8+25.1 & 266.84 & 25.08 & 8.19 & A3444 & 0.25 & 27.46 & 0.0027 & 0.0004 & NaN & NaN & NaN \\
PLCKG216.6+47.0 & 216.62 & 47.02 & 7.48 & RXC J0949.8+1707 & 0.38 & 19.99 & 0.0021 & 0.0004 & NaN & NaN & NaN \\
PLCKG228.2+75.2 & 228.16 & 75.19 & 7.13 & MACS J1149.5+2223 & 0.55 & 15.93 & 0.0016 & 0.0003 & NaN & NaN & NaN \\
PLCKG342.3-34.9 & 342.32 & -34.91 & 7.24 & RXCJ2023.4-5535 & 0.23 & 24.13 & 0.0029 & 0.0004 & NaN & NaN & NaN \\
PLCKG342.8-30.5 & 342.82 & -30.46 & 6.01 & A3651 & 0.06 & 54.80 & 0.0044 & 0.0009 & NaN & NaN & NaN \\
PLCKG124.2-36.5 & 124.22 & -36.49 & 7.74 & A0115 & 0.20 & 31.15 & 0.0050 & 0.0007 & NaN & NaN & NaN \\
PLCKG257.3-22.2 & 257.34 & -22.18 & 7.13 & A3399 & 0.20 & 25.71 & 0.0019 & 0.0003 & NaN & NaN & NaN \\
PLCKG118.4+39.3 & 118.45 & 39.34 & 6.33 & RXCJ1354.6+7715 & 0.40 & 17.24 & 0.0016 & 0.0003 & NaN & NaN & NaN \\
PLCKG118.6+28.6 & 118.60 & 28.56 & 6.41 & A2294 & 0.18 & 29.83 & 0.0022 & 0.0004 & NaN & NaN & NaN \\
PLCKG229.6+78.0 & 229.64 & 77.96 & 7.45 & A1443 & 0.27 & 21.40 & 0.0027 & 0.0004 & NaN & NaN & NaN \\
PLCKG180.6+76.7 & 180.62 & 76.65 & 7.48 & A1423 & 0.21 & 25.40 & 0.0027 & 0.0004 & NaN & NaN & NaN \\
PLCKG2.7-56.2 & 2.75 & -56.18 & 6.48 & A3856 & 0.14 & 35.56 & 0.0031 & 0.0005 & NaN & NaN & NaN \\
PLCKG347.2-27.4 & 347.19 & -27.35 & 8.19 & ACO S0821 & 0.24 & 24.78 & 0.0022 & 0.0004 & NaN & NaN & NaN \\
PLCKG71.6+29.8 & 71.61 & 29.80 & 7.47 & Zw 8284 & 0.16 & 27.18 & 0.0024 & 0.0004 & NaN & NaN & NaN \\
PLCKG36.7+14.9 & 36.72 & 14.92 & 6.98 & RXCJ1804.4+1002 & 0.15 & 35.28 & 0.0035 & 0.0006 & NaN & NaN & NaN \\
PLCKG18.5-25.7 & 18.53 & -25.72 & 7.30 & RXCJ2003.5-2323 & 0.32 & 20.71 & 0.0027 & 0.0004 & NaN & NaN & NaN \\
PLCKG237.0-26.7 & 236.96 & -26.67 & 7.03 & A3364 & 0.15 & 35.80 & 0.0030 & 0.0005 & NaN & NaN & NaN \\
PLCKG273.6+63.3 & 273.64 & 63.28 & 7.30 & A1437 & 0.13 & 37.96 & 0.0051 & 0.0007 & NaN & NaN & NaN \\
PLCKG46.1+27.2 & 46.08 & 27.18 & 7.34 & MACS J1731.6+2252 & 0.39 & 17.79 & 0.0021 & 0.0003 & NaN & NaN & NaN \\
PLCKG49.7-49.5 & 49.67 & -49.51 & 6.88 & A2426 & 0.10 & 47.77 & 0.0038 & 0.0007 & NaN & NaN & NaN \\
PLCKG143.2+65.2 & 143.25 & 65.22 & 7.34 & A1430 & 0.21 & 24.27 & 0.0023 & 0.0003 & NaN & NaN & NaN \\
PLCKG296.4-32.5 & 296.41 & -32.49 & 7.20 & ACO S0405 & 0.06 & 62.06 & 0.0044 & 0.0007 & NaN & NaN & NaN \\
PLCKG269.3-49.9 & 269.31 & -49.88 & 6.51 & A3126 & 0.09 & 47.30 & 0.0040 & 0.0007 & NaN & NaN & NaN \\
PLCKG83.3-31.0 & 83.29 & -31.03 & 6.19 & RXC J2228.6+2036 & 0.41 & 19.89 & 0.0021 & 0.0004 & NaN & NaN & NaN \\
PLCKG304.7-31.7 & 304.67 & -31.67 & 6.37 & A4023 & 0.19 & 25.68 & 0.0020 & 0.0004 & NaN & NaN & NaN \\
PLCKG39.9-40.0 & 39.86 & -39.99 & 6.32 & A2345 & 0.18 & 29.89 & 0.0031 & 0.0005 & NaN & NaN & NaN \\
PLCKG56.0-34.9 & 55.98 & -34.89 & 7.03 & A2355 & 0.12 & 33.85 & 0.0036 & 0.0005 & NaN & NaN & NaN \\
PLCKG303.8+33.7 & 303.76 & 33.66 & 6.05 & A3528S & 0.05 & 68.17 & 0.0085 & 0.0014 & NaN & NaN & NaN \\
PLCKG163.7+53.5 & 163.72 & 53.53 & 7.46 & A0980 & 0.16 & 32.07 & 0.0030 & 0.0004 & NaN & NaN & NaN \\
PLCKG318.1-29.6 & 318.13 & -29.58 & 6.63 & RXCJ1947.3-7623 & 0.22 & 25.55 & 0.0031 & 0.0005 & NaN & NaN & NaN \\
PLCKG244.7+32.5 & 244.70 & 32.49 & 6.27 & A0868 & 0.15 & 31.25 & 0.0029 & 0.0005 & NaN & NaN & NaN \\
PLCKG284.5+52.4 & 284.46 & 52.44 & 7.27 & RXCJ1206.2-0848 & 0.44 & 18.33 & 0.0029 & 0.0004 & NaN & NaN & NaN \\
PLCKG260.0-63.4 & 260.03 & -63.44 & 7.29 & RXCJ0232.2-4420 & 0.28 & 23.09 & 0.0024 & 0.0004 & NaN & NaN & NaN \\
PLCKG253.0-56.1 & 252.97 & -56.05 & 6.79 & A3112 & 0.08 & 65.98 & 0.0047 & 0.0007 & NaN & NaN & NaN \\
PLCKG234.6+73.0 & 234.59 & 73.02 & 6.39 & A1367 & 0.02 & 169.59 & 0.0146 & 0.0029 & NaN & NaN & NaN \\
PLCKG278.6+39.2 & 278.61 & 39.17 & 7.57 & A1300 & 0.31 & 23.38 & 0.0035 & 0.0005 & NaN & NaN & NaN \\
PLCKG246.5-26.1 & 246.52 & -26.06 & 6.52 & A3376 & 0.05 & 77.30 & 0.0053 & 0.0010 & NaN & NaN & NaN \\
PLCKG114.3+64.9 & 114.34 & 64.87 & 6.18 & A1703 & 0.28 & 20.89 & 0.0020 & 0.0003 & NaN & NaN & NaN \\
\hline
\hline
\end{tabular}
\end{table*}

 \begin{table*}[!h]
\caption{The ESZ sample (continued) }\label{tab:eszfits}
\begin{tabular}{|l|r|r|r|r|r|r|r|r|r|r|r|r|r|r|r|r|}
\hline
\tiny
Name & GLON & GLAT & S/N & ID & z & $\Theta_X$ & $Y_{PSX}$ & $Y_{PSX}^{ERR}$ & $\Theta$ & $Y$ & $Y^{ERR}$ \\ \hline
PLCKG80.4-33.2 & 80.38 & -33.20 & 6.06 & A2443 & 0.11 & 39.40 & 0.0039 & 0.0006 & NaN & NaN & NaN \\
PLCKG249.9-39.9 & 249.88 & -39.87 & 6.25 & A3292 & 0.15 & 31.40 & 0.0018 & 0.0004 & NaN & NaN & NaN \\
PLCKG182.6+55.8 & 182.64 & 55.82 & 6.81 & A0963 & 0.21 & 27.40 & 0.0019 & 0.0004 & NaN & NaN & NaN \\
PLCKG62.4-46.4 & 62.42 & -46.41 & 6.33 & A2440 & 0.09 & 47.33 & 0.0041 & 0.0007 & NaN & NaN & NaN \\
PLCKG8.4-56.4 & 8.45 & -56.36 & 6.39 & A3854 & 0.15 & 33.94 & 0.0024 & 0.0005 & NaN & NaN & NaN \\
PLCKG229.2-17.2 & 229.22 & -17.25 & 6.18 & RXCJ0616.3-2156 & 0.17 & 28.31 & 0.0031 & 0.0005 & NaN & NaN & NaN \\
PLCKG341.0+35.1 & 340.96 & 35.12 & 6.61 & ACO S0780 & 0.24 & 30.43 & 0.0030 & 0.0007 & NaN & NaN & NaN \\
PLCKG218.9+35.5 & 218.86 & 35.51 & 6.87 & A0750 & 0.18 & 31.52 & 0.0027 & 0.0005 & NaN & NaN & NaN \\
PLCKG165.1+54.1 & 165.09 & 54.12 & 6.34 & A0990 & 0.14 & 35.91 & 0.0027 & 0.0005 & NaN & NaN & NaN \\
PLCKG161.4+26.2 & 161.44 & 26.23 & 6.63 & A0576 & 0.04 & 88.79 & 0.0076 & 0.0012 & NaN & NaN & NaN \\
PLCKG295.3+23.3 & 295.33 & 23.34 & 6.11 & RXCJ1215.4-3900 & 0.12 & 36.31 & 0.0042 & 0.0008 & NaN & NaN & NaN \\
PLCKG280.2+47.8 & 280.20 & 47.82 & 7.06 & A1391 & 0.16 & 30.98 & 0.0042 & 0.0006 & NaN & NaN & NaN \\
PLCKG0.4-41.8 & 0.44 & -41.84 & 6.55 & A3739 & 0.17 & 31.54 & 0.0025 & 0.0005 & NaN & NaN & NaN \\
PLCKG195.6+44.1 & 195.62 & 44.05 & 6.88 & A0781 & 0.30 & 19.18 & 0.0017 & 0.0003 & NaN & NaN & NaN \\
PLCKG241.9+51.5 & 241.86 & 51.53 & 6.96 & A1066 & 0.07 & 48.70 & 0.0024 & 0.0007 & NaN & NaN & NaN \\
PLCKG81.0-50.9 & 81.00 & -50.91 & 6.76 & A2552 & 0.30 & 22.98 & 0.0026 & 0.0005 & NaN & NaN & NaN \\
PLCKG304.5+32.4 & 304.50 & 32.44 & 6.86 & A3532 & 0.06 & 69.91 & 0.0068 & 0.0015 & NaN & NaN & NaN \\
PLCKG306.8+58.6 & 306.80 & 58.61 & 6.81 & A1651 & 0.08 & 59.16 & 0.0077 & 0.0012 & NaN & NaN & NaN \\
PLCKG172.9+65.3 & 172.89 & 65.32 & 6.30 & A1190 & 0.08 & 47.15 & 0.0030 & 0.0006 & NaN & NaN & NaN \\
PLCKG99.0+24.9 & 98.95 & 24.86 & 6.49 & A2312 & 0.09 & 41.02 & 0.0022 & 0.0004 & NaN & NaN & NaN \\
PLCKG247.2-23.3 & 247.17 & -23.33 & 6.19 & ACO S0579 & 0.15 & 31.86 & 0.0019 & 0.0004 & NaN & NaN & NaN \\
PLCKG176.3-35.1 & 176.28 & -35.05 & 6.38 & 2A0335+096 & 0.03 & 126.35 & 0.0117 & 0.0025 & NaN & NaN & NaN \\
PLCKG57.9+27.6 & 57.93 & 27.64 & 6.13 & ZwCl 1742.1+3306 & 0.08 & 58.65 & 0.0037 & 0.0008 & NaN & NaN & NaN \\
PLCKG275.2+43.9 & 275.22 & 43.92 & 6.29 & A1285 & 0.11 & 43.45 & 0.0044 & 0.0008 & NaN & NaN & NaN \\
PLCKG96.9+52.5 & 96.85 & 52.47 & 6.12 & A1995 & 0.32 & 20.73 & 0.0015 & 0.0003 & NaN & NaN & NaN \\
PLCKG72.8-18.7 & 72.80 & -18.72 & 10.10 & ZwCl2120.1+2256 & 0.14 & NaN  & NaN & NaN & 37.22 & 0.0052 & 0.0010 \\
PLCKG239.3-26.0 & 239.29 & -26.00 & 8.64 & MACS J0553.4-3342 & 0.41 & NaN  & NaN & NaN & 17.22 & 0.0026 & 0.0006 \\
PLCKG8.3-64.8 & 8.30 & -64.76 & 8.47 & AC114Northern & 0.31 & NaN  & NaN & NaN & 43.83 & 0.0048 & 0.0010 \\
PLCKG94.0+27.4 & 94.02 & 27.43 & 6.92 & H1821+643 & 0.30 & NaN  & NaN & NaN & 40.25 & 0.0030 & 0.0014 \\
PLCKG157.4+30.3 & 157.43 & 30.34 & 6.18 & RXJ0748.7+5941 & NaN & NaN  & NaN & NaN & 22.43 & 0.0025 & 0.0014 \\
PLCKG345.4-39.3 & 345.41 & -39.34 & 7.10 & ABELL3716S & 0.04 & NaN  & NaN & NaN & 118.59 & 0.0109 & 0.0032 \\
PLCKG53.4-36.3 & 53.44 & -36.27 & 6.88 & MACS J2135.2-0102 & 0.32 & NaN  & NaN & NaN & 8.07 & 0.0018 & 0.0003 \\
PLCKG271.5-56.6 & 271.50 & -56.56 & 6.71 & ACO S0295 & 0.30 & NaN  & NaN & NaN & 20.26 & 0.0025 & 0.0007 \\
PLCKG86.0+26.7 & 86.00 & 26.71 & 6.55 & A2302 & 0.18 & NaN  & NaN & NaN & 56.62 & 0.0043 & 0.0019 \\
PLCKG96.9+24.2 & 96.88 & 24.22 & 6.24 & ZwCl1856.8+6616 & NaN & NaN  & NaN & NaN & 20.64 & 0.0015 & 0.0005 \\
PLCKG164.6+46.4 & 164.61 & 46.39 & 6.06 & ZwCl0934.8+5216 & NaN & NaN  & NaN & NaN & 16.50 & 0.0018 & 0.0006 \\
PLCKG285.0-23.7 & 284.99 & -23.71 & 11.48 & null & 0.44 & 19.50 & 0.0023 & 0.0002 & NaN & NaN & NaN \\
PLCKG287.0+32.9 & 286.99 & 32.92 & 10.62 & null & 0.39 & 25.33 & 0.0061 & 0.0006 & NaN & NaN & NaN \\
PLCKG171.9-40.7 & 171.95 & -40.66 & 10.61 & null & 0.39 & 29.56 & 0.0062 & 0.0006 & NaN & NaN & NaN \\
PLCKG271.2-31.0 & 271.20 & -30.97 & 8.48 & null & 0.27 & 20.06 & 0.0020 & 0.0002 & NaN & NaN & NaN \\
PLCKG262.7-40.9 & 262.71 & -40.91 & 8.27 & ACT-CLJ0438-5419 & 0.37 & 18.71 & 0.0021 & 0.0002 & NaN & NaN & NaN \\
PLCKG308.3-20.2 & 308.32 & -20.23 & 8.26 & null & 0.39 & NaN  & NaN & NaN & 32.81 & 0.0049 & 0.0013 \\
PLCKG277.8-51.7 & 277.75 & -51.73 & 7.40 & null & 0.21 & 17.51 & 0.0027 & 0.0003 & NaN & NaN & NaN \\
PLCKG286.6-31.3 & 286.59 & -31.25 & 6.89 & null & 0.30 & 28.12 & 0.0026 & 0.0004 & NaN & NaN & NaN \\
PLCKG292.5+22.0 & 292.52 & 21.99 & 6.88 & null & 0.35 & 25.66 & 0.0037 & 0.0006 & NaN & NaN & NaN \\
PLCKG337.1-26.0 & 337.09 & -25.97 & 6.59 & null & 0.48 & NaN  & NaN & NaN & 31.56 & 0.0034 & 0.0008 \\
PLCKG285.6-17.2 & 285.64 & -17.25 & 6.35 & null & 0.12 & 17.68 & 0.0016 & 0.0003 & NaN & NaN & NaN \\
PLCKG225.9-20.0 & 225.93 & -20.00 & 8.07 & null & NaN & NaN  & NaN & NaN & 28.21 & 0.0040 & 0.0011 \\
PLCKG255.6-46.2 & 255.63 & -46.17 & 8.46 & null & NaN & NaN  & NaN & NaN & 31.23 & 0.0026 & 0.0006 \\
PLCKG304.8-41.4 & 304.84 & -41.42 & 7.58 & null & NaN & NaN  & NaN & NaN & 21.68 & 0.0022 & 0.0006 \\
PLCKG121.1+57.0 & 121.12 & 57.01 & 6.66 & null & NaN & NaN  & NaN & NaN & 17.99 & 0.0016 & 0.0004 \\
PLCKG283.2-22.9 & 283.16 & -22.93 & 6.03 & null & NaN & NaN  & NaN & NaN & 26.73 & 0.0018 & 0.0008 \\
PLCKG139.6+24.2 & 139.60 & 24.19 & 7.21 & null & NaN & NaN  & NaN & NaN & 24.52 & 0.0032 & 0.0013 \\
PLCKG189.8-37.2 & 189.85 & -37.24 & 6.71 & null & NaN & NaN  & NaN & NaN & 62.50 & 0.0080 & 0.0021 \\
PLCKG264.4+19.5 & 264.42 & 19.48 & 6.15 & null & NaN & NaN  & NaN & NaN & 32.25 & 0.0028 & 0.0010 \\
PLCKG115.7+17.5 & 115.72 & 17.53 & 6.78 & null & NaN & NaN  & NaN & NaN & 17.48 & 0.0025 & 0.0008 \\
\hline
\hline
\end{tabular}
\end{table*}

\end{appendix}
\raggedright

\end{document}